\documentclass[a4paper,usenatbib]{mnras}

\usepackage{newtxtext,newtxmath}
\usepackage[T1]{fontenc}
\usepackage{ae,aecompl}

\usepackage{graphicx} % Including figure files
\usepackage{float}
\usepackage{amsmath}	% Advanced maths commands
\usepackage{amssymb}	% Extra maths symbols
\usepackage{dcolumn}
\usepackage{epsfig}
\usepackage{color}
\usepackage{pdflscape}
\usepackage{upgreek}
\usepackage{mathptmx}
\usepackage{grffile}
\usepackage{array,float}
\usepackage{multirow}
\usepackage{lscape}
\usepackage{hyperref}
\usepackage{xargs}
\usepackage{xcolor}
\usepackage[normalem]{ulem}
\usepackage{booktabs, caption}
\usepackage{threeparttable}
\newcolumntype{d}[1]{D{.}{\cdot}{#1}}

\newcolumntype{.}{D{.}{.}{-1}}

\newcommand{\lsun}{L$_\odot$}
\newcommand{\msun}{M$_\odot$}

\newcommand{\kms}{km\,s$^{-1}$}

\newcommand{\hii}{H{\sc ii}~}

\newcommand{\htwoo}{H$_2$O}
\newcommand{\nhthree}{NH$_3$~}

\newcommand{\fw}{\texttt{FellWalker}}

\title[The RMS Survey: Ammonia Mapping]{The RMS Survey: Ammonia mapping of the environment of young massive stellar objects II\thanks{The full version of Tables\,\ref{table:fellwalker} and \ref{table:fitted_parameters_table_sub} are only available in electronic form at the CDS via anonymous ftp to cdsarc.u-strasbg.fr (130.79.125.5) or via http://cdsweb.u-strasbg.fr/cgi-bin/qcat?J/MNRAS/.}}

\author[S.\,J.\,Billington et al.]{
S.\,J.\,Billington,$^{1}$\thanks{E-mail: sjbb2@kent.ac.uk} J.\,S.\,Urquhart,$^{1}$ C.\,Figura,$^{2}$ D.\,J.\,Eden,$^{3}$ T.\,J.\,T.\,Moore$^{3}$\\
\\
$^{1}$ Centre for Astrophysics and Planetary Science, University of Kent, Canterbury CT2\,7NH, UK \\
{\color{black}$^{2}$ Wartburg College, Waverly, IA 50677, USA}\\
$^{3}$ Astrophysics Research Institute, Liverpool John Moores University, IC2, Liverpool Science Park, 146 Brownlow Hill, Liverpool L3 5RF, UK\\}
\date{Accepted XXX. Received YYY; in original form ZZZ}

\pubyear{2018}

\begin{document}
\label{firstpage}
\pagerange{\pageref{firstpage}--\pageref{lastpage}}
\maketitle

\begin{abstract}

We present the results from \nhthree mapping observations towards 34 regions identified by the Red MSX Source (RMS) survey. We have used the Australia Telescope Compact Array to map ammonia (1,1) and (2,2) inversion emission spectra at a resolution of 10\arcsec\ with velocity channel resolution of 0.4\,\kms\ towards the positions of embedded massive star formation. Complementary data have been used from the ATLASGAL and GLIMPSE Legacy Surveys in order to improve the understanding of the regions and to estimate physical parameters for the environments. The fields have typical masses of $\sim$1000\,\msun, radii of $\sim$0.15\,pc and distances of $\sim$3.5\,kpc. Luminosities range between $\sim$10$^{3}$ to $\sim$10$^{6}$\,\lsun\ and kinetic temperatures between 10 and 40\,K. We classify each field into one of two subsets in order to construct an evolutionary system for massive star formation in these regions based on the morphology and relative positions of the \nhthree emission, RMS sources and ATLASGAL thermal dust emission. Differences in morphology between \nhthree emission and ATLASGAL clumps are shown to correspond to evolutionary stages of ongoing massive star formation in these regions. The study has been further refined by including the positions of known methanol and water masers in the regions to gain insight into possible protostellar regions and triggered star formation.

\end{abstract}


%%%%%%%%%%%%%%%%%%%%%%%%%%%%%%%%%%%%%%%%%%%%%%%%%%

%%%%%%%%%%%%%%%%% BODY OF PAPER %%%%%%%%%%%%%%%%%%

\section{Introduction}

Massive stars ($>8$\,M$_{\sun}$) play a vital role in the evolution of the Galaxy due to their powerful outflows, strong stellar winds, large amounts of UV radiation and chemical enrichment of material of the interstellar medium (ISM). These feedback processes can have a dramatic impact on the local environment by changing the chemistry and injecting huge amounts of energy into the ISM. They are also responsible for driving strong shocks into the surrounding molecular clouds, which can have a direct effect on future generations of stars through the compression and subsequent collapse of molecular structures often referred to as triggering, e.g. (\citealt{Urquhart2007b,Deharveng2010,thompson2012}), or by disrupting clouds before the star formation has begun. Massive stars therefore play an important role in regulating future generations of stars and driving the evolution of their host galaxies (\citealt{kennicutt2005}).

Even though massive stars have a profound influence on the universe, our knowledge and understanding of how these celestial objects form is still fairly poor for a number of reasons.  Massive stars form almost exclusively in compact OB clusters or large OB associations, making it difficult to differentiate between the properties of a single star and the cluster itself. Their relative rarity compared to lower-mass stars results in regions of high-mass star formation being located at greater distances (generally farther than a few kpc away), exacerbating source confusion. Furthermore, these objects evolve rapidly (their Kelvin-Helmholtz timescale is much shorter than their free-fall collapse timescale).  They reach the main sequence while still deeply embedded in their natal cloud, so the earliest stages in their evolution take place behind many hundreds of magnitudes of visual extinction and can only be studied at far-infrared and submillimetre wavelengths.

Two of the earliest stages in the formation of OB stars are massive young stellar objects (MYSOs) (\citealt{Lumsden2013}) and the compact \hii regions (\citealt{wood1989a,kurtz1994,urquhart2013_cornish}). During the MYSO stage the newly-formed protostar begins to heat up its envelope and becomes detectable at mid-infrared wavelengths. This stage ends when the star joins the main sequence and begins to ionize its natal molecular cloud, leading to the creation of an ultra compact (UC) \hii region. Both of these stages are physically distinct from the earlier hot molecular core stage, which is generally not detectable at mid-infrared wavelengths (\citealt{de-buizer2002}).

The Red MSX Source (RMS) survey \citep{Lumsden2013}, has identified $\sim$3000 MYSOs and \hii regions candidates located throughout the Galactic plane. These sources were initially identified from their mid-infrared colours using the MSX point source catalog \citep{Price2001} and 2MASS data (\citealt{cutri2003}). The nature of these candidates were later confirmed through an extensive multi-wavelength follow-up campaign (e.g., 
\citealt{urquhart_radio_south,urquhart_13co_south,Mottram2007c,urquhart_13co_north,Urquhart2009b,urquhart_radio_north,urquhart2011_nh3,Cooper2013}). This survey has identified $\sim$1700 MYSOs and \hii regions and is the largest and most well-characterised catalogue of these stages of massive star formation that has been compiled with determined distances (\citealt{urquhart2012_hiea,urquhart2014_rms}) and luminosities (\citealt{mottram2011b,mottram2011a,urquhart2014_rms}) and is an order of magnitude larger than any previous study. The RMS catalogue sample is therefore an ideal starting point for more detailed studies of these massive star forming regions.

This is the second RMS paper that used ammonia mapping observations to investigate the structure and properties of high-mass star forming clumps. The first paper (\citealt{Urquhart2015}; hereafter Paper\,I) followed up 66 high-mass star forming clumps with the Green Bank Telescope (GBT) with an angular resolution of $\sim30$\arcsec.  In this paper we present ammonia observations towards another 34 massive star forming regions located in the fourth quadrant. This subsample of MYSOs and \hii regions drawn from the RMS survey has been mapped using the lowest inversion emission spectra of interstellar ammonia (i.e. \nhthree (J,K) = (1,1) and (2,2)). 
Ammonia does not deplete from the gas phase in the cold (T$\sim$10-40\,K; \citealt{Ho1983,Mangum1992}) and high-density (n$>$10$^4$\,cm$^{-3}$) conditions that are typical (20-30\,K; \citealt{urquhart2011_nh3}, Paper\,I) of these natal clumps, making it an excellent probe of the physical conditions at this stage of star formation.

Ammonia observations have been used by the H$_2$O Southern Galactic Plane Survey  (HOPS; \citealt{walsh2011,purcell2012,longmore2017}) to map the distribution of dense star forming gas; other studies have used it to investigate the physical properties and kinematics of different kinds of star formation environments in the Galaxy, such as infrared dark clouds (e.g., \citealt{Perault1996,pillai2006, ragan2011, chira2013}) and high-mass star forming regions (e.g., \citealt{dunham2011b}, \citealt{urquhart2011_nh3}, Paper\,I). The main reason for the wide-spread use of ammonia transitions is that the hyperfine structure of the emission can be used to derive multiple free parameters allowing for an in-depth analyses of the kinematics, morphology and thermodynamics of such regions.

The main aims of this study are to compare the physical properties of different stages of high-mass star formation and investigate how feedback from outflows and UV-radiation can affect the local environment as the embedded high-mass stars evolve. The structure of the paper is as follows: Section~\ref{sect:sample} outlines our sample selection along with archival data sets used to complement this study, in Section~\ref{sect:obs} we describe the observational setup and reduction procedures used to process the data. We outline the source extraction method used to identify sources and the classification system which has been applied in Section~\ref{sect:method}. We describe our spectral line analysis and fitting procedure and how the various physical parameters are derived in Section~\ref{sect:physical_properties}. In Section~\ref{sect:physical_discussion} we discuss the differences between the classification groups with respect to multiple parameters from the spectral line analysis and results reported by other surveys (Methanol Multibeam (MMB) survey (\citealt{caswell2010_mmb}), HOPS and ATLASGAL (\citealt{Schuller2009a})) and also how the regions are being affected by their local environments. The study is then summarised and concluded in Section~\ref{sect:conclusions}.

\section{Sample selection and archival data}
\label{sect:sample}

The sample selection for this study was based on RMS objects towards which both strong \nhthree (1,1) and (2,2) inversion transition emission has previously been detected. The observational data set along with data reduction techniques is outline in the Section~\ref{sect:obs}. The majority of objects that were selected lie within the inner Galactic plane and so are complemented by a wealth of additional data provided by other surveys such as ATLASGAL, HOPS, GLIMPSE and MMB. We have therefore taken complementary data from these Galactic surveys in order to improve the understanding of the surrounding environments for the regions that are presented herein.

\subsubsection{GLIMPSE Images}
\label{sect:glimpse_images}

We have used images from the GLIMPSE Legacy Survey \citep{benjamin2003,churchwell2009} in order to investigate the surrounding environments at mid-infrared wavelengths. Three-colour images have been created using data at IRAC 3.4, 4.5 and 8\,\micron\ wavelength bands from the Spitzer Space Telescope \citep{Fazio1998}.

\begin{figure*}
	\includegraphics[width=0.45\textwidth]{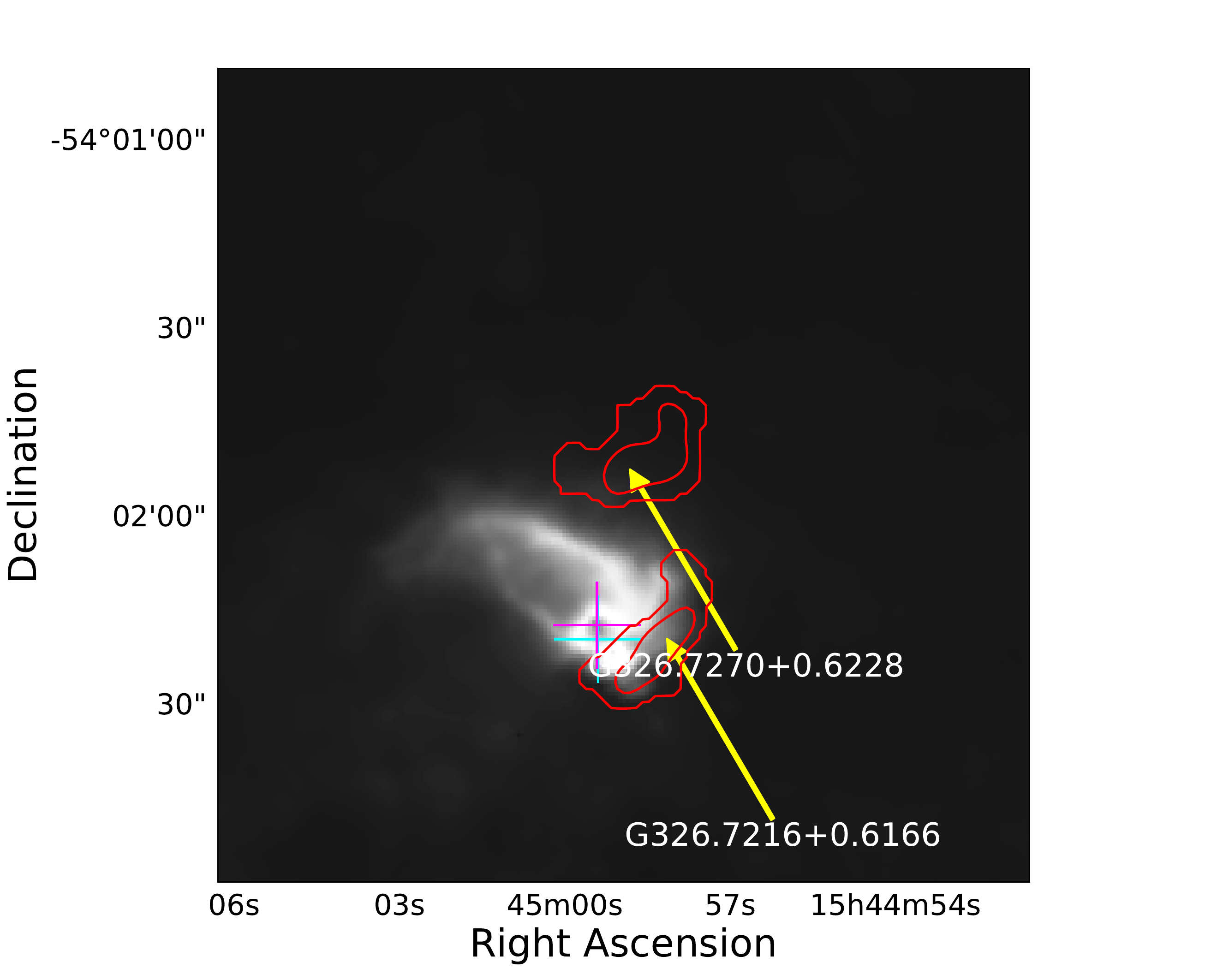}
	\includegraphics[width=0.45\textwidth]{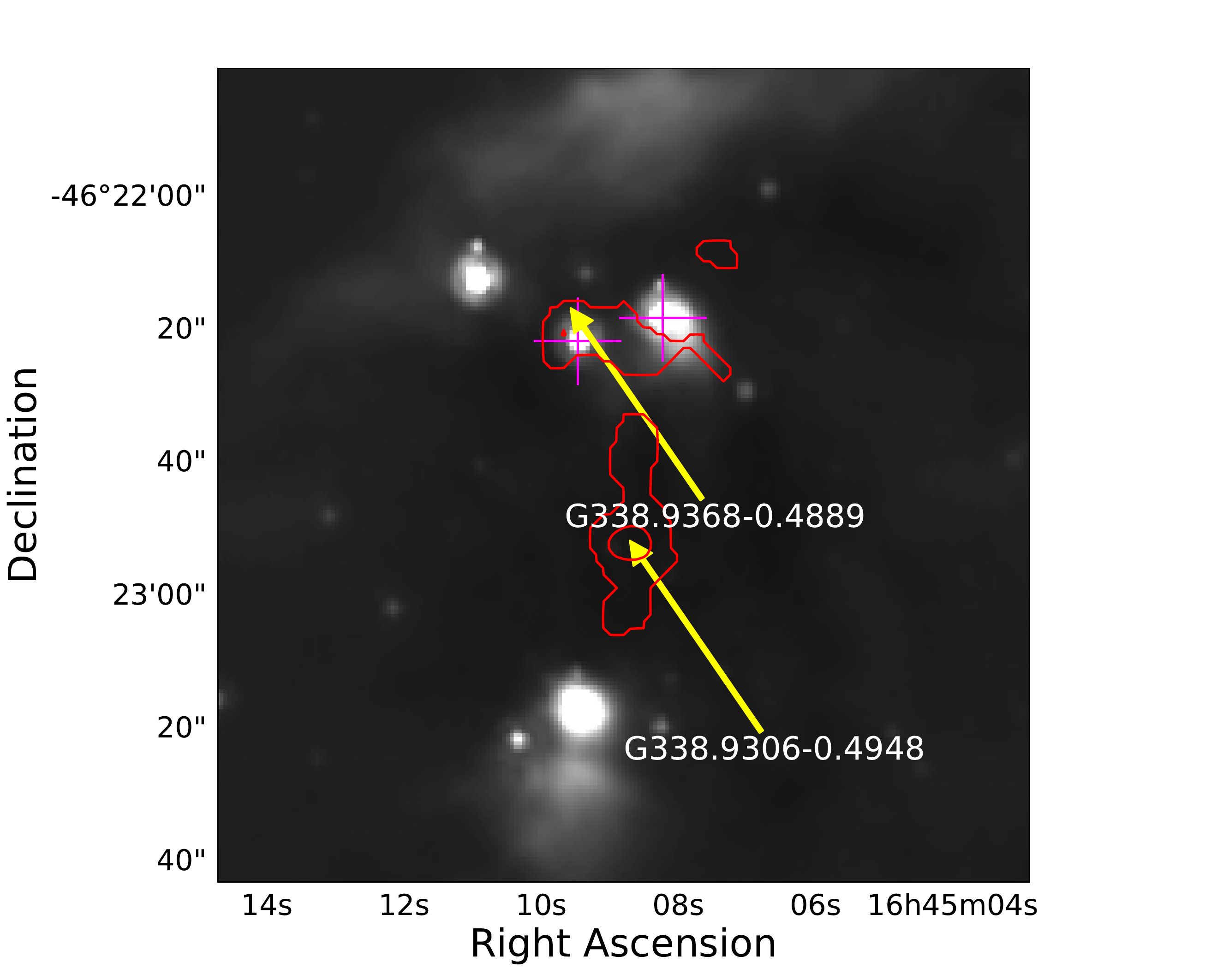} \\
	\includegraphics[width=0.45\textwidth]{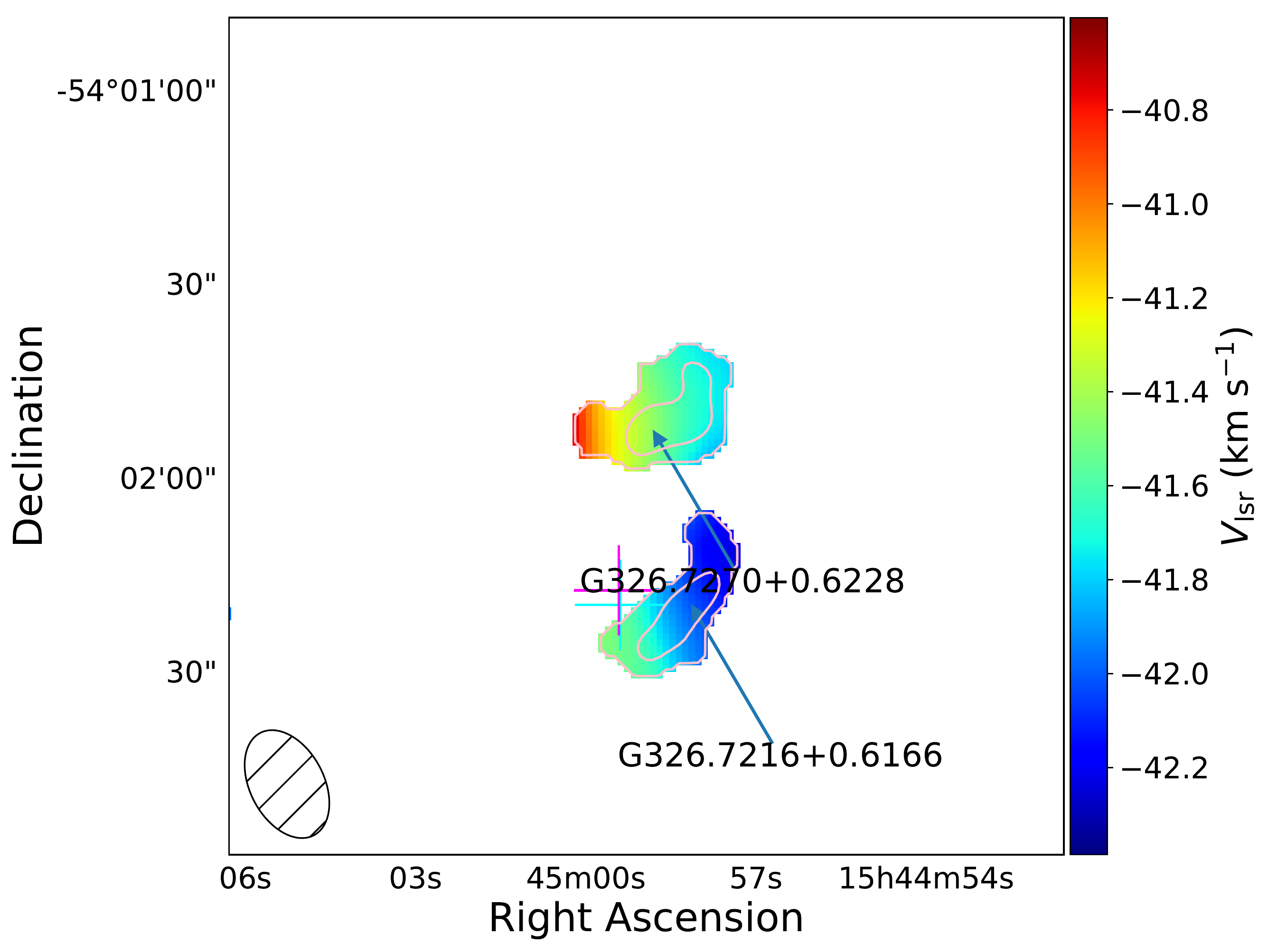}
	\includegraphics[width=0.45\textwidth]{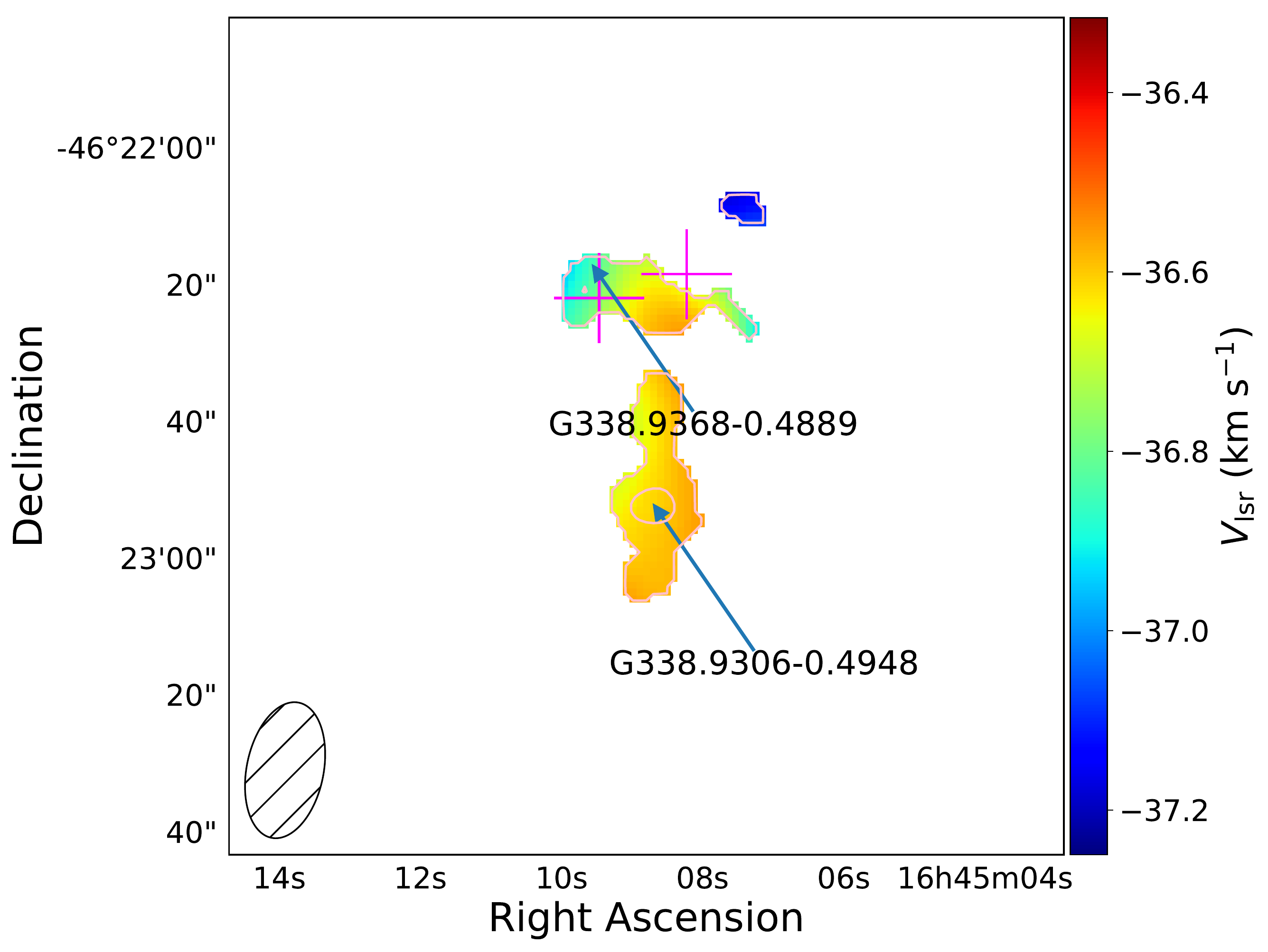}
	\caption{The upper panels present example 8\,\micron\ images taken from the GLIMPSE Legacy survey, while the lower panels show the velocity maps for the same regions. Magenta and cyan crosses mark the positions of any MYSOs or \hii regions respectively.  \nhthree (1,1) integrated emission contours are overplotted on both sets of images, the contours begin at 3$\sigma$ and increase in steps of 0.5$\sigma$ (Values for $\sigma$/r.m.s noise can be found in Table~\ref{table:field_parameters_sub}).}
	\label{fig:8mu_examples}
\end{figure*}

These images are capable of revealing the position of embedded objects with respect to the ammonia emission, such as MYSOs and compact \hii regions under investigation here. The 8\,\micron\ band is also sensitive to the emission from polycyclic aromatic hydrocarbons (PAHs) that have been excited by UV radiation of embedded or nearby \hii regions, making it an excellent tracer of the boundaries between molecular and ionized gas \citep{Urquhart2007b} (right panel of Figure\,\ref{fig:8mu_examples}). Additionally, \cite{Cyganowski2008} have used the 4.5\,\micron\ band images in order to create a catalog of extended green objects. These objects have an excess of 4.5\,\micron\ emission, which is thought to be result of shock-excited H$_2$\,($v=0-$0)\,S(9, 10, 11) lines and/or CO\,($v=1-$0) bandhead \citep{churchwell2009} associated with molecular outflows from MYSOs, and therefore considered to be a good indicators for ongoing massive star formation. 

These mid-infrared images can therefore provide a useful overview of the position of the embedded MYSOs and \hii regions, the structure of their host clumps and their local environment.

GLIMPSE images are presented throughout the paper, and for each region which lies within the GLIMPSE survey coverage, their corresponding RGB image is given in the appendix. The wavelengths used for the three-colour images are 8, 4.5 and 3.4\micron\ for the red, green and blue channels respectively.

\subsubsection{ATLASGAL Dust Emission}
\label{sect:atlas_dust_emission}

The APEX Large Area Survey of the Galaxy (ATLASGAL) survey \citep{Schuller2009a,Contreras2013a,csengeri2014} has surveyed 420 sq. degrees of the Galactic plane, between $-80\degr \leq \ell \leq +60\degr$, and $-2\degr \leq b \leq +1.5\degr$ at 870\,\micron\ (345\,GHz). ATLASGAL has traced dust emission across the Galactic plane and the Compact Source Catalogue (CSC; \citealt{Contreras2013a,urquhart2014_csc}) produced from these maps consists of $\sim$10,000 compact clumps. The ATLASGAL survey covers 33 out of 34 of our observed fields (G261.6429$-$02.0922 lies outside of the ATLASGAL region). We have used emission contours overlaid onto the three-colour GLIMPSE images to compare the distribution of interstellar dust with the integrated ammonia emission and the mid-infrared environment for each observed field. These data are used to build a detailed picture of these star-forming environments and examine their physical properties, such as clump mass and luminosity (see \citealt{Urquhart2018} for details of how these parameters are determined).

The clump masses and luminosities for 32 of the 33 fields covered by ATLASGAL range between $\sim$ 10$^{2.5}$-10$^{4}$\,M$_{\sun}$ and $\sim$ 10$^{3}$-10$^{6}$\,L$_{\sun}$, respectively. Although G286.2086+00.1694 is located in the ATLASGAL region it was not included in the study by \citet{Urquhart2018} and so the luminosity and clump masses were not available. These extracted data complement the sample by providing masses and luminosities of the entire regions as opposed to the small, denser substructures mapped by the \nhthree emission.

The GLIMPSE images show a range of environments as shown in Fig.\,\ref{fig:8mu_examples}. The upper left panel presents a region with an offset \hii region, which appears to be affecting the nearby dense gas, while the rest of the field appears to be quiescent. This source highlights the distribution of 8\,\micron\ emission produced by polycyclic aromatic hydrocarbons (PAHs) excited in the interaction layers between ionization fronts and molecular gas. The upper right panel shows a much more evolved region with a number of evolved background stars, which some diffuse extended emission.

\subsubsection{Methanol MultiBeam Survey}

The Methanol MultiBeam (MMB) Survey \citep{Green2009} is an unbiased survey of the Galactic plane for 6668\,MHz methanol masers. The coverage for the survey is $186\degr \leq \ell \leq 60\degr$ with the majority of detections found between $|b|$ $\leq$ 1\degr. This species of astronomical maser is one of the most frequently detected, and is considered to be  exclusively associated with the early stages of high-mass star formation (e.g. \citealt{Minier2003}). Comparison between the MMB catalogue and the ATLASGAL CSC (\citealt{urquhart2013_methanol}) and a set of dedicated dust continuum observations  (\citealt{urquhart2015_mathanol}) found that 99\,per\,cent of methanol masers are associated with dense star-forming clumps. There is an MMB source present in 20 of the 34 fields observed ($\sim$59\,per\,cent). 

\subsubsection{HOPS}

The H$_2$O Southern Galactic Plane Survey (HOPS; \citealt{walsh2011}) has mapped 100 square degrees of the southern Galactic plane, between $290\degr \leq \ell \leq 30\degr$ and $|b|$ $\leq$ 0.5\degr, at 19.5 to 27.5\,GHz. They have detected over 540 groups of H$_2$O masers in this region, which have been followed up at high resolution with the ATCA to get accurate positions ($<1\arcsec$; \citealt{walsh2014}).

H$_2$O masers are excellent tracers of shocked gas often associated with molecular outflows from protostellar objects. They are known to occur in all regions of star formation \citep{Claussen1996, Forster1996}, within both high-mass and low-mass environments. While the majority of the currently known H$_2$O masers are associated with star-forming regions, they can also be found in other environments, including evolved stars \citep{Dickinson1976} and planetary nebulae \citep{Miranda2001}. H$_2$O masers are present in 11 of our fields: this constitutes 52\,per\,cent of the 21 fields that are covered by HOPS.

\subsection{Extracted properties}
\label{sect:extracted_parameters}

\setlength{\tabcolsep}{3pt}
\begin{table}
	\caption{Extracted field parameters from the ATLASGAL survey.}
	\begin{tabular}{clccc}
		\hline
		\hline
		Field    &         Field  &  Log[$L_{bol}$] &  Log[$M_{field}$] &  Log[$N$(H$_2$)] \\
		id    &         name   &  (L$_{\sun}$)     &  (M$_{\sun}$)       &  (cm$^{-2}$)   \\
		\hline
		1 &  G261.6429-02.0922 &  $\dots$ &     $\dots$ &  $\dots$ \\
		2 &  G286.2086+00.1694 &  $\dots$ &     $\dots$ &  $\dots$ \\
		3 &  G305.2017+00.2072 &    5.139 &       3.527 &   23.362 \\
		4 &  G309.4230-00.6208 &    3.327 &       3.075 &   22.810 \\
		5 &  G312.5963+00.0479 &    4.827 &       3.175 &   22.836 \\
		6 &  G314.3197+00.1125 &    4.077 &       3.176 &   22.748 \\
		7 &  G318.9480-00.1969 &    3.871 &       2.523 &   23.092 \\
		8 &  G322.1729+00.6442 &    5.466 &       3.736 &   23.416 \\
		9 &  G323.4584-00.0787 &    5.074 &       3.106 &   22.873 \\
		10 &  G326.4755+00.6947 &    3.733 &       2.926 &   23.513 \\
		11 &  G326.7249+00.6159 &    4.545 &       2.686 &   23.079 \\
		12 &  G327.3941+00.1970 &    3.938 &       3.246 &   22.928 \\
		13 &  G327.4014+00.4454 &    4.761 &       3.587 &   23.321 \\
		14 &  G328.2523-00.5320 &    4.665 &       3.597 &   23.310 \\
		15 &  G328.3067+00.4308 &    5.828 &       3.705 &   23.076 \\
		16 &  G328.8074+00.6324 &    5.128 &       3.225 &   23.511 \\
		17 &  G330.9288-00.4070 &    3.371 &       2.816 &   22.894 \\
		18 &  G332.2944-00.0962 &    4.303 &       3.150 &   23.133 \\
		19 &  G332.9868-00.4871 &  $\dots$ &     $\dots$ &  $\dots$ \\
		20 &  G333.0058+00.7707 &    4.460 &       3.653 &   23.331 \\
		21 &  G333.0682-00.4461 &  $\dots$ &     $\dots$ &  $\dots$ \\
		22 &  G333.1075-00.5020 &    4.601 &       3.251 &   22.557 \\
		23 &  G336.3684-00.0033 &    4.794 &       3.736 &   23.072 \\
		24 &  G338.9196+00.5495 &    4.963 &       3.975 &   23.571 \\
		25 &  G338.9377-00.4890 &    3.475 &       2.863 &   22.668 \\
		26 &  G339.5836-00.1265 &    3.441 &       2.829 &   23.016 \\
		27 &  G339.9267-00.0837 &    3.605 &       3.064 &   22.900 \\
		28 &  G340.7455-01.0021 &    3.770 &       2.848 &   22.822 \\
		29 &  G341.2182-00.2136 &    4.030 &       2.775 &   22.933 \\
		30 &  G342.7057+00.1260 &    4.440 &       3.369 &   23.112 \\
		31 &  G343.5024-00.0145 &    4.466 &       3.050 &   23.016 \\
		32 &  G343.9033-00.6713 &    3.132 &       2.707 &   22.556 \\
		33 &  G344.4257+00.0451 &    5.389 &       3.619 &   22.857 \\
		34 &  G345.5043+00.3480 &  $\dots$ &     $\dots$ &  $\dots$ \\
		\hline
	\end{tabular}
	\label{table:atlas_values}
\end{table}

Distances have been determined for all of the fields either by \cite{urquhart2014_rms} or from the \citet{Reid2016} Bayesian model (see discussion presented in Section~\ref{sect:dists_radii}).

Due to the large uncertainties in the derived column densities, we have used the dust clump masses derived from the ATLASGAL survey \citep{Urquhart2018} in our analysis. These values are naturally higher than expected for the \nhthree emission itself as the thermal dust emission encompasses a larger area.
It is assumed, when using these values, that the densities of the regions are sufficiently high that the gas and dust are well-coupled and in local thermodynamic equilibrium. 

Values for both the mass and luminosity for each region have been taken from the ATLASGAL compact-source catalogue \citep{Urquhart2018}, and are used to give a global overview of the regions. While the \nhthree emission can be used to derive parameters for individual clumps, the dust emission values are representative of the entire regions and surrounding environmental material. Bolometric luminosities for the sample range from $\sim$1300 to $\sim$670,000\,L$_{\sun}$ and masses range from $\sim$330 to $\sim$9500\,M$_{\sun}$.

There is no associated value for either mass or luminosity for five of our fields from the ATLASGAL survey, as these regions either lay outside the survey's coverage or due to a non-detection in the compact source catalogue.

\subsubsection{Uncertainties in extracted parameters}

The distances that have been assigned to each field are kinematic and so have an associated uncertainty of $\pm$1\,kpc.  This uncertainty is mainly caused by peculiar motions of the clouds through the spiral arms of the Galaxy, which causes them to deviate from the rotation models.  These are commonly referred to as ``streaming motions'' and can lead to perturbation from the expected radial velocities of $\pm7$\,\kms\ (\citealt{reid2009}).

The ATLASGAL dust masses are estimated to be correct to within a factor of 2-3: this is mainly due to the fact that many of the parameters involved in the calculation are poorly constrained, such as the dust-to-gas ratio and the dust absorption coefficient (\citealt{urquhart2013_methanol}). Two other parameters additionally affect the mass uncertainty: the kinetic or dust temperatures as derived from the spectral line analysis, with an approximate error of $\pm$1.5\,K, and the uncertainty in the distance measurements as mentioned in the previous paragraph.

The main source of uncertainty for the luminosity values arises from the distance uncertainties and bolometric flux calculations which results from the fitting of the spectral energy distributions for each region: these are estimated to be no more than a factor of two.

While the uncertainties may be large for the absolute values for these parameters, the whole sample are uniformly affected and so the properties should provide statistically robust results.

\section{Observational data}
\label{sect:obs}

\begin{table*}
\begin{threeparttable}
\caption{\label{table:field_parameters_sub}Observed field parameters. Distances appended with $\star$ identify values obtained using a Bayesian distance estimator describe in \citet{Reid2016}.}
\centering
\begin{tabular}{cccccccccccc}
\hline
\hline
Field & Field & RA & Dec &  Number &   $v_{\rm{lsr}}$ & Distance & r.m.s noise & Beam$_{\rm{maj}}$ & Beam$_{\rm{min}}$ \\
id & name & (J2000) & (J2000) & of clumps & (km s$^{-1}$) & (kpc) & (Jy\,beam$^{-1}$) & (\arcsec) & (\arcsec) \\
\hline
        1 &  G261.6429-02.0922 &  08h32m07.46s &  -43d13m48.70s &          1 &   14.5 &       2.0 &  0.25 &      9.36 &      5.57 \\
        2 &  G286.2086+00.1694 &  10h38m32.70s &  -58d19m14.30s &          2 &  -21.1 &       3.0 &  0.30 &     10.28 &      4.94 \\
        3 &  G305.2017+00.2072 &  13h11m10.45s &  -62d34m38.60s &          1 &  -42.4 &       3.8 &  0.45 &      6.74 &      5.97 \\
        4 &  G309.4230-00.6208 &  13h48m38.86s &  -62d46m09.50s &          1 &  -42.4 &       3.5 &  0.38 &      7.66 &      5.40 \\
        5 &  G312.5963+00.0479 &  14h13m14.12s &  -61d16m48.90s &          1 &  -63.7 &       6.0 &  0.32 &      8.70 &      5.07 \\
        6 &  G314.3197+00.1125 &  14h26m26.28s &  -60d38m31.50s &          2 &  -47.5 &       4.2 &  0.33 &      7.61 &      5.47 \\
        7 &  G318.9480-00.1969 &  15h00m55.10s &  -58d59m06.00s &          1 &  -34.4 &       2.1 &  0.41 &      6.88 &      5.83 \\
        8 &  G322.1729+00.6442 &  15h18m38.29s &  -56d37m30.90s &          1 &  -57.7 &       3.3 &  0.49 &      7.90 &      5.34 \\
        9 &  G323.4584-00.0787 &  15h29m19.36s &  -56d31m21.70s &          1 &  -66.8 &       4.0 &  0.29 &      7.72 &      6.19 \\
        10 &  G326.4755+00.6947 &  15h43m18.94s &  -54d07m35.40s &          1 &  -41.2 &       1.8 &  0.58 &      8.78 &      5.77 \\
        11 &  G326.7249+00.6159 &  15h44m59.39s &  -54d02m19.60s &          2 &  -41.7 &       1.8 &  0.33 &      8.95 &      5.74 \\
        12 &  G327.3941+00.1970 &  15h50m20.07s &  -53d57m07.10s &          1 &  -89.3 &       5.2 &  0.39 &      9.08 &      5.62 \\
        13 &  G327.4014+00.4454 &  15h49m19.36s &  -53d45m14.40s &          1 &  -78.4 &       5.0 &  0.52 &      8.80 &      5.66 \\
        14 &  G328.2523-00.5320 &  15h57m59.82s &  -53d58m00.40s &          3 &  -45.1 &       2.7 &  0.44 &      8.08 &      6.01 \\
        15 &  G328.3067+00.4308 &  15h54m06.34s &  -53d11m39.20s &          1 &  -93.2 &       5.8 &  0.33 &      8.44 &      5.93 \\
        16 &  G328.8074+00.6324 &  15h55m48.36s &  -52d43m06.80s &          2 &  -41.7 &       2.7 &  0.52 &      9.30 &      5.55 \\
        17 &  G330.9288-00.4070 &  16h10m45.07s &  -52d05m50.20s &          1 &  -41.2 &       2.6 &  0.30 &      9.22 &      5.40 \\
        18 &  G332.2944-00.0962 &  16h15m45.86s &  -50d56m02.40s &          1 &  -48.9 &       3.1 &  0.33 &      9.52 &      5.47 \\
        19 &  G332.9868-00.4871 &  16h20m37.81s &  -50d43m49.60s &          1 &  -52.8 &       3.6 &  0.42 &      9.66 &      5.44 \\
        20 &  G333.0058+00.7707 &  16h15m13.79s &  -49d48m52.00s &          1 &  -49.2 &       3.0 &  0.62 &      9.71 &      5.62 \\
        21 &  G333.0682-00.4461 &  16h20m48.95s &  -50d38m40.30s &          1 &  -53.3 &       3.6 &  0.59 &      9.37 &      5.50 \\
        22 &  G333.1075-00.5020 &  16h21m14.22s &  -50d39m12.60s &          2 &  -56.5 &       3.6 &  0.38 &     10.10 &      5.27 \\
        23 &  G336.3684-00.0033 &  16h32m56.46s &  -47d57m52.30s &          1 & -126.7 &       6.7 &  0.54 &     10.48 &      5.74 \\
        24 &  G338.9196+00.5495 &  16h40m34.04s &  -45d42m07.90s &          1 &  -62.8 &       4.2 &  1.21 &      9.91 &      5.67 \\
        25 &  G338.9377-00.4890 &  16h45m08.80s &  -46d22m17.00s &          2 &  -36.6 &       2.9 &  0.36 &     10.08 &      5.64 \\
        26 &  G339.5836-00.1265 &  16h45m58.48s &  -45d38m41.40s &          1 &  -34.3 &       2.6 &  0.37 &      9.75 &      5.28 \\
        27 &  G339.9267-00.0837 &  16h47m03.94s &  -45d21m20.50s &          1 &  -52.8 &       3.6 &  0.34 &      9.90 &      5.56 \\
        28 &  G340.7455-01.0021 &  16h54m04.05s &  -45d18m50.00s &          1 &  -29.1 &       2.4 &  0.36 &     10.08 &      5.71 \\
        29 &  G341.2182-00.2136 &  16h52m17.93s &  -44d26m53.00s &          1 &  -42.9 &       3.3 &  0.50 &      9.84 &      5.45 \\
        30 &  G342.7057+00.1260 &  16h56m02.91s &  -43d04m43.90s &          1 &  -41.0 &       3.4 &  0.46 &     10.05 &      5.50 \\
        31 &  G343.5024-00.0145 &  16h59m20.90s &  -42d32m38.40s &          2 &  -27.9 &       2.6 &  0.45 &     10.20 &      5.57 \\
        32 &  G343.9033-00.6713 &  17h03m30.11s &  -42d37m48.60s &          1 &  -29.4 &       2.2 &  0.36 &     10.69 &      6.04 \\
        33 &  G344.4257+00.0451 &  17h02m09.35s &  -41d46m44.30s &          2 &  -66.6 &       4.9 &  0.45 &     10.61 &      5.35 \\
        34 &  G345.5043+00.3480 &  17h04m22.87s &  -40d44m23.50s &          1 &  -17.7 &       2.4 &  0.66 &     13.27 &      4.81 \\

\hline
\end{tabular}
\end{threeparttable}
\end{table*}

\subsection{ATCA observations and data reduction}
\label{sect:atca_obs}

Observations were made of the \nhthree (1,1) and (2,2) inversion transitions towards 34 RMS identified MYSOs and \hii regions (see Table\,\ref{table:field_parameters_sub} for details of observed fields). These observations were made between 15-22 February 2011 (Project Id: C2369; \citealt{2010atnf}). These observations were conducted using the Australia Telescope National Facilities' (ATNF) Australia Telescope Compact Array (ATCA)\footnote{The
Australia Telescope Compact Array is part of the Australia Telescope National Facility, which is funded by the Commonwealth of
Australia for operation as a National Facility managed by CSIRO.}.
The ATCA comprises of six 22\,m diameter antennas, with five lying on a 3\,km long east-west track with the sixth antenna being in a fixed position 3\,km west of the track.

The array was set up in an east-west 352 configuration, utilising five of the antennae in a compact configuration with shortest and longest baselines of 31 and 352\,metres respectively. The sixth antenna was not used for this study due to the large gap in $uv$-coverage. The observations were made with the Australia Telescope Compact Array Broad-band Backend (CABB; see \citealt{wilson2011} for details).

This provides a primary beam size of $\sim$2\arcmin\ (FWHM field of view) and a synthesised beam of 5-10\arcsec\ (FWHM resolution of the observations). We used a 64\,MHz spectral window covering 23.6945 - 23.7226\,GHz so as to include the \nhthree (1,1) and (2,2) transition in the same bandpass. Each source was observed for approximately 60 minutes, providing a velocity resolution of 0.4\,\kms\,channel$^{-1}$ and a sensitivity of 2.3\,K\,channel$^{-1}$\,beam$^{-1}$.

We observed 1934$-$638 and 1253$-$055 once per day for absolute flux and bandpass calibration. The target sources were separated into groups of $\sim$8 closely located sources to allow the sharing of phase calibrators and minimise observing overheads. An appropriate phase calibrator was selected for each group and the observations of the target sources were sandwiched in between observations of the phase calibrator. These observation blocks arranged to be less than an hour to allow the phase calibrators to be observed at regular intervals (typically 2-3 minutes every hour) throughout the observing session. These allow us to correct for fluctuations in the phase and amplitude of the data caused by atmospheric and instrumental effects throughout the observations.

The calibration and reduction of these data were performed using the MIRIAD reduction package (\citealt{sault1995}) following standard ATCA procedures. We initially imaged a region twice the size of the primary beam, choosing a pixel size to provide $\sim$10 pixels across the synthesised beam (1\arcsec\ pixels). We imaged a velocity range of 120\,\kms\ centered on the systemic velocity of the RMS source using the native velocity resolution of the spectrometer. This resulted in spectral line cubes of $256\arcsec\times256\arcsec\times300$\,\kms\ for each transition.

These cubes were deconvolved using a robust weighting of 0.5 using a couple of hundred cleaning components per velocity channel, or until the first negative component was encountered. Weighting values less than -2 correspond to minimising sidelobe levels only (uniform weighting), whereas values greater than +2 minimise noise levels (natural weighting). A value of 0.5 gives nearly the same sensitivity as natural weighting, but with a significantly better beam. These cubes were inspected to identify bright ammonia peaks in the field, and when detected in a map, these were integrated along the velocity axis in order to produce a high signal-to-noise ratio (SNR) map of the ammonia emission. These maps allow the peak position of the emission (taken to be the centre of the clump) and morphology of the dense gas traced by the ammonia emission to be discerned and spatially compared with the position of their embedded MYSOs and/or \hii regions.

The largest well-imaged structure possible at this frequency from these snapshot observations is limited to approximately 1\arcmin\ due to the limited $uv$-coverage and integration time. However, many of our maps displayed evidence of large-scale emission, which, when undersampled, can distort the processed images and lead to prominent imaging artifacts and confusion in the processed maps. This can reduce the SNR in the maps and over-resolve large-scale extended emission, breaking it up into irregular and/or multiple-component structures.

\section{Methods and Source Identification}
\label{sect:method}

\subsection{Source extraction}

\begin{table*}
\begin{threeparttable}
\caption{\label{table:fellwalker}\fw\ source catalogue for detected clumps. All parameters that have been derived in the table have been obtained using the \nhthree emission maps. Columns: (1) Field identification number; (2) Clump name derived from Galactic coordinates of the peak emission of the source; (3)-(4) Right ascension and declination of source in J2000 coordinates; (5)-(6) Semi-major and semi-minor axis of clumps; (7) Aspect ratio ($\sigma_{maj}$ / $\sigma_{min}$); (8) Radius for each clump; (9)-(10) Sum and peak emission values; (11) Signal to noise ratio.}
\begin{tabular}{ccccccccccc}
\hline
\hline
 Field &        Clump name &            RA &            Dec &    $\sigma_{maj}$ &   $\sigma_{min}$ &  Aspect & Radius &   Sum &   Peak &  SNR \\
 id & & (J2000) & (J2000) & (\arcsec) & (\arcsec) & Ratio & (pc) & (K \kms) & (K \kms) & \\
(1) & (2) & (3) & (4) & (5) & (6) & (7) & (8) & (9) & (10) & (11)\\

\hline
        1 &  G261.6454$-$2.0884 &  08h32m09.00s &  $-$43d13m47.57s &   5.43 &  3.75 &    1.45 & 0.07 & 409.24 &   1.90 &        7.21 \\
        2 &  G286.2124+0.1697   &  10h38m34.18s &  $-$58d19m17.65s &   4.13 &  2.62 &    1.57 & $\dots$ & 370.88 &   2.96 &        9.06 \\
        2 &  G286.2052+0.1706   &  10h38m31.49s &  $-$58d19m01.99s &   3.27 &  1.86 &    1.76 & $\dots$ & 137.05 &   2.16 &        6.24 \\
        3 &  G305.2084+0.2061   &  13h11m13.80s &  $-$62d34m41.23s &   4.65 &  3.87 &    1.20 & 0.14 & 1423.72 &  10.29 &       22.03 \\
        4 &  G309.4208$-$0.6204 &  13h48m37.80s &  $-$62d46m09.98s &   7.45 &  6.57 &    1.13 & 0.26 & 950.19 &   2.58 &        6.63 \\
        5 &  G312.5987+0.0449   &  14h13m15.38s &  $-$61d16m54.37s &   3.64 &  2.25 &    1.62 & $\dots$ & 231.04 &   2.33 &        6.53 \\
        6 &  G314.3214+0.1146   &  14h26m26.47s &  $-$60d38m20.51s &   4.31 &  3.37 &    1.28 & 0.12 & 262.18 &   2.11 &        6.18 \\
        6 &  G314.3197+0.1092   &  14h26m26.64s &  $-$60d38m40.85s &   2.33 &  1.45 &    1.61 & 0.09 &  53.54 &   1.64 &        4.08 \\
        7 &  G318.9477$-$0.1959 &  15h00m55.30s &  $-$58d58m52.14s &   5.94 &  4.99 &    1.19 & 0.12 & 1298.33 &   5.00 &       11.97 \\
        8 &  G322.1584+0.6361   &  15h18m34.63s &  $-$56d38m24.79s &   6.50 &  3.79 &    1.72 & 0.15 & 1096.10 &   5.67 &       11.18 \\
\hline
\end{tabular}
\begin{tablenotes}
\centering
\item Notes: Only a small portion of the data is provided here, the full table is only available in electronic format.
\end{tablenotes}
\end{threeparttable}
\end{table*}

For each source we have created \nhthree\ (1,1) and (2,2) emission intensity maps by integrating the velocity channels between $\pm$25\,\kms\ of the peak above a 3$\sigma$ threshold. These maps reveal the presence of high SNR clumps.

The \fw\ algorithm (\citealt{Berry2015,Moore2015}) has been applied to the \nhthree (1,1) integrated emission maps due to their higher SNR since these are more likely to trace the full extent of individual \nhthree clumps and their associated emission peaks. We have chosen to use the \fw\ algorithm as it has been widely used in recent studies (Paper\,I, \citealt{Eden2017}), and is robust against a wide choice of input parameters. It was required that all clumps be above a detection threshold of 3$\sigma$, where in this case $\sigma$ refers to the image r.m.s. level as determined from emission-free regions of the maps. We also required that clumps be larger than the beam size ($\sim$10 pixels) in order to avoid spurious detections. Detections towards the edges of the fields were also excluded due to lower SNRs, and because source parameters are likely to be poorly constrained; this did not prove to be a significant constraint, however, as emission is generally concentrated towards the centre of the fields. There appear to be ten fields which contain imaging artifacts that resulted from the poor sensitivity of the interferometric snapshot to large-scale extended emission (as mentioned in Section~\ref{sect:atca_obs}). This has been mitigated through the two thresholds required for a detected clump described above. An example integrated emission map is shown in Figure~\ref{fig:example_int_emission}.

\setlength{\tabcolsep}{3pt}
\begin{table}
\caption{Table of RMS associations with corresponding angular offset values for all detected clumps. All RMS names appended with * relate to RMS objects deeply embedded within the \nhthree clump shown.}
\begin{tabular}{clllr}
\hline
\hline
 Field    &        Clump &          RMS &   RMS  &  Angular \\
 id       &         name &         name &   Type &  Offset (\arcsec) \\
\hline
1 &  G261.6454$-$2.0884 &   G261.6429$-$02.0922 &  \hii region &   25 \\
2 &  G286.2124+0.1697 &   G286.2086+00.1694 &         YSO &   22 \\
2 &  G286.2052+0.1706 &   G286.2086+00.1694 &         YSO &   22 \\
3 &  G305.2084+0.2061 &  G305.2017+00.2072A &         YSO &   49 \\
4 &  G309.4208$-$0.6204 &   G309.4230$-$00.6208 &         YSO &    7 \\
5 &  G312.5987+0.0449 &   G312.5963+00.0479 &  \hii region &   21 \\
6 &  G314.3214+0.1146 &   G314.3197+00.1125 &         YSO &   11 \\
6 &  G314.3197+0.1092 &   G314.3197+00.1125 &         YSO &   14 \\
7 &  G318.9477$-$0.1959 &  G318.9480$-$00.1969A* &         YSO &    0 \\
8 &  G322.1584+0.6361 &   G322.1729+00.6442 &         YSO &   79 \\
9 &  G323.4537$-$0.0830 &   G323.4584$-$00.0787 &  \hii region &   27 \\
10 &  G326.4753+0.7030 &   G326.4755+00.6947 &         YSO &   43 \\
11 &  G326.7270+0.6228 &  G326.7249+00.6159B &         YSO &   26 \\
11 &  G326.7216+0.6166 &  G326.7249+00.6159A &  \hii region &   18 \\
12 &  G327.3933+0.1987 &   G327.3941+00.1970* &         YSO &   20 \\
13 &  G327.4030+0.4447 &   G327.4014+00.4454* &  \hii region &    3 \\
14 &  G328.2544$-$0.5318 &  G328.2523$-$00.5320A* &         YSO &    0 \\
14 &  G328.2610$-$0.5278 &  G328.2523$-$00.5320A &         YSO &   29 \\
14 &  G328.2607$-$0.5206 &  G328.2523$-$00.5320B &         YSO &   46 \\
15 &  G328.3021+0.4377 &   G328.3067+00.4308 &  \hii region &   48 \\
16 &  G328.8057+0.6347 &   G328.8074+00.6324 &  \hii region &   18 \\
17 &  G328.8109+0.6336 &   G328.8074+00.6324 &  \hii region &   17 \\
18 &  G330.9259$-$0.4066 &   G330.9288$-$00.4070* &  \hii region &   10 \\
19 &  G332.2962$-$0.0927 &   G332.2944$-$00.0962 &  \hii region &   11 \\
20 &  G332.9836$-$0.4885 &   G332.9868$-$00.4871* &         YSO &    9 \\
21 &  G333.0179+0.7654 &   G333.0162+00.7615 &  \hii region &   12 \\
22 &  G333.0674$-$0.4464 &   G333.0682$-$00.4461* &         YSO &    3 \\
23 &  G333.1038$-$0.5026 &   G333.1075$-$00.5020 &         YSO &   19 \\
23 &  G333.1120$-$0.4996 &   G333.1075$-$00.5020 &         YSO &   10 \\
24 &  G336.3696$-$0.0045 &  G336.3684$-$00.0033A &  \hii region &    9 \\
25 &  G338.9231+0.5523 &   G338.9196+00.5495* &         YSO &   17 \\
25 &  G338.9306$-$0.4948 &  G338.9377$-$00.4890B* &         YSO &   32 \\
25 &  G338.9368$-$0.4889 &  G338.9377$-$00.4890B* &         YSO &    5 \\
26 &  G339.5845$-$0.1267 &   G339.5836$-$00.1265* &         YSO &   15 \\
27 &  G339.9257$-$0.0825 &   G339.9267$-$00.0837* &         YSO &   10 \\
28 &  G340.7460$-$1.0005 &   G340.7455$-$01.0021* &         YSO &   14 \\
29 &  G341.2166$-$0.2113 &   G341.2182$-$00.2136* &         YSO &    1 \\
30 &  G342.7069+0.1250 &  G342.7057+00.1260B* &         YSO &    2 \\
31 &  G343.5027$-$0.0133 &   G343.5024$-$00.0145* &  \hii region &    1 \\
31 &  G343.5058$-$0.0172 &   G343.5024$-$00.0145* &  \hii region &   21 \\
32 &  G343.9043$-$0.6705 &   G343.9033$-$00.6713* &         YSO &    2 \\
33 &  G344.4279+0.0514 &  G344.4257+00.0451A &  \hii region &   20 \\
33 &  G344.4165+0.0455 &  G344.4257+00.0451C &         YSO &   12 \\
34 &  G345.5044+0.3484 &   G345.5043+00.3480* &         YSO &    0 \\
\hline
\end{tabular}
\label{table:rms_offset_sub}
\end{table}
\setlength{\tabcolsep}{6pt}

In total, \fw\ has identified 44 clumps in the 34 fields observed. The number of clumps detected in each field is given in Column~5 of Table~\ref{table:field_parameters_sub}. The clump parameters determined by \fw\ are given in Table~\ref{table:fellwalker}. The clump names are derived from the coordinates of the centre of the clump which can be found in Columns 3 and 4 of this table. A total of 9 fields contain more than a single clump, with the maximum number of three clumps per field. Clumps lying within the same field have coherent velocities ($\delta v$ < 5\,\kms), consistent with the hypothesis that they are all associated with the same giant molecular cloud (GMC), and many appear to be a part of the same mid-infrared structure within individual fields. In Figure~\ref{fig:8mu_examples} we present a few examples of the \nhthree (1,1) inversion emission contours overlaid on 8\,micron\ IRAC images of the same region along with the associated velocity maps.

\begin{figure}
\includegraphics[width = 0.45\textwidth]{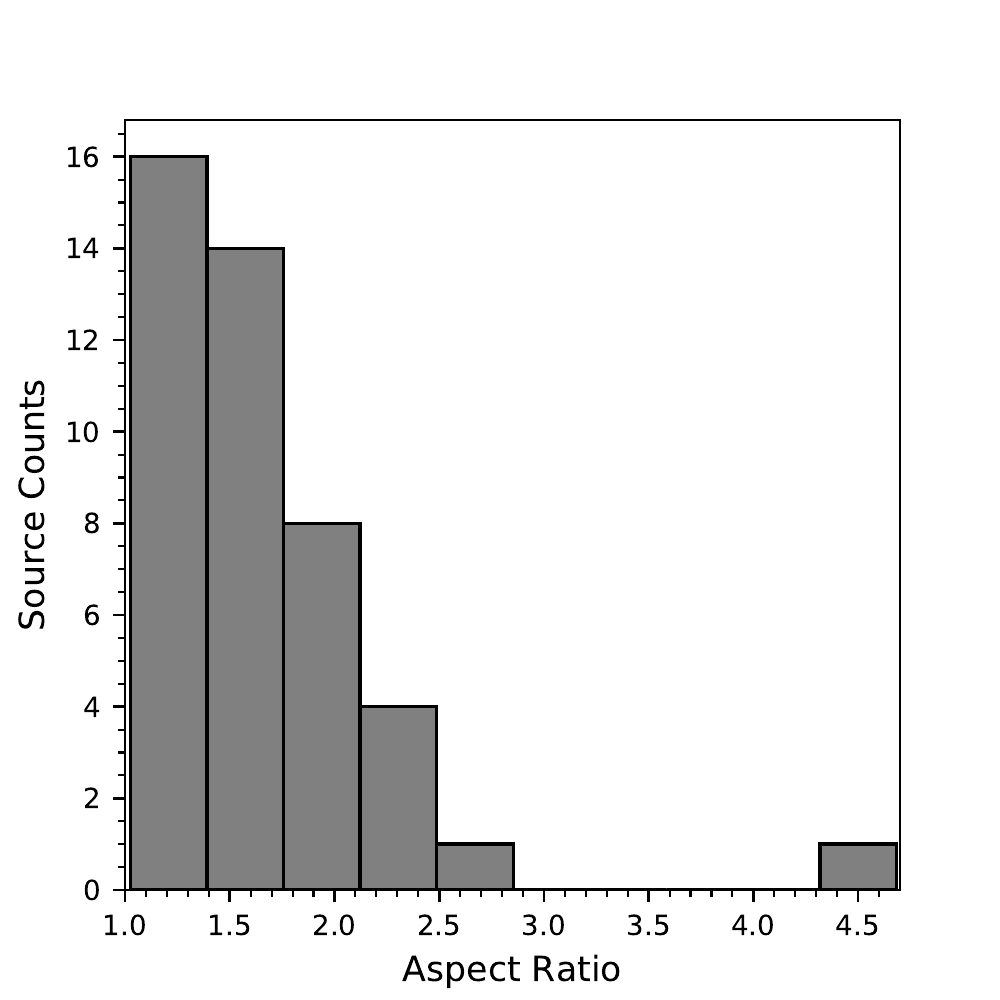}
\caption{Histogram presenting the aspect ratios for all detected clumps (i.e. $\sigma_{\rm{maj}}$ / $\sigma_{\rm{min}}$).}
\label{fig:rms_aspects}
\end{figure}

\begin{figure}
\includegraphics[width = 0.45\textwidth]{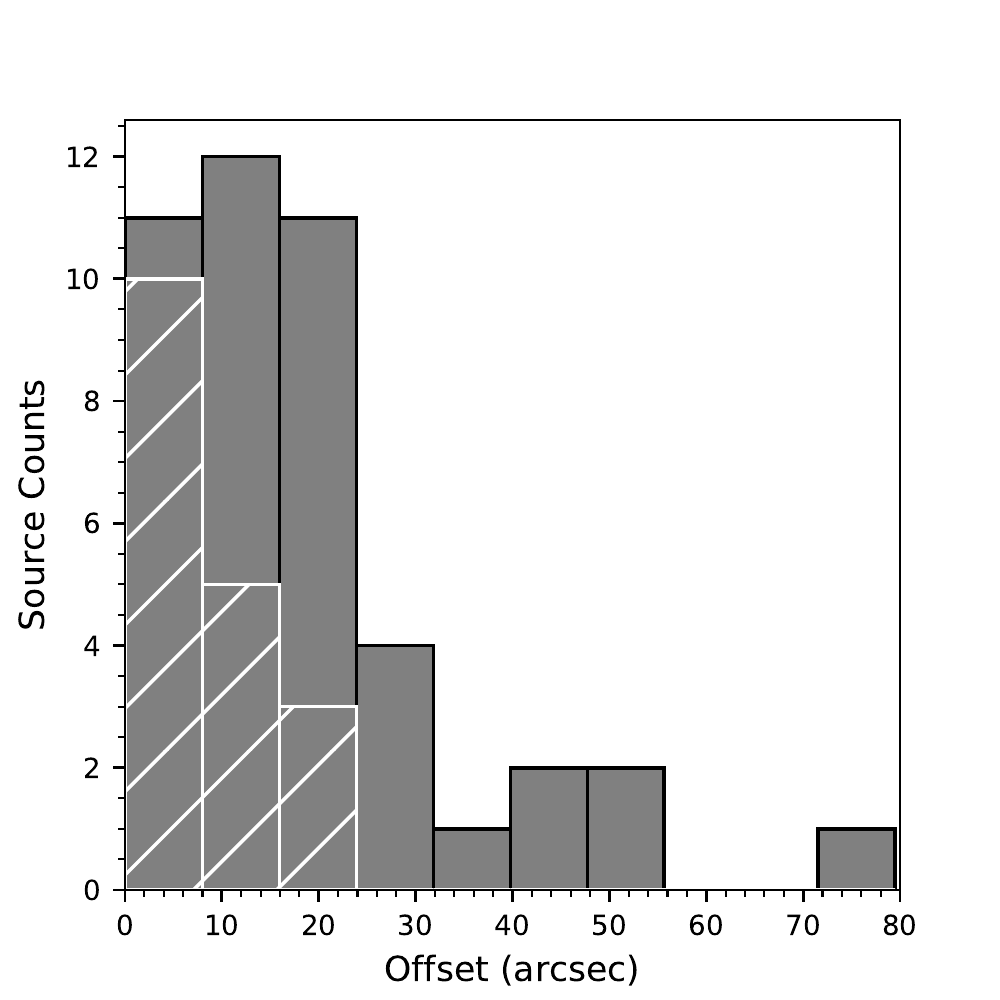}
\caption{Histogram presenting the distribution of the angular offsets between associated RMS sources and peak clump positions, embedded sources are shown with hatched markings.}
\label{fig:rms_offsets}
\end{figure}

\begin{figure}
\includegraphics[width=0.49\textwidth]{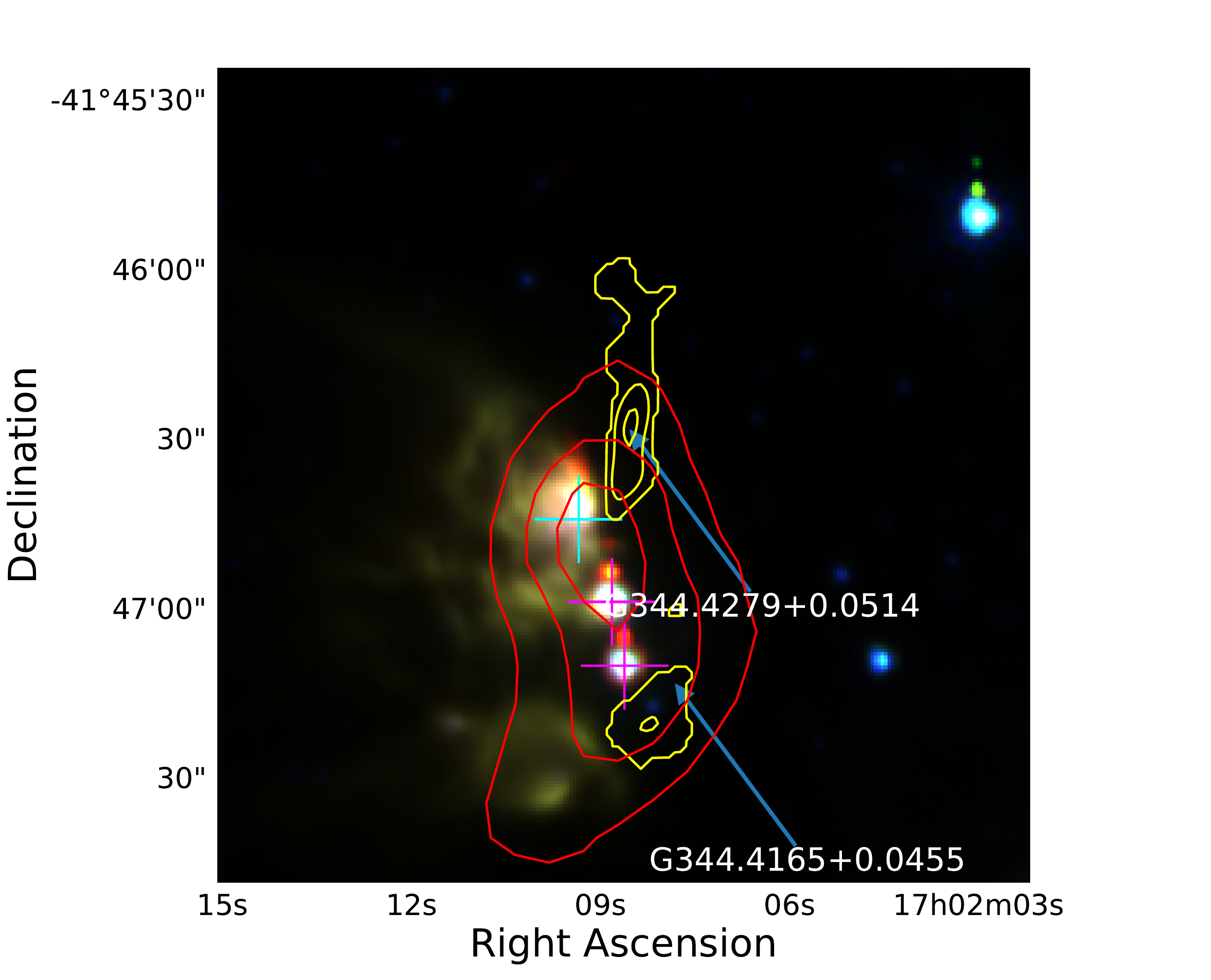}
\caption{Example RGB GLIMPSE image of region (G344.4257+00.0451), overlaid with \nhthree and ATLASGAL contours in yellow and red respectively, the contours levels are equivalent to those in Figure~\ref{fig:8mu_examples}. This field contains two clumps, one of which (G344.4279+0.0514) presents the most elongated structure in the sample, with a filamentary morphology.}
\label{filamentary_clump_example}
\end{figure}

The \fw\ algorithm also fits the semi-major and semi-minor axes of each clump, the ratio of which is used to find the corresponding aspect ratio: these are given in Columns 5-7 of Table\,\ref{table:fellwalker}. Almost every clump is extended with respect to the beam and are typically elongated: the mean and standard deviation values of the aspect ratio are 1.66 and 0.58 respectively. The upper panel of Figure\,\ref{fig:rms_aspects} shows a histogram of the aspect ratios for the entire sample. One clump has an aspect ratio of 4.68 (G344.4279+0.0514). This particular source seems to be filamentary in nature as shown in Figure~\ref{filamentary_clump_example}. Due to the low aspect ratios, any projection elements are unlikely to impact on the morphologies of the \nhthree clumps.

The angular radius for each clump has been estimated from the geometric mean of the deconvolved major and minor axis of the \nhthree emission, multiplied by a factor $\eta$ that relates the r.m.s. size of the emission distribution of the source to its angular radius (Eqn.\,6 of \citealt{Rosolowsky2010}):

\begin{equation}
\theta_{\rm{R}} = \eta \Bigg[ \big(\sigma_{\rm{maj}}^2 - \sigma_{\rm{bm}}^2\big) \big(\sigma_{\rm{min}}^2 - \sigma_{\rm{bm}}^2 \big) \Bigg]^{\frac{1}{4}}
\end{equation}

\noindent where $\sigma_{\rm{bm}}$ is the r.m.s. size of the beam (i.e., $\sigma_{\rm{bm}} = \frac{\theta_{\rm{FWHM}}}{\sqrt{8 ln 2})}$). The value of $\eta$ is taken to be 2.4 following \cite{Rosolowsky2010}. An angular size has been obtained for 31 out of the 44 clumps (70\%): the remaining ten clumps have one axis smaller than the beam size and so a size cannot be accurately determined. This is an artifact of the detection threshold of the \fw\ algorithm; only \nhthree emission above this threshold is used for this analysis, although the \nhthree emission may be extended at levels below this threshold. This can result in the measured source sizes underestimating the size of the clumps and can result in sizes that are smaller than the beam \citep{Rosolowsky2010}. The obtained angular radius values range from 2.8\arcsec\ to 19.4\arcsec\ with a mean of 9.5\arcsec.

With the exception of one field not included in the ATLASGAL survey coverage (G261.6429$-$02.0922), all detected clumps in every field are enveloped by the thermal dust emission of the region, meaning that 43 of the 44 of the \nhthree clumps are associated with 32 dust sources. For the 32 ATLASGAL sources, 24 (75\%) are associated with a single \nhthree clump, 6 (19\%) with two \nhthree clumps, and 2 (6\%) with three \nhthree clumps. The association of multiple ammonia clumps with a single dust clump, and the fact that ammonia is tracing high volume densities allows us to investigate the substructure of these regions.

\subsection{RMS and Maser associations}

Each detected clump has been matched with a source from the RMS survey.
RMS sources that lie towards the centre of the \nhthree (1,1) emission are classified as `embedded', while sources located further are classified as `associated'. Every clump has at least one associated source due to the targeted nature of the observations. There are a total 48 RMS sources in the sample, and we find that there are only 20 RMS sources (49\%) classified as either MYSO (15 sources) or \hii region (5 sources) embedded with the clumps identified in our 34 fields.

There are therefore 14 MYSOs and 14 \hii regions which are not embedded.  The offsets between the RMS sources and each clump's peak position can be found in Table~\ref{table:rms_offset_sub}, and a histogram plot of angular offsets is shown in the lower panel of Figure~\ref{fig:rms_offsets}. The angular offsets between the position of the targeted RMS sources and the centre of the \nhthree emission for each clump ranges between 0.04\arcsec\ and 79.5\arcsec\ with a mean offset of 18.19\arcsec. The maximum offset between the source and the \nhthree emission for deeply-embedded RMS sources is 21\arcsec. These values are similar to those reported in previous studies (e.g. Paper\,I).

\subsubsection{Distances and clump radii}
\label{sect:dists_radii}

Distances to the RMS sources in each field have been taken from \cite{urquhart2014_rms}, with only two fields (G261.6429$-$02.0922 and G286.2086+00.1694) having no reliable distance value. We use the \cite{Reid2016} model to estimate the distances for these two sources. This method uses a Bayesian approach to assign sources to a particular spiral arm based on their (l,b,v) coordinates while taking into account kinematic distances, displacement from the Galactic plane, and proximity to individual parallax sources. We find the distance to G261.6429$-$02.0922 to be $2.02\pm 0.78$\,kpc and $3.0\pm 1.23$\,kpc for G286.2086+00.1694. This additional information gives a complete list of distances for every observed field; these are given in the last column of Table\,1. Overall, the fields have a minimum and maximum distance of 1.80 and 6.70\,kpc respectively, with a mean distance of $3.46\pm 1.22$\,kpc. We find no difference between the distances towards MYSOs or \hii regions. As mentioned in the previous section we have calculated the angular radius for 31 out of the 44 (70\%) detected clumps, and with the determined distances, it is possible to calculated the physical clump size. We find that the clumps have a mean radius of 0.15\,pc with a range between 0.04\,pc and 0.36\,pc (see Table~\ref{table:fellwalker}), which is on the size scale of individual cores or small stellar systems.

\subsubsection{Image Analysis and Classification}

The GLIMPSE three-colour images provide a view into the infrared properties of the environments, and can be used to detect multiple phenomena as discussed in Section~\ref{sect:glimpse_images}. We present three-colour IRAC images of the observed regions in Figures~\ref{filamentary_clump_example}, \ref{fig:esf_example_maps} and \ref{fig:lsf_example_maps} (GLIMPSE images for all regions can be found in the appendix), and have overplotted contours of the ammonia and dust emission in order to examine the morphological correlation of molecular gas and mid-infrared emission and the embedded star formation.

We have constructed a classification system based on the morphologies of the \nhthree emission and how it relates to the RMS sources and thermal dust emission from an inspection of these composite images. Fields which contain a single well-defined clump that is well-correlated with the ATLASGAL emission and which have an RMS source located centrally towards the \nhthree emission are classified as \textit{early star forming} (ESF).  Fields that show a more unstructured and broken morphology, generally containing multiple \nhthree clumps and RMS sources that are not coincident with an \nhthree clump are classified as \textit{late star forming} (LSF), as it is expected that the feedback from forming massive star has a significant disruptive impact on its surroundings. All regions show signs of ongoing star formation processes, and Figures~\ref{fig:esf_example_maps} and \ref{fig:lsf_example_maps} show some examples of regions that are classified as ESF and LSF, respectively.

Three of the fields in this sample were classified as `quiescent' as they shown no signs of current star formation.
These quiescent clumps are well aligned with the dust emission in the area but show no signs of having undergone any star formation (i.e., no evidence of an embedded mid-infrared point source, which is usually taken as evidence of the presence of a proton ongoing stellar object). While all observations were targeted towards RMS sources, the offset between the \nhthree emission and the nearest RMS object for these three regions is sufficiently large to conclude that they are not associated. An example of a quiescent clump is shown in the left panel of Figure\,\ref{fig:8mu_examples} (G326.7270+0.6228).

\begin{figure*}
\includegraphics[width=0.45\textwidth]{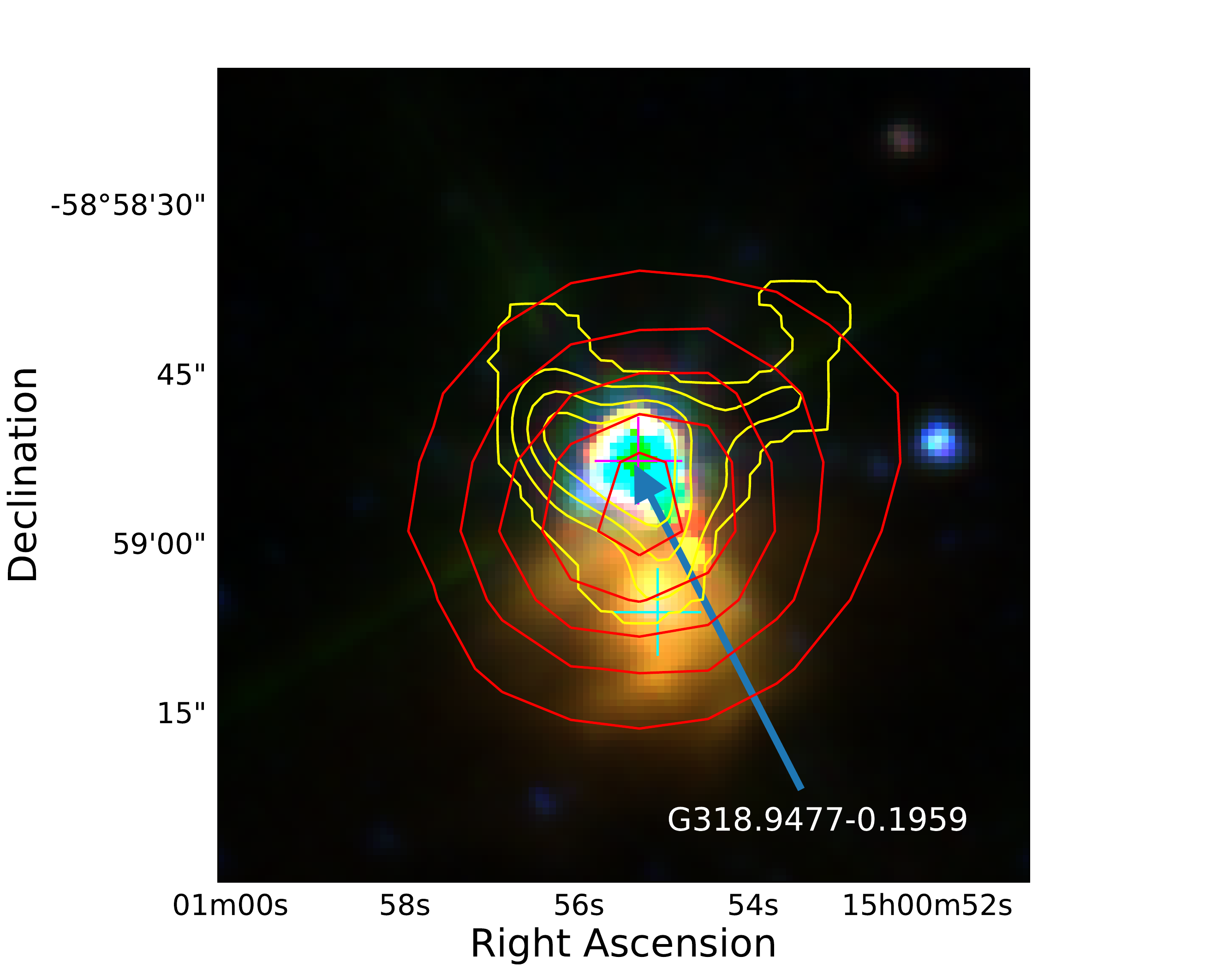}
\includegraphics[width=0.45\textwidth]{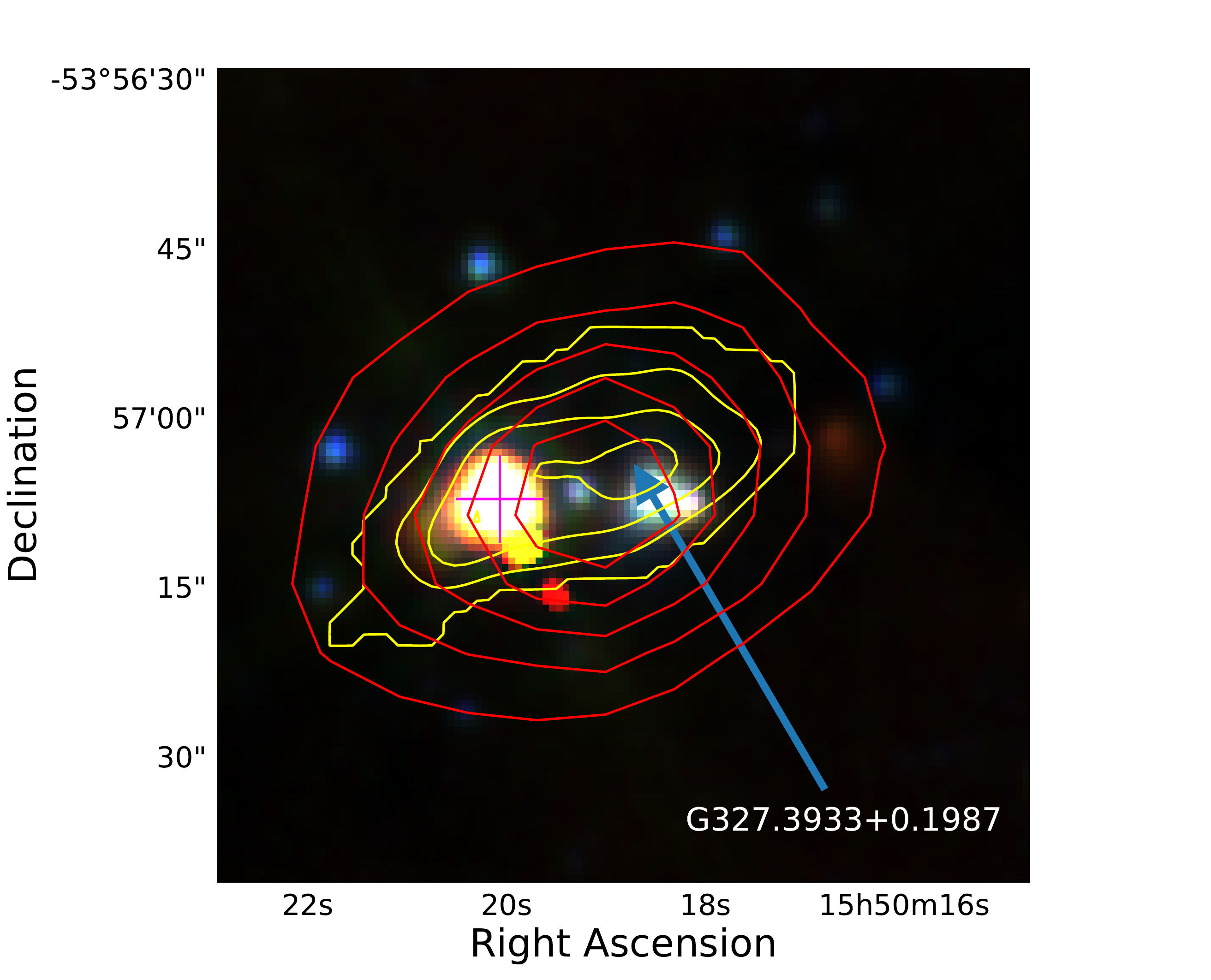}
\includegraphics[width=0.45\textwidth]{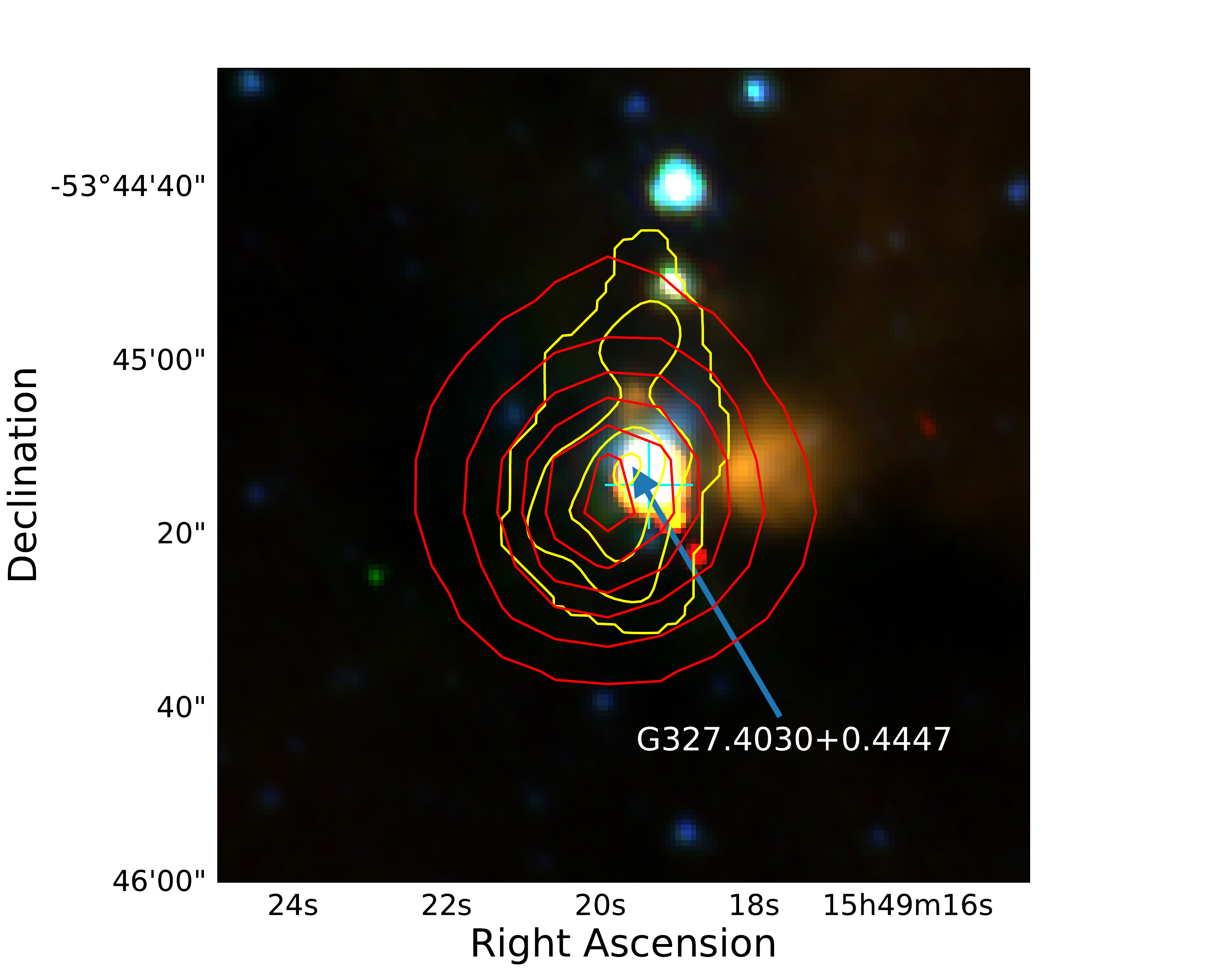}
\includegraphics[width=0.45\textwidth]{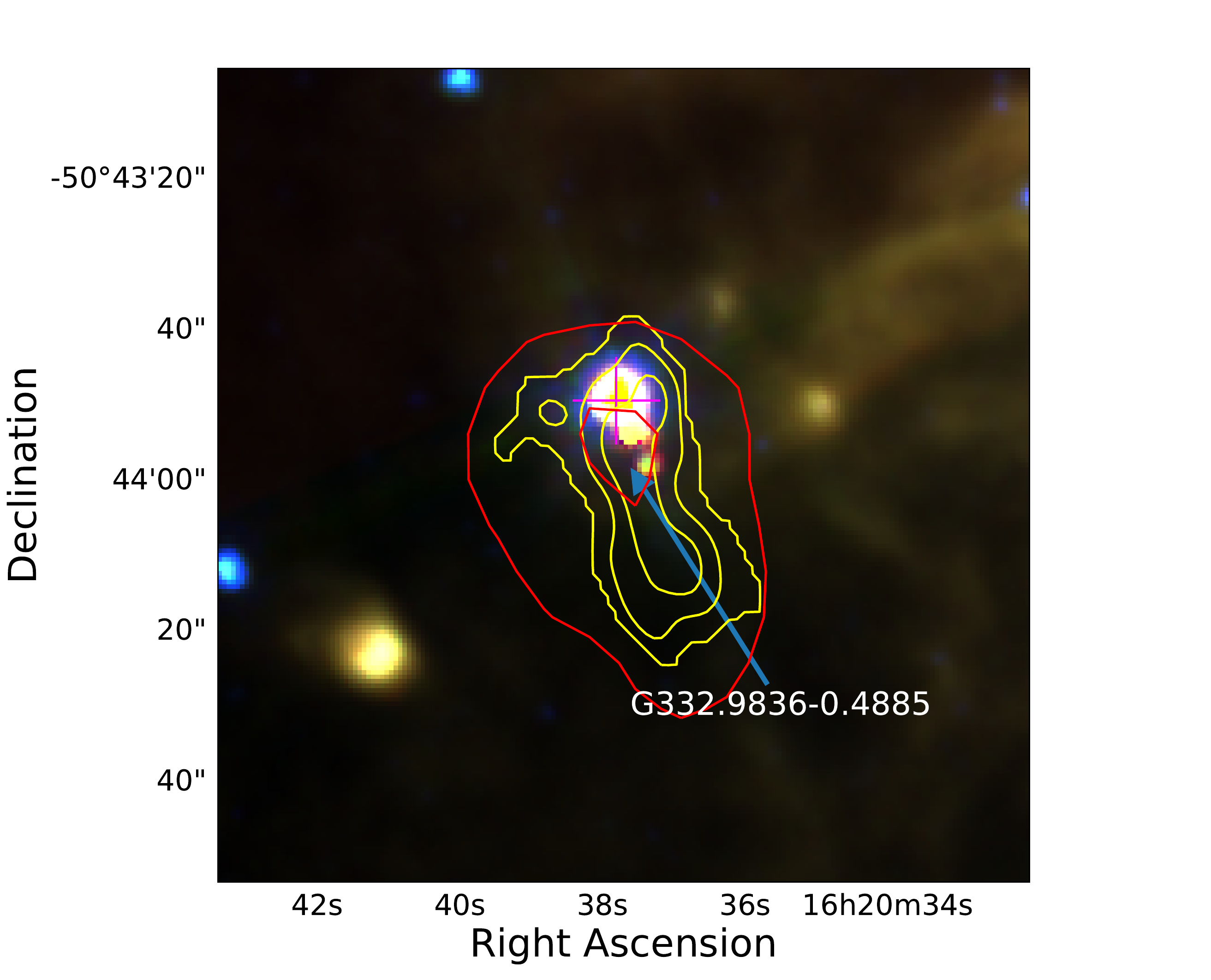}
\includegraphics[width=0.45\textwidth]{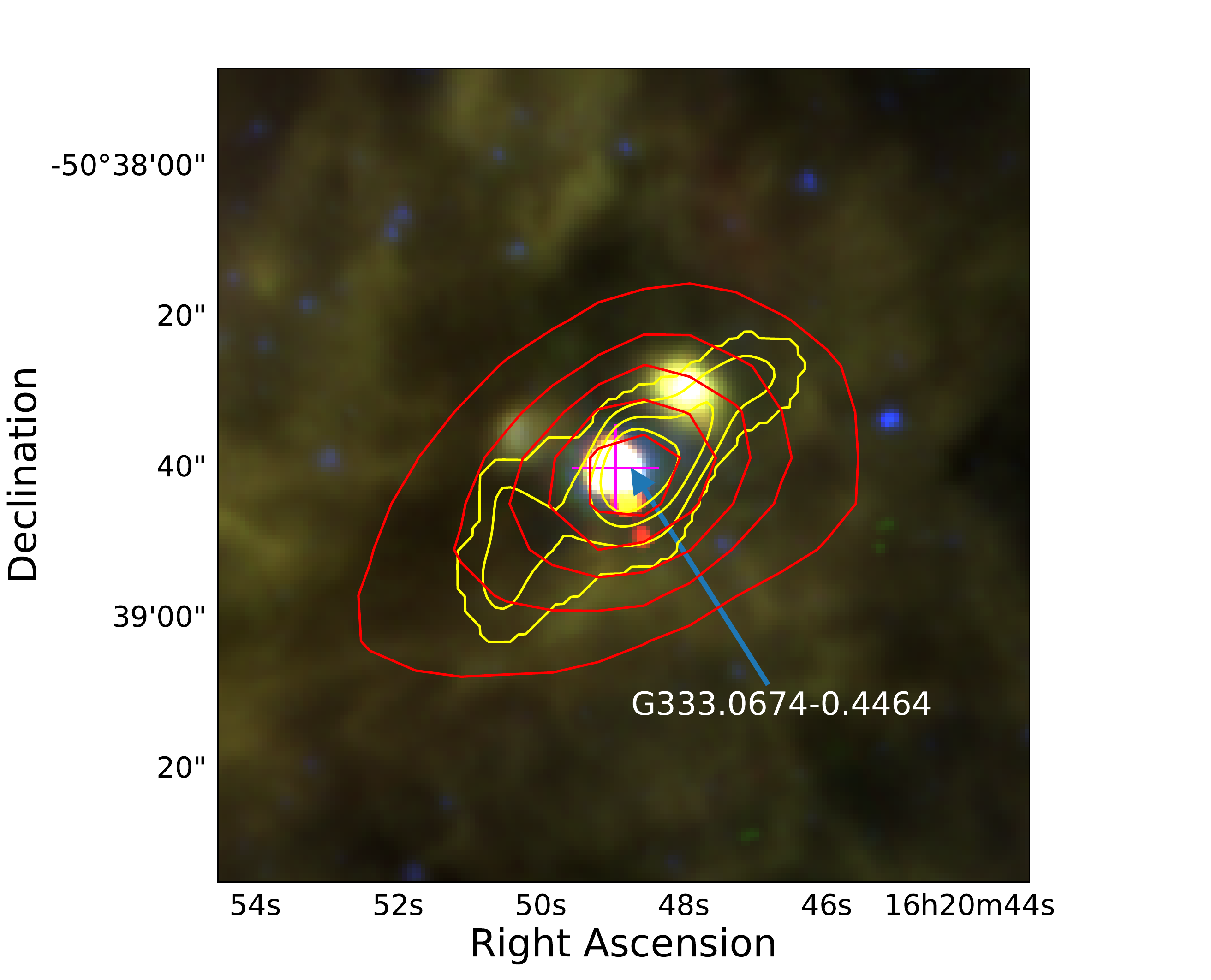}
\includegraphics[width=0.45\textwidth]{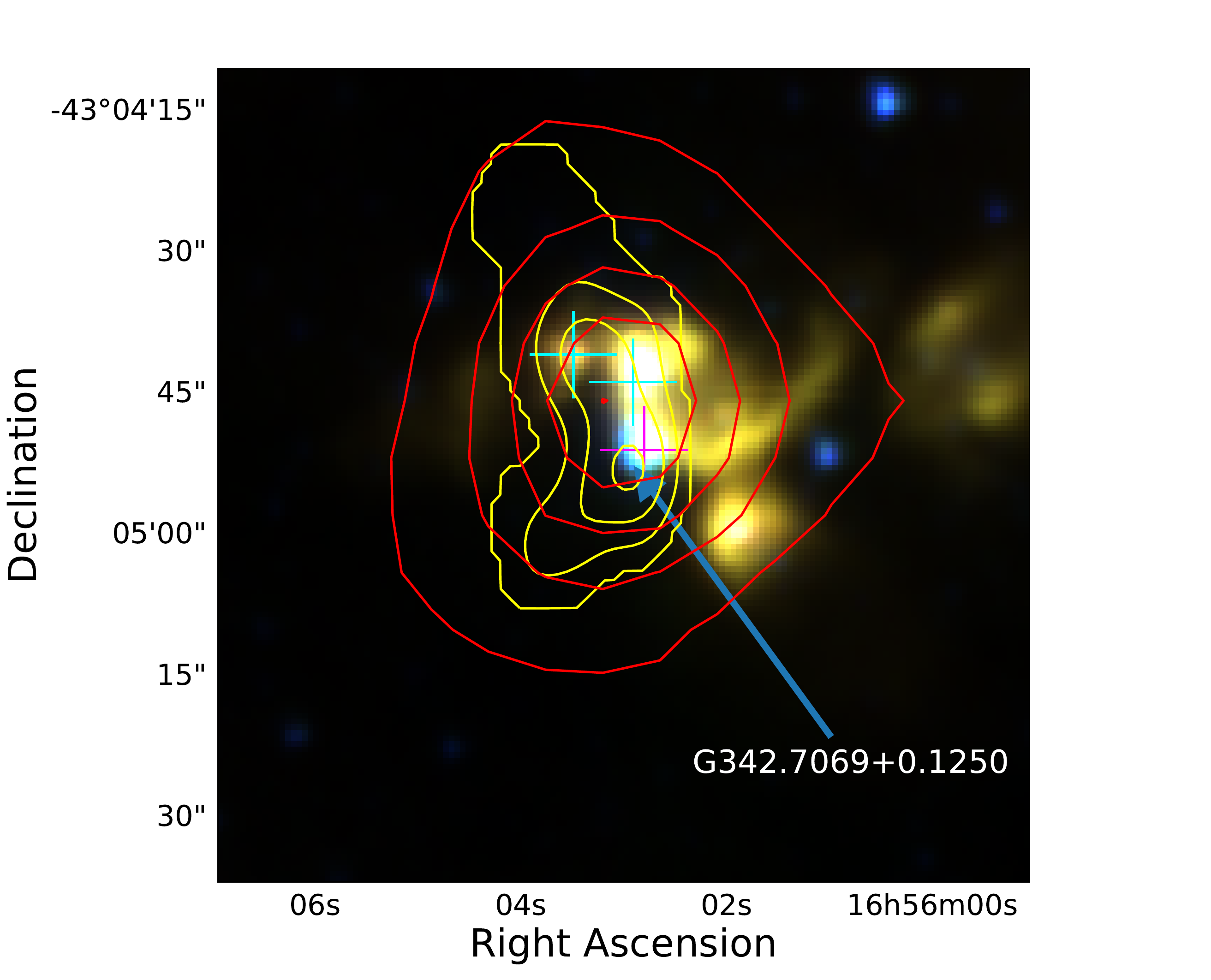}

\caption{Example maps of early star forming regions, presented in each panel is a three-colour GLIMPSE image overlaid with \nhthree emission and ATLASGAL dust emission contours in yellow and red respectively, the contours levels are equivalent to those in Figure~\ref{fig:8mu_examples}. The wavelengths for the RGB channels are 8, 4.5 and 3.4\,\micron\ respectively. MYSOs and \hii regions identified by the RMS survey are shown with magenta and cyan crosses respectively. Clump positions and names are shown with blue arrows and white labels. GLIMPSE images for the remaining 12 ESF fields can be found in the appendix.}
\label{fig:esf_example_maps}
\end{figure*}

\begin{figure*}
\includegraphics[width=0.45\textwidth]{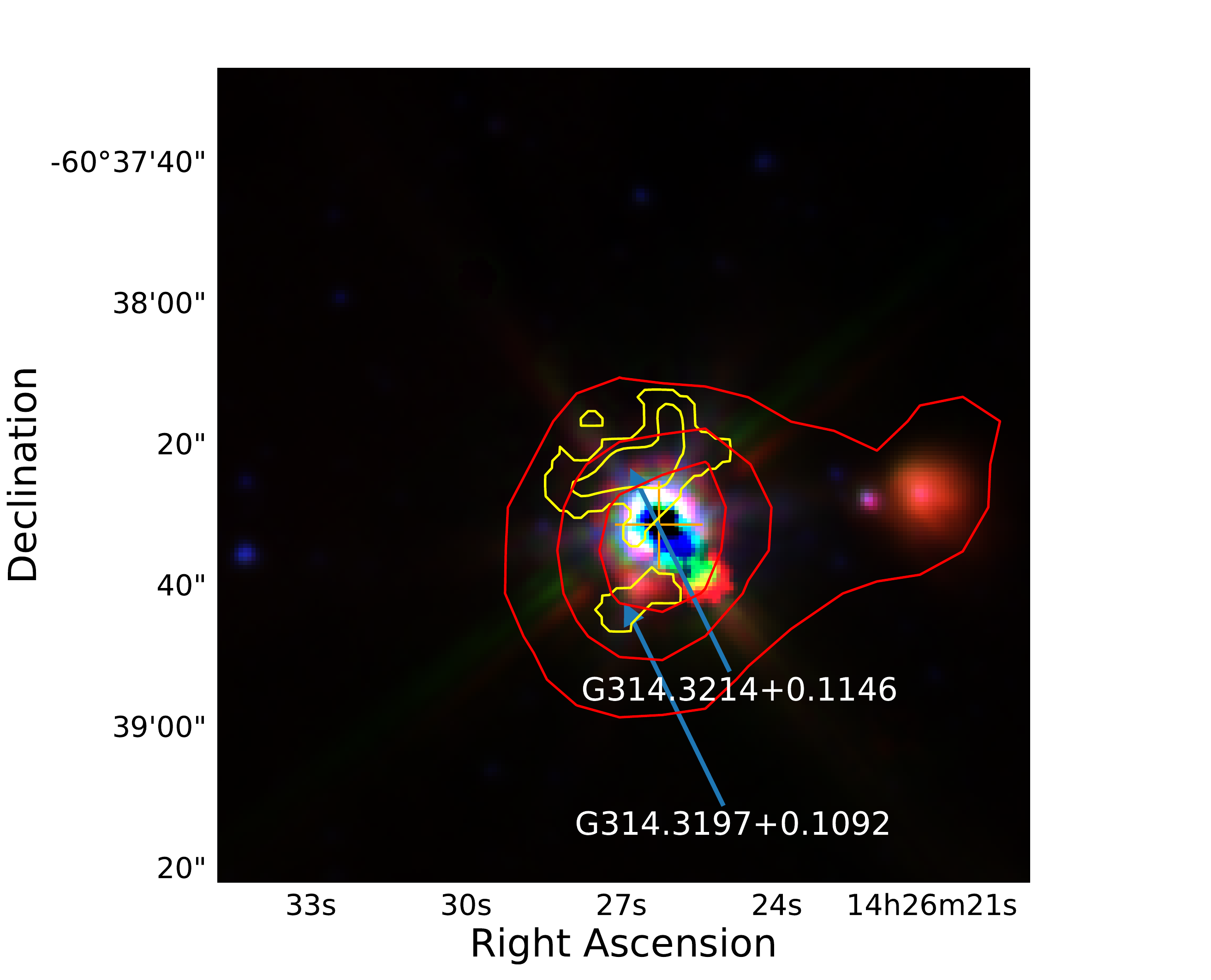}
\includegraphics[width=0.45\textwidth]{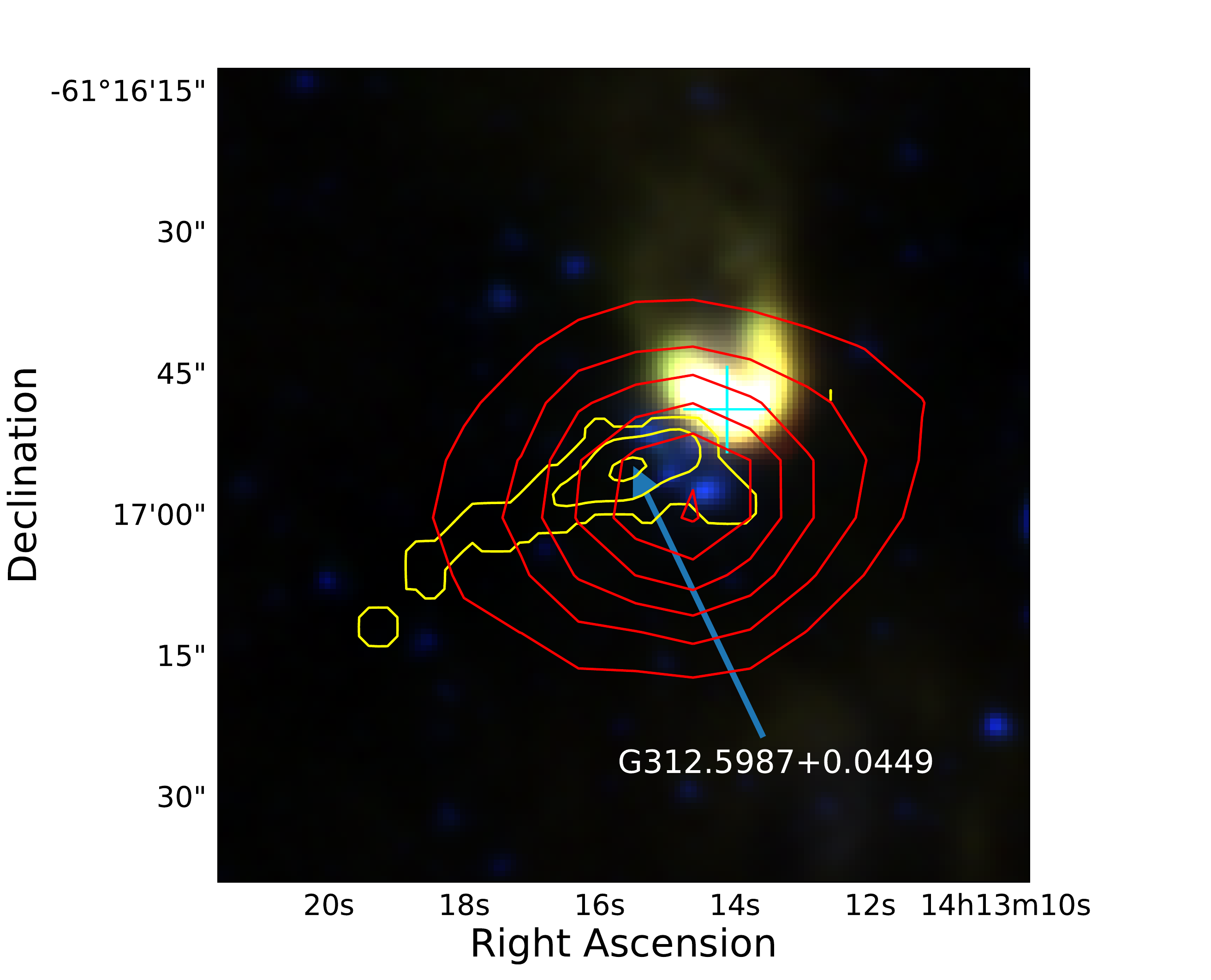}
\includegraphics[width=0.45\textwidth]{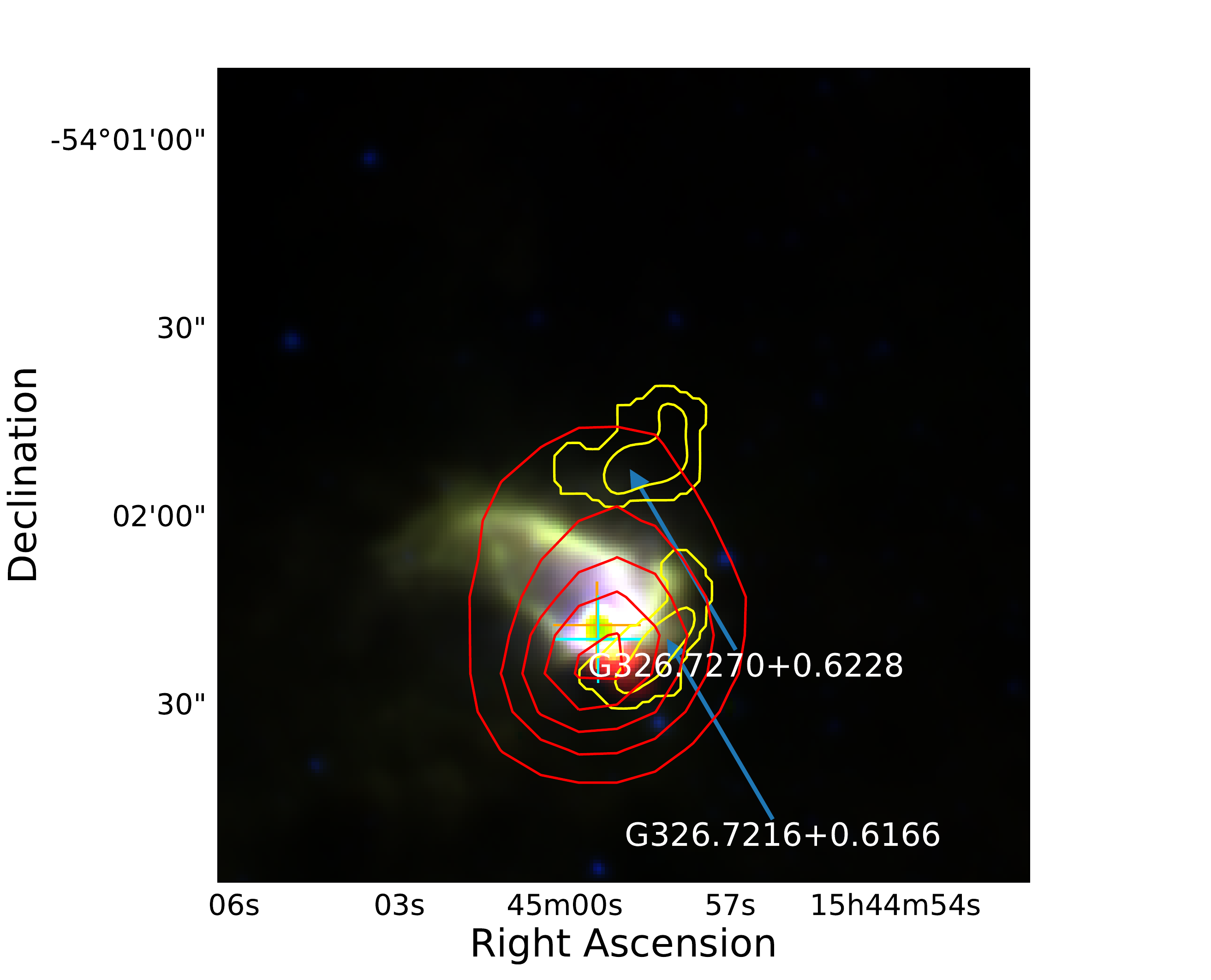}
\includegraphics[width=0.45\textwidth]{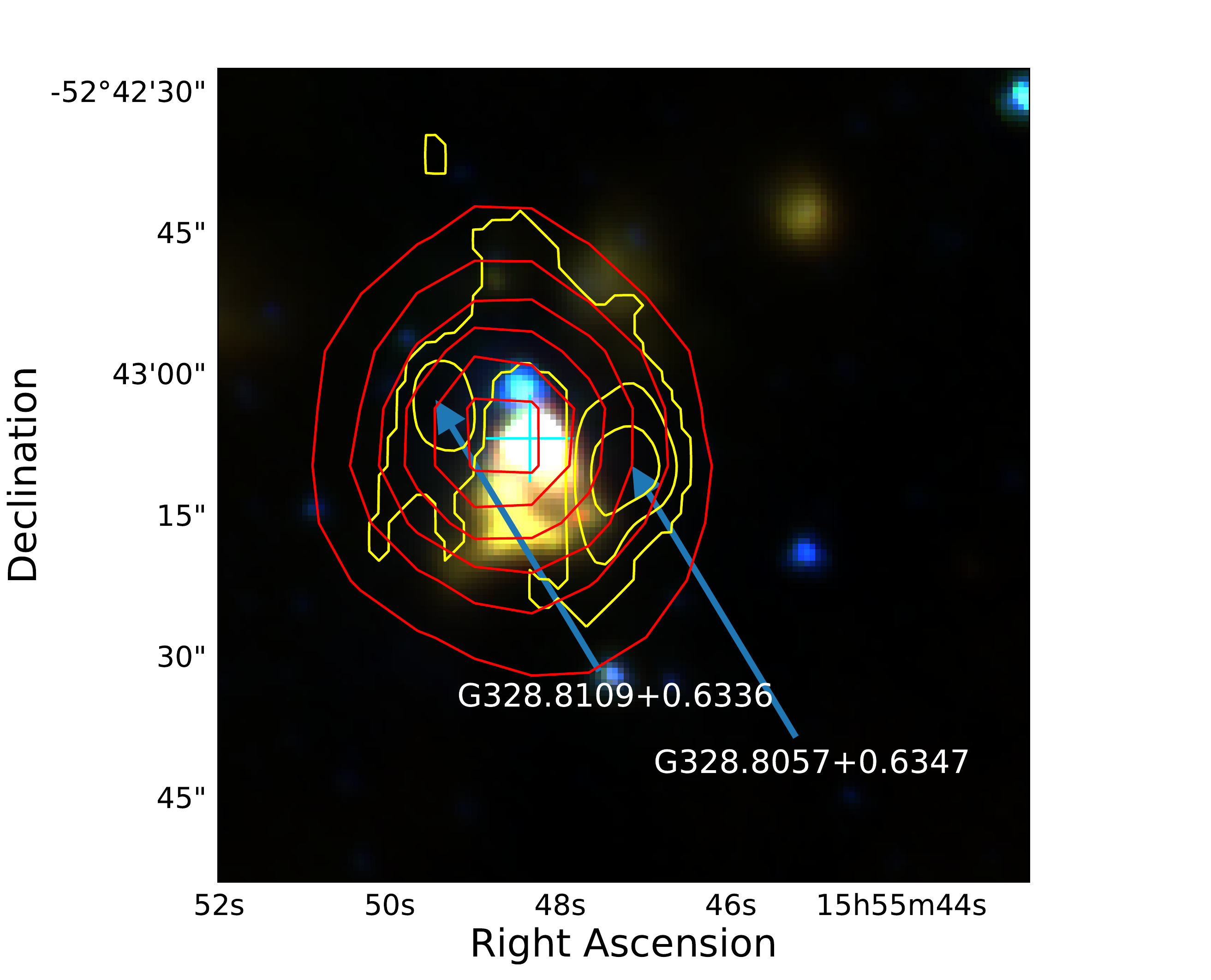}
\includegraphics[width=0.45\textwidth]{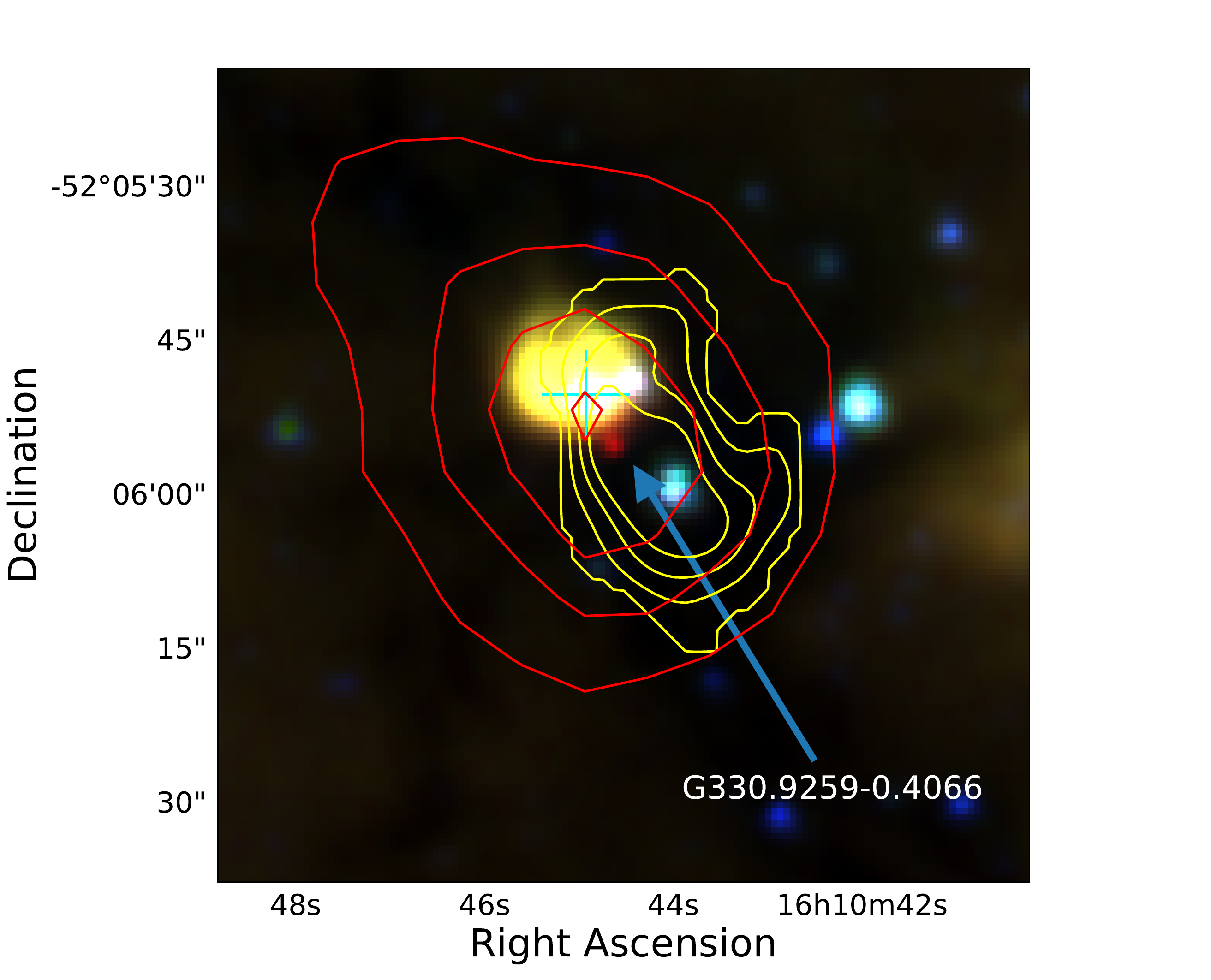}
\includegraphics[width=0.45\textwidth]{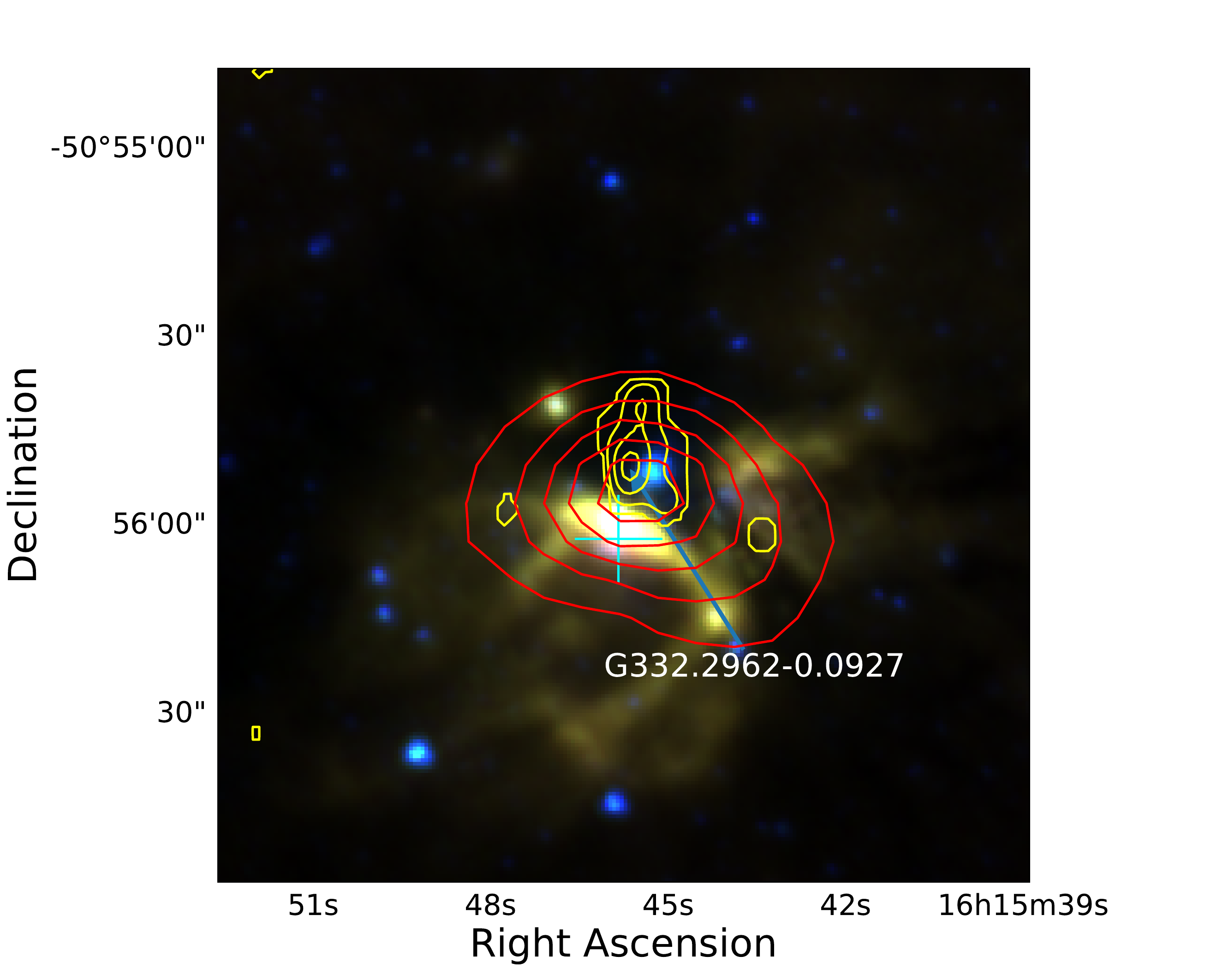}

\caption{Image details are the same as for Figure~\ref{fig:esf_example_maps}, but for a sample of late star forming regions. GLIMPSE images for the remaining 10 LSF regions can be found in the appendix.}
\label{fig:lsf_example_maps}
\end{figure*}

If a field contains both an embedded MYSO and an \hii region (or multiple of either), then the most centrally-located source is taken as the primary association (10 fields, 6 of which are ESF and 4 are LSF). There are 15 fields in the sample that have been classified as ESF fields. Of these fifteen, 13 are associated with MYSOs (as defined by the RMS survey), with the remaining two associated with an \hii region. Sixteen fields have been classified as LSF.  Only six of these fields contain an embedded MYSO, while the majority of fields are associated with an \hii region. An inspection of the images indicates that the majority of early star forming regions are associated with MYSOs (87\%), while later regions are dominated by \hii regions (63\%). MYSOs, therefore, appear to be still very embedded, while the majority of the \hii regions appear to be actively disrupting their natal clump and breaking out of their dust cocoons (as shown in Figure~\ref{fig:8mu_examples}). This is not obvious in the ATLASGAL data (which is sensitive to the whole column of gas along the line of sight, and also has lower spatial resolution than the \nhthree observations) but can clearly be seen by the \nhthree emission, which has a critical density of $n \sim 1.8\times 10^3$\,cm$^{-3}$ and so only traces the high volume-density substructure within the clumps.

Our visual examination of the dust- and ammonia-overlaid three-colour IRAC maps has resulted in the identification of three visually distinct stages (quiescent, ESF and LSF). We need to demonstrate that this sequence is reliable, and one such approach is to compare the physical properties of these different types of regions and determine whether there are any physical trends that support our visual classification.

We have used data from the Methanol MultiBeam (MMB) survey and the H$_2$O Southern Galactic Plane Survey (HOPS) in order to provide more information on the regions presented in this study. Overall, we find that 52 H$_2$O and 31 methanol masers are associated with our 34 fields. A breakdown of this sample is presented in Table~\ref{table:maser_counts} for each classification and central RMS object.

\begin{figure*}
	\includegraphics[width=0.45\textwidth]{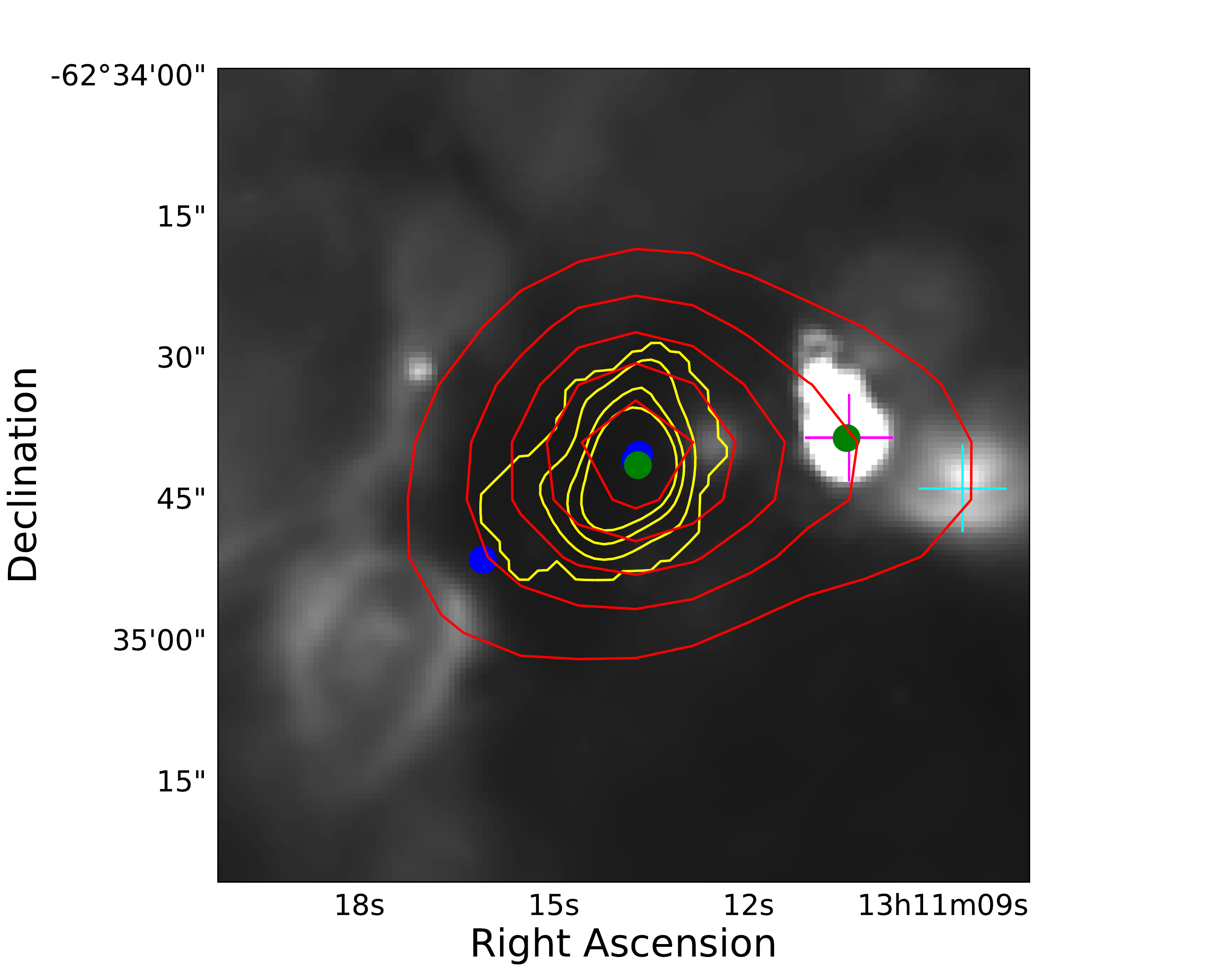}
	\includegraphics[width=0.45\textwidth]{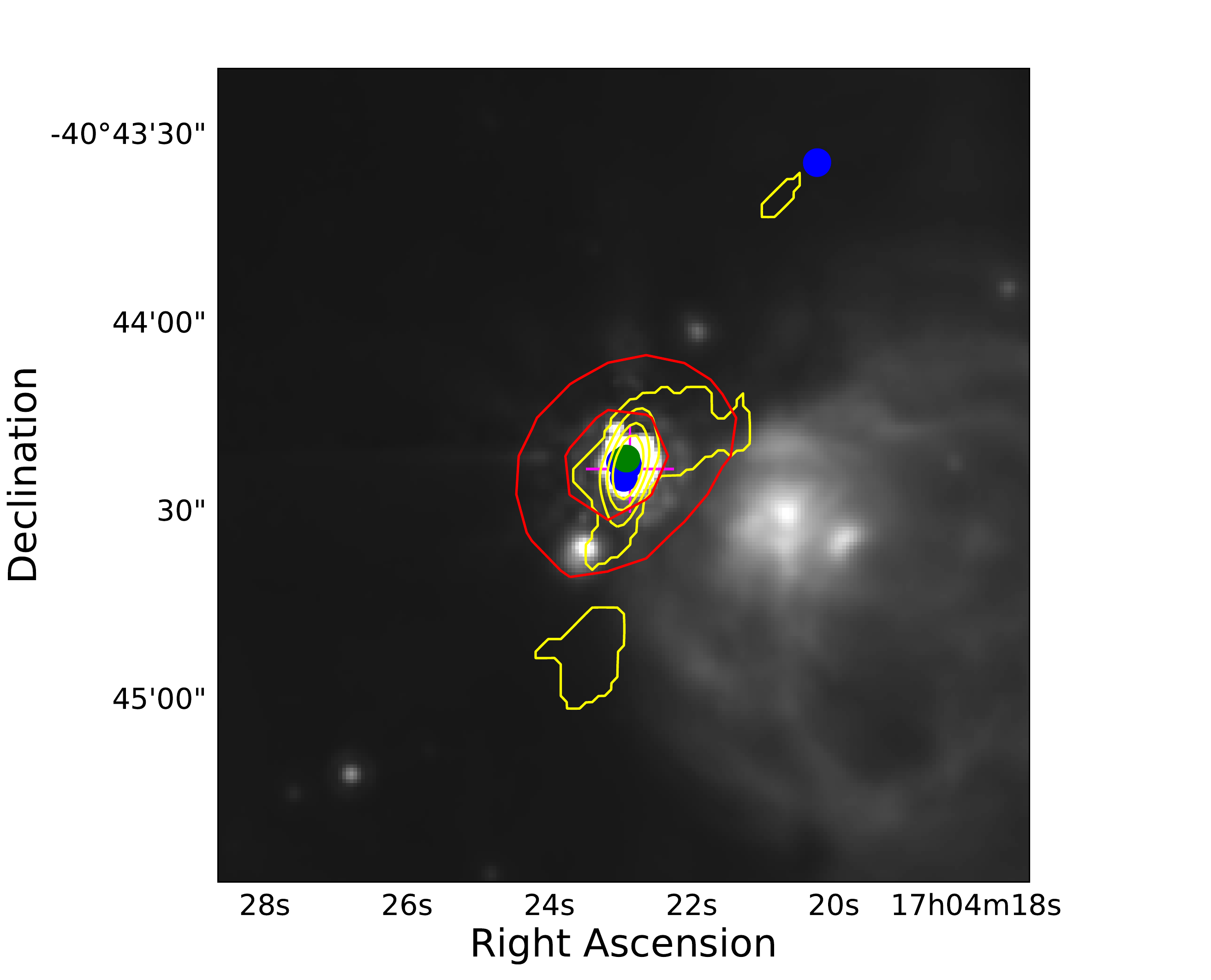}
	\caption{GLIMPSE 8\,\micron\ images showing two regions from the sample (G305.2017+00.2072 and G345.5043+00.3480). The red contours outline the ATLASGAL dust emission and the yellow contours outline the \nhthree emission, while the green and blue filled circles represent the methanol and H$_2$O maser detections respectively, the contours levels are equivalent to those in Figure~\ref{fig:8mu_examples}. Magenta markers represents the position of identified MYSOs, while cyan markers show the positions of \hii regions.}
	\label{figure:maser_example_maps}
\end{figure*} 

\setlength{\tabcolsep}{3pt}
\begin{table}
\begin{center}
\caption{\label{table:maser_counts}Distribution of masers across the sample.}
\begin{tabular}{cccc}
\hline
\hline
Classification &  Methanol & H$_2$O \\
               &  Masers   & Masers \\
\hline
%Quiescent      & 5        & 9  \\
ESF            & 20       & 41  \\
LSF            & 11       & 11  \\
\hline
YSO            & 23       & 32 \\
\hii region     & 8        & 15 \\
\hline
\end{tabular}
\end{center}
\end{table}
\setlength{\tabcolsep}{6pt}

GLIMPSE 8\,\micron\ images have been used to provide context for the maser distribution with \nhthree and ATLASGAL contours: two example maps of these can be found in Fig.\,\ref{figure:maser_example_maps}. The image shown in the left panel shows a region which has been defined as quiescent with an MYSO and \hii region in close proximity to the west; however, there are also coincident masers (8 H$_2$O and 1 methanol) towards the centre of the gas and dust emission. Therefore this region is unlikely to be quiescent and is probably in an early protostellar stage of star formation, as the presence of maser emission indicates that a central core is likely to have already developed. This is the case for all three regions which have been identified as quiescent (G305.2017+00.2072, G322.1729+00.6442 \& G326.4755+00.6947), all of which have at least one methanol maser towards the centre of the field. These quiescent fields have been reclassified as early star forming, leaving no quiescent regions in the sample. We have, therefore, classified 18 fields as being ESF and 16 fields as LSF. The right panel of Fig.\,\ref{figure:maser_example_maps} presents another region with a relatively high number of maser detections. This field is classified as early star forming and is associated with a total of 12 masers (10 H$_2$O and 2 methanol). All of these are found towards the central RMS object. This source appears to be located on the edge of an evolved \hii region and so this may be an example of triggered star formation (see further discussion in Sect.\,5.4.4).

\section{Determination of Physical Properties}
\label{sect:physical_properties}

\subsection{Ammonia line fitting}
\label{sect:spectral_analysis}

We have conducted a spectral line analysis for all 44 clumps which have been identified by \fw.

This study makes use of the method presented in \cite{Rosolowsky2008} for estimating physical parameters from ammonia spectra, using a nonlinear least-squares minimisation code to determine the optimal fit to the observed spectra. This technique simultaneously fits observed the \nhthree inversion transitions (in our case, the (1,1) \& (2,2) transitions) using the physical parameters (such as kinetic temperature and optical depth) as fitting parameters. This avoids potential systematic errors that may be encountered when transitions are fitted independently, and automatically determines uncertainties for the physical parameters of the system. We use \texttt{pyspeckit} \citep{Ginsburg2011}, a \texttt{Python} implementation of this method to fit the eighteen individual (1,1) and twenty-one (2,2) hyperfine transitions simultaneously, determining the free parameters which include the kinetic ($T_{\rm{kin}}$) and excitation ($T_{\rm{ex}}$) temperatures, FWHM line-width ($\Delta v$), radial velocity ($v_{\rm{LSR}}$) and the \nhthree column density. While the satellite lines of the (2,2) emission are generally undetected in our observations, a well-detected (2,2) main line sufficiently constrains the solution, allowing good determination of the free parameters.

This model assumes that the kinetic temperature is less than $T_0$ = 41.5\,K, which is the energy difference associated with these lowest two inversion transitions, implying that only these first two energy states are significantly populated. The reduced data cubes are $\sim$4.3\arcmin\ in diameter and gridded using 1\arcsec\ pixels with contours levels starting at 3$\sigma$ and increasing in levels of 0.5$\sigma$.

For each of the data cubes, a pixel-by-pixel line analysis has been performed in order to extract spectra which can then be fit using the aforementioned technique. The moment maps of the ammonia spectra were used to provide initial guesses at the velocity and line-width parameters, which were then refined by a pixel-by-pixel fitting process. Only pixels above a 3$\sigma$ threshold were fitted to avoid contamination of the results due to the inclusion of pixels with low SNRs. The observed maps have been spatially smoothed to twice the size of the original beam ($\sim$20\arcsec) in order to improve the SNR of individual spectra. The derived peak and median values for the fitted parameters for all of the clumps are given in Table~\ref{table:fitted_parameters_table_sub}.

\begin{figure*}
	\includegraphics[width=0.49\textwidth]{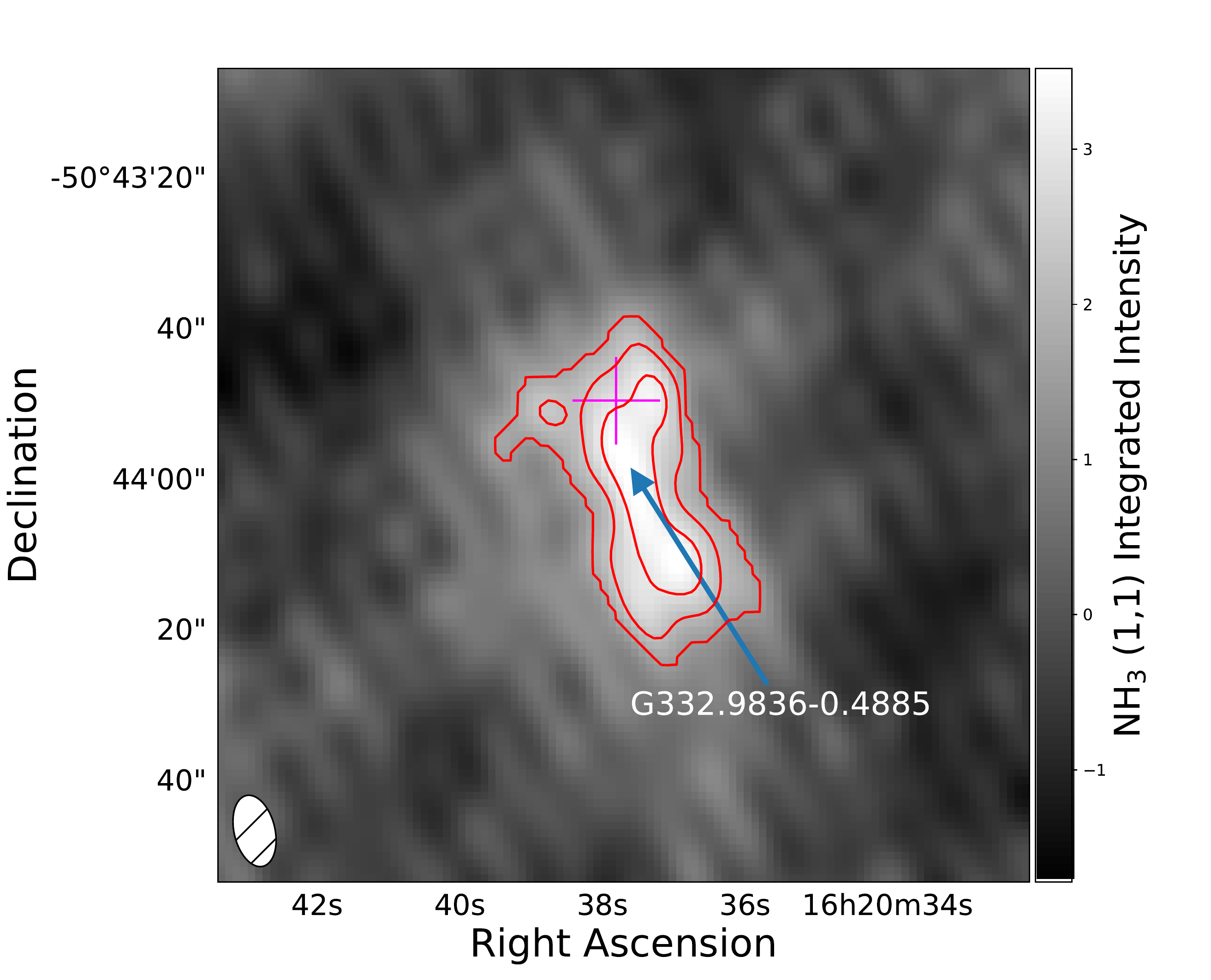}
	\includegraphics[width=0.49\textwidth]{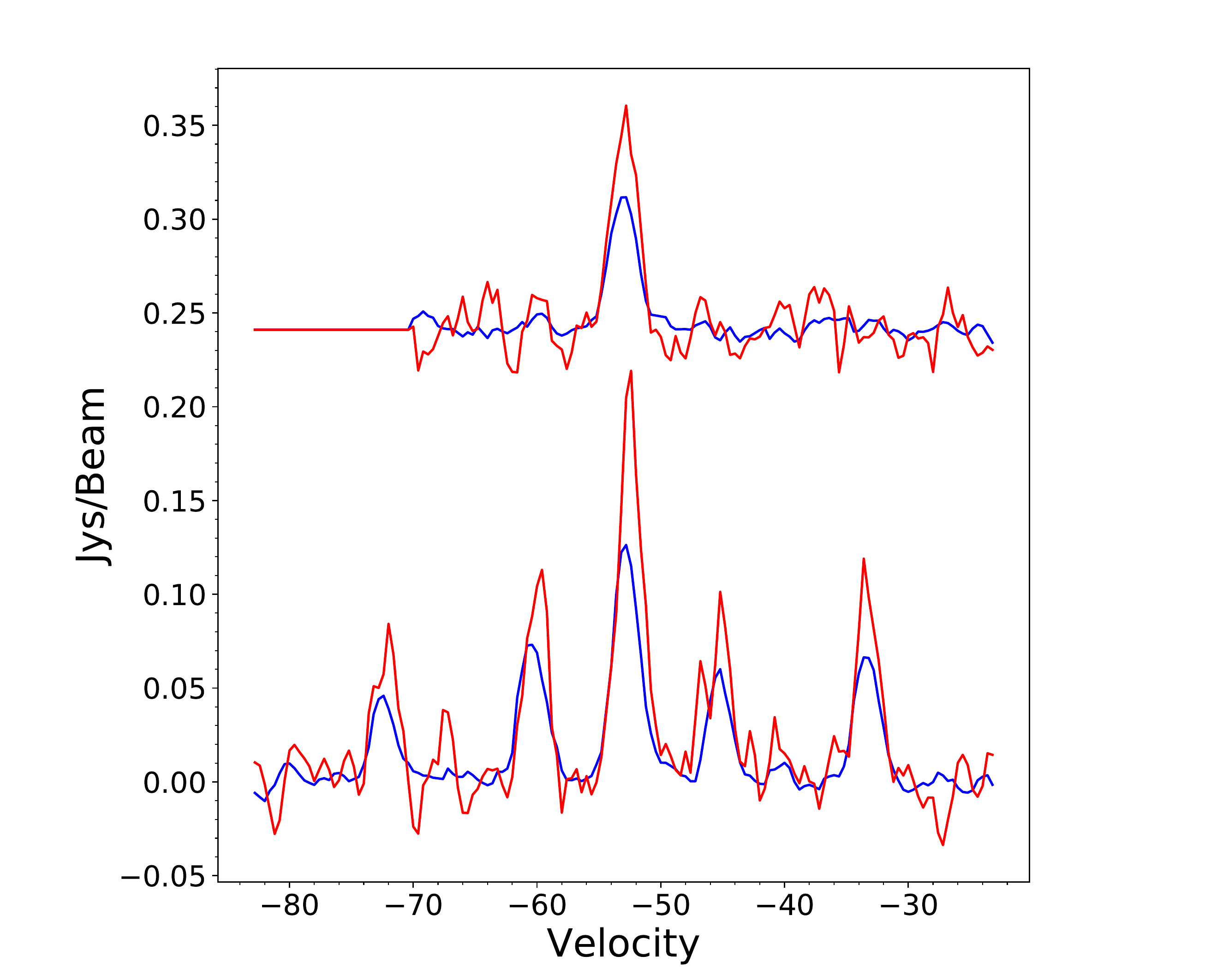}
	\caption{The left panel presents an example region of an integrated \nhthree (1,1) emission map. The central RMS source (MYSO) is marked with a magenta cross. \nhthree contours are shown in red, starting at a 3$\sigma$ threshold and increasing in steps of 0.5$\sigma$, equivalent to those in Figure~\ref{fig:8mu_examples}. The right panel presents associated \nhthree (1,1) and (2,2) spectra for this region, the average spectrum across the emission above the 3$\sigma$ threshold is shown in blue, while the spectrum from the peak integrated position is shown in red.}
	\label{fig:example_int_emission}
\end{figure*}

\subsubsection{Thermal and non-thermal line-widths}

The observed FWHM line-width ($\Delta v_{\rm{obs}}$, a free parameter determined by the fitting procedure) is a convolution of the intrinsic line-width of the source ($\Delta v_{\rm{int}}$) and the velocity resolution of the observations. We remove the 0.4\,\kms\ spectrometer channel width by subtracting this value from the measured FWHM line-width in quadrature:

\begin{equation}
\Delta v_{\rm{int}} = \sqrt{\left(\Delta v_{\rm{obs}}^2 - (0.4)^2\right)}
\end{equation}

\noindent The peak intrinsic linewidths have mean and median values of 3.85 $\pm$ 0.3 and 3.28\,\kms\ respectively. These values are similar to values found in previous studies \citep{Sridharan2002,Wienen2012}.

The line-widths of the ammonia spectra consist of thermal and non-thermal components, where the thermal component can be estimated using:

\begin{equation} \label{eq:therm_linewidth}
\Delta v_{\rm{th}} = \sqrt{\left(\frac{8\ln 2 k_{\rm{B}}T_{\rm{kin}}}{m_{\rm{NH_3}}}\right)}
\end{equation}

\noindent where 8$\ln$2 is the conversion between the velocity dispersion $\sigma v$ and the FWHM line-width $\Delta v$, $k_{\rm{B}}$ is the Boltzmann constant, and $m_{\rm{NH_3}}$ is the mass of an ammonia molecule (17.03 AMU). The measured line-widths themselves are significantly broader than this estimation (for gas temperatures of 20\,K, $\Delta$v$_{\rm{th}}$ is $\sim$0.22\,\kms), meaning there is a large contribution from non-thermal components such as supersonic turbulent motions, outflows, shocks and magnetic fields (\citealt{Elmegreen2004}). The impact on the data from these mechanisms can be derived by subtracting the thermal component in quadrature:

\begin{equation} \label{eq:nontherm_linewidth}
\Delta v_{\rm{nt}} = \sqrt{\left(\Delta v_{\rm{int}}^2 - \frac{8ln2k_{\rm{B}}T_{\rm{kin}}}{m_{\rm{NH_3}}} \right)}
\end{equation}

The gas pressure ratio can be estimated from the ratio of the thermal and non-thermal line-widths, which are equivalent to the thermal and non-thermal pressures within the gas:

\begin{equation} \label{eq:gaspress_ratio}
R_p = \left( \frac{\Delta v_{\rm{th}}^2}{\Delta v_{\rm{nt}}^2}\right). 
\end{equation}

These ratio values are quite low (the mean and median values of the mean pressure for each clump is 0.012 and 0.01 respectively, with a range of 0.002 to 0.03, see Table~\ref{table:total_stats}), indicating that the pressure in the gas is dominated by non-thermal motion.

\subsubsection{Beam Filling Factor}

The previously-mentioned assumption of kinetic temperature limit mentioned above ($T_{\rm{kin}} < T_0 = 41.5$\,K) provides a significant restriction on this study.  

For $T_{\rm{kin}} < T_0$, the calculation of the rotation temperature can be completed using the following equation \citep{Walmsley1983,Swift2005}:

\begin{equation} \label{eq:trot_tex}
T_{\rm{rot}} = T_{\rm{kin}} \left\{ 1 + \frac{T_{\rm{kin}}}{T_0} \ln \left[1 + 0.6\,\exp\left(-15.7\,T_{\rm{kin}} \right) \right]\right\}^{-1}
\end{equation}

We have used the calculated values of rotational temperature in order to determine the beam filling factor for the entire sample:

\begin{equation}
B_{\rm{ff}} = \frac{T_{\rm{ex}}}{T_{\rm{{rot}}}}
\end{equation}

The calculated excitation temperatures (2.77 to 7.10\,K) and beam filling factors are relatively low (0.09 to 0.21 with a mean value of 0.15), which is similar to other studies (\citealt{urquhart2011_nh3}, Paper\,I, \citealt{Friesen2009}). The less-than-unity values of the filling factor suggest that although the emission is extended with respect to the beam, clumps are likely to consist of a significant number of smaller dense substructures (cores), which when convolved with the beam results in the appearance of an extended emission region.

\begin{figure*}
	\includegraphics[width=0.45\textwidth]{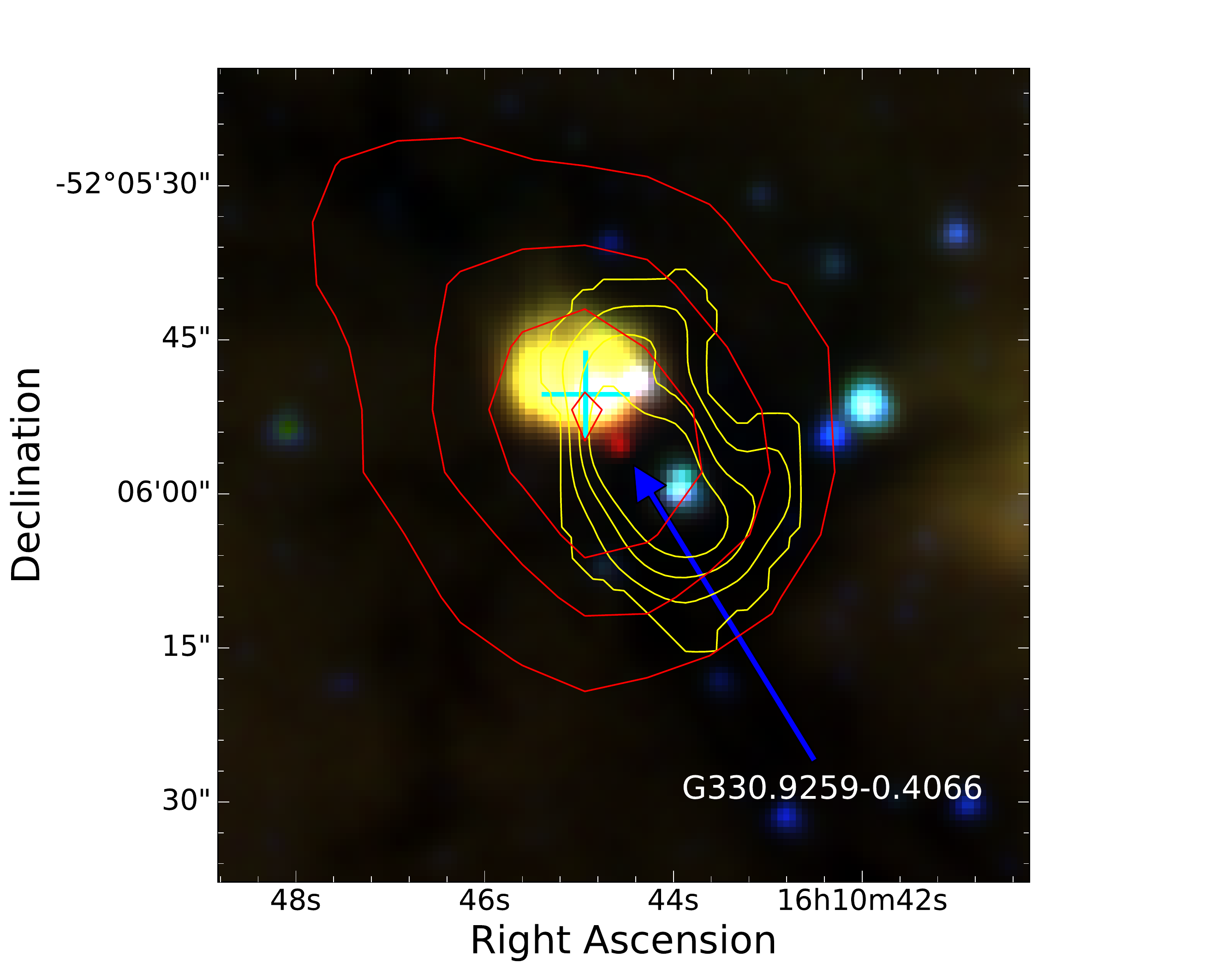}
	\includegraphics[width=0.45\textwidth]{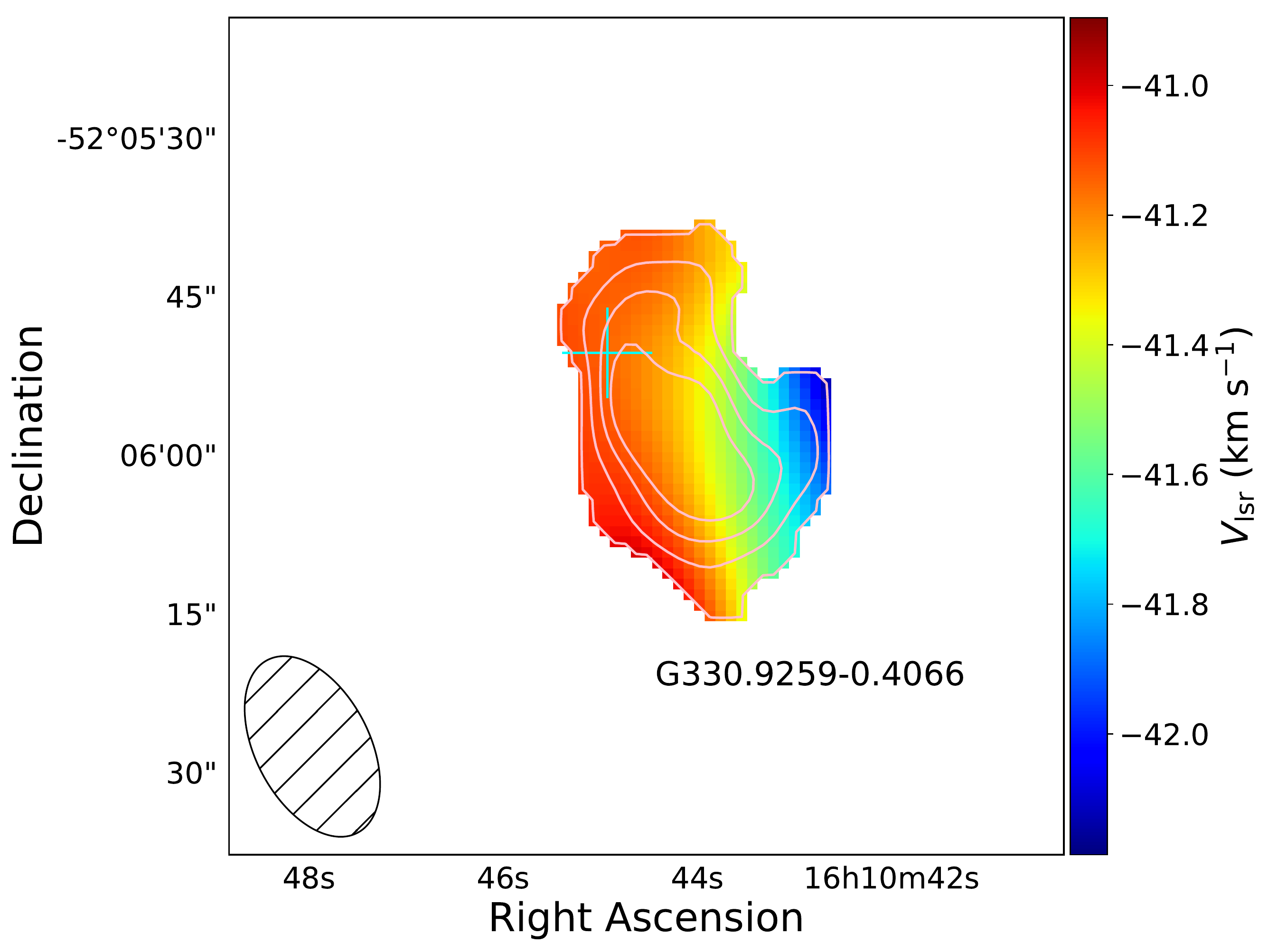} \\
	\includegraphics[width=0.45\textwidth]{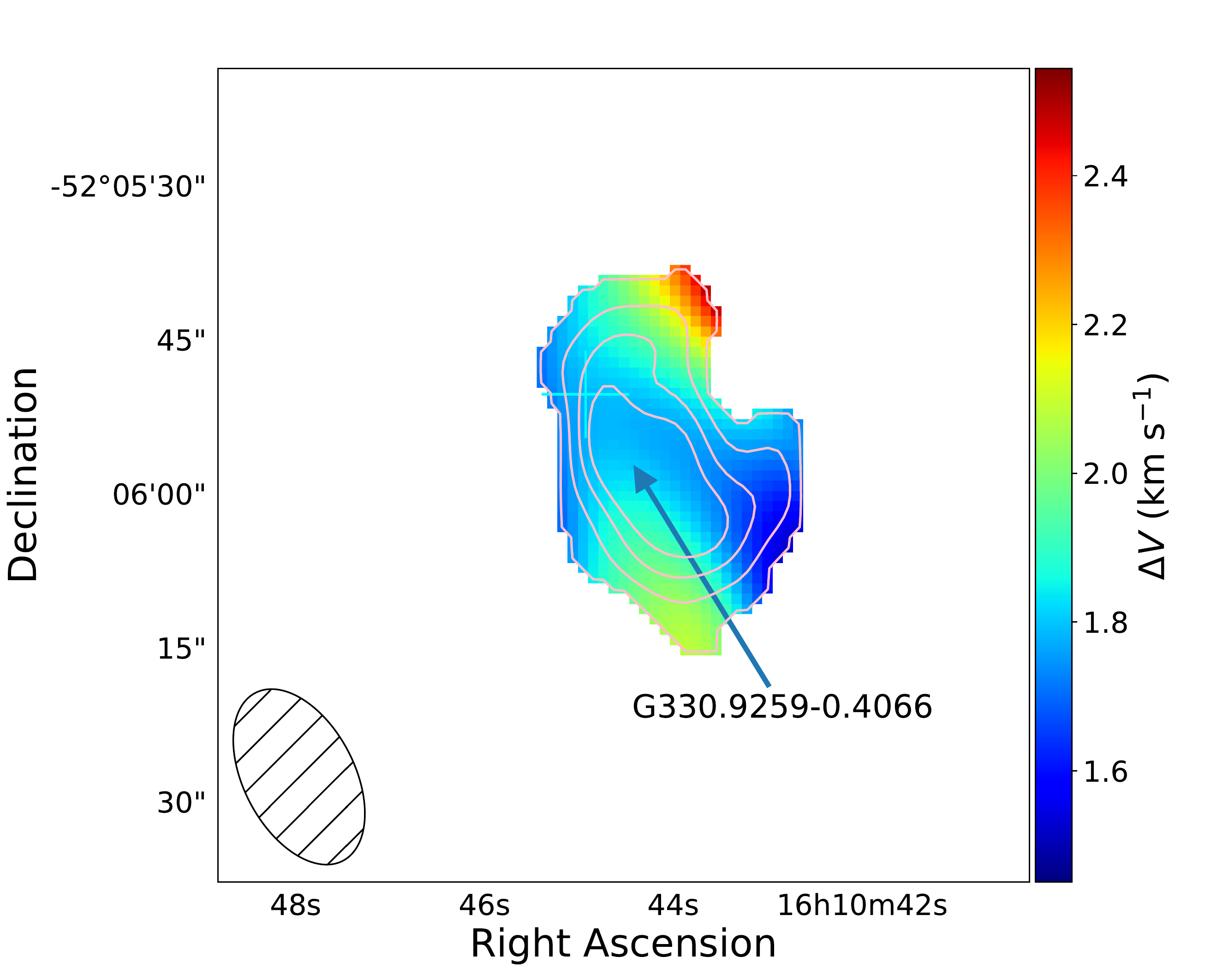} 
	\includegraphics[width=0.45\textwidth]{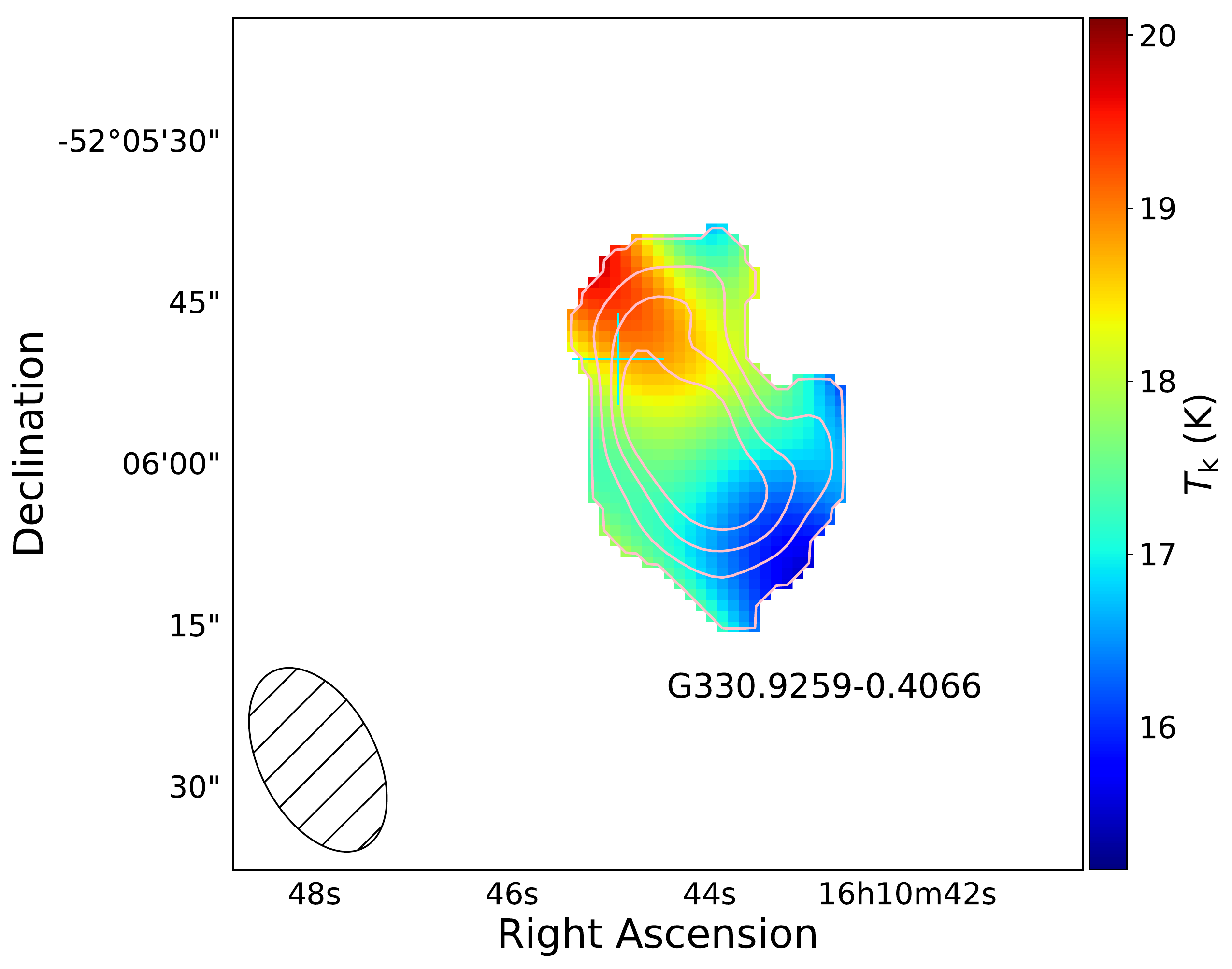}
	\caption{Sample maps of the various parameters described in Sections~\ref{sect:method} and \ref{sect:spectral_analysis} for region G330.9288-00.4070. In the upper left panel we present a three-colour RGB image created from the GLIMPSE data, overplotted with the \nhthree and ATLASGAL emission in yellow and red contours respectively. The contours levels are equivalent to those in Figure~\ref{fig:8mu_examples}. From the upper right clockwise to the bottom left, the maps presented are the velocity, velocity dispersion, and the kinetic temperature. The blue arrows and black labels identify ammonia clumps identified using the \fw\ algorithm. Orange and cyan markers identify the positions of any MYSOs or \hii regions respectively. Beam sizes are shown as a hatched black ellipsoids in the lower right corner of each plot.}
	\label{fig:example_plots_309}
\end{figure*}

\subsubsection{Column Densities}
\label{sect:column_densities}

The \nhthree column density is calculated as a free parameter using the inversion transition hyperfine structure. This essentially uses the derived rotation temperature, and so has already taken into account the beam filling factor. The optical depth is calculated from the ratio between the observed brightness of the main and inner satellite lines, but low signal-to-noise ratios for the satellite lines produce large uncertainty in this ratio, resulting in a very poorly constrained optical depth for weakly-detected pixels. As the column density of the \nhthree\ gas is directly derived from the optical depth, the majority of column density values are also unreliable. Pixels with this feature were located based on their high relative errors (>100\%) from the associated column density error maps produced by the fitting routine and removed from the analysis. 

The statistical values for each parameter is given in Table~\ref{table:fitted_parameters_table_sub} and a summary of these for the whole sample is given in Table~\ref{table:total_stats}.

\begin{table*}
	\begin{threeparttable}
		\caption{\label{table:fitted_parameters_table_sub}Detected \nhthree clump parameters. The columns are as follows: (1) field ID given in Table 1; (2) Name of clump derived from Galactic coordinates of peak emission of each clump; (3)-(4) radial velocity and intrinsic FWHM line width; (5-7) show different calculated temperatures for each clump; (8) beam filling factor ($T_{\rm{ex}}$/$T_{\rm{rot}}$); (10) \nhthree column density. (For each cell, initial values are peak values, with median values shown in parenthesis for the pixel by pixel fitting routine.) Regions appended with * identified regions with contain dual spectra components.}
		\begin{tabular}{cccccccccccc}
			\hline
			\hline
			Field & Clump name & $v_{\rm{lsr}}$ & $\Delta v$ & $T_{\rm ex}$ & $T_{\rm rot}$ & $T_{\rm kin}$ &  $B_{\rm ff}$ & log(column density) \\
			id &  & (\kms) & (\kms) & (K) & (K) & (K) & & (cm$^{-2}$) & \\
			
			(1)& (2) & (3) & (4) & (5) & (6) & (7) & (8) & (9) \\
			\hline
			1 &  G261.6454$-$2.0884 &   14.55 &  2.97 (2.45) &  3.15 (2.82) &  23.89 (21.07) &  28.86 (24.49) &  0.15 (0.13) &  16.41 (15.96) \\
			2 &  G286.2124+0.1697 &  $-$21.15 &  2.84 (1.53) &  2.86 (2.83) &  17.28 (16.50) &  19.15 (18.12) &  0.17 (0.17) &  15.94 (14.79) \\
			2 &  G286.2052+0.1706 &  $-$18.09 &  2.91 (2.13) &  2.94 (2.81) &  23.98 (19.43) &  29.00 (22.10) &  0.17 (0.15) &  14.70 (14.35) \\
			3 &  G305.2084+0.2061 &  $-$41.94 &  8.69 (6.50) &  2.93 (2.82) &  27.81 (25.20) &  35.60 (31.01) &  0.14 (0.11) &  17.60 (15.45) \\
			4 &  G309.4208$-$0.6204 &  $-$42.51 &  1.66 (1.46) &  2.84 (2.81) &  17.88 (15.90) &  19.96 (17.34) &  0.22 (0.18) &  18.87 (17.42) \\
			5 &  G312.5987+0.0449 &  $-$63.55 &  3.08 (2.69) &  2.85 (2.82) &  24.09 (20.97) &  29.18 (24.34) &  0.15 (0.13) &  15.27 (14.74) \\
			6 &  G314.3214+0.1146 &  $-$47.44 &  1.74 (1.62) &  2.81 (2.80) &  18.85 (16.95) &  21.29 (18.71) &  0.20 (0.16) &  18.78 (15.95) \\
			6 &  G314.3197+0.1092 &  $-$48.26 &  2.11 (1.85) &  2.77 (2.77) &  17.15 (16.31) &  18.98 (17.88) &  0.19 (0.17) &  17.42 (17.11) \\
			7 &  G318.9477$-$0.1959 &  $-$34.54 &  3.98 (2.82) &  2.88 (2.83) &  22.26 (19.99) &  26.29 (22.91) &  0.17 (0.14) &  18.57 (17.41) \\
			8 &  G322.1584+0.6361 &  $-$57.91 &  4.47 (3.83) &  2.87 (2.81) &  29.43 (23.28) &  38.65 (27.87) &  0.14 (0.12) &  19.02 (16.69) \\
			\hline
		\end{tabular}
		\begin{tablenotes}
			\centering
			\item Notes: Only a small portion of the data is provided here, the full table is only available in electronic format.
		\end{tablenotes}
	\end{threeparttable}
\end{table*}

\begin{figure}
	\includegraphics[width=0.5\textwidth]{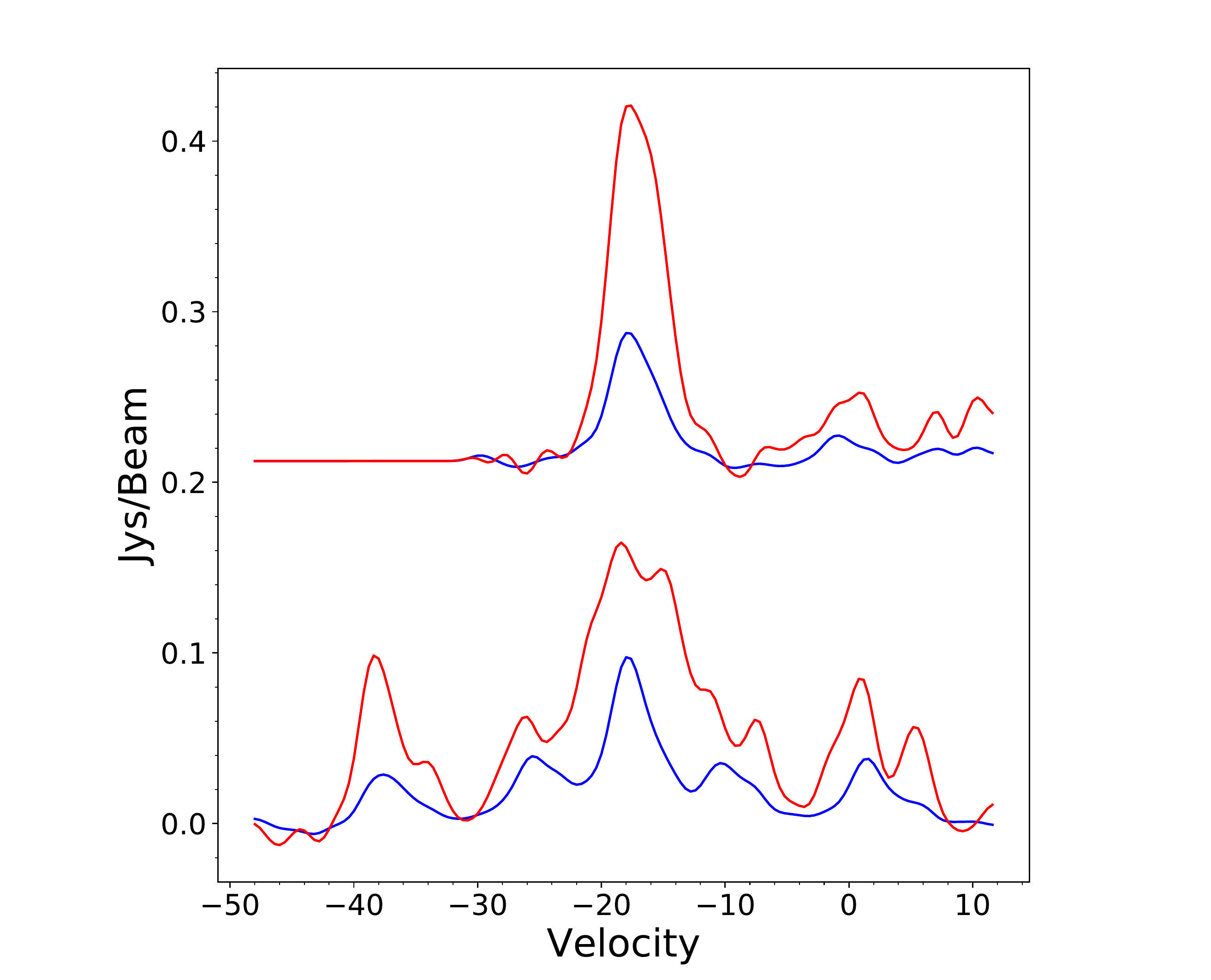}
	\caption{Example \nhthree spectra of field G345.5043+00.3480. The peak and average spectra across the clump are shown in red and blue respectively. The spectra have been smoothed with a 1D gaussian kernel ($\sigma$ = 2).}
	\label{fig:dual_spectra}
\end{figure}

\begin{table*}
	\begin{threeparttable}
		\caption{\label{table:total_stats}Statistical properties for the whole sample.}
		\begin{tabular}{lccccccc}
			\hline
			\hline
			Parameter &  Number &    Mean &  Standard Error &  Standard Deviation &  Median &     Min &     Max \\
			\hline
			Aspect Ratio   & 44 & 1.66 & 0.09 & 0.60 & 1.57 & 1.03 & 4.68 \\
			Angular Offset & 44 & 18.19 & 2.41 & 15.97 & 15.32 & 0.04 & 79.5 \\
			Distance (kpc) (fields) & 34 & 3.46 & 0.21 & 1.22 & 3.30 & 1.80 & 6.70 \\
			Radius (pc) (fields)     & 34 & 0.15 & 0.01 & 0.09 & 0.12 & 0.04 & 0.36 \\
			\hline
			Tkin (Mean) (K) & 44 & 22.31 & 0.74 & 4.90 & 21.47 & 14.43 & 36.67 \\
			FWHM line width (Mean) (\kms) & 44 & 2.69 & 0.17 & 1.15 & 2.47 & 1.28 & 6.09 \\ 
			Pressure Ratio (Mean) & 44 & 0.01 & 0.001 & 0.01 & 0.01 & 0.002 & 0.02 \\ 
			Beam filling factor (Mean)  & 44 & 0.15 & 0.004 & 0.02 & 0.15 & 0.09 & 0.21 \\ 
			N(NH3) (Mean)   & 44 & 16.72 & 0.26 & 1.71 & 16.58 & 14.33 & 22.36 \\ 
			\hline
			Log[Field mass]  (M$_{\sun}$) & 29 & 3.21 & 3.18 & 0.07 & 3.18 & 2.52 & 3.98 \\
			Log[Field luminosity] (L$_{\sun}$) & 29 & 4.37 & 0.13 & 0.72 & 4.46 & 3.13 & 5.83 \\
			\hline
		\end{tabular}
	\end{threeparttable}
\end{table*}

\subsubsection{Uncertainties on the fitted parameters}
\label{sect:fitted_uncertainties}

The \fw\ algorithm provides no estimation of the uncertainties for the position or size of the detected clumps (which is a function of beam size and SNR); we calculate the mean error of this to be $\sim$3.7\arcsec\ (beam size / $\sqrt{SNR}$).

The procedure used during the fitting process automatically computes error maps for each individual free parameter which is derived from the ammonia spectra, the uncertainties for which are the derived uncertainties from the nonlinear least-squares fitting using the covariance matrix. These uncertainties include statistical and systematic errors arising from the model and also error propagation, a detailed explanation of the errors for this fitting routine can be found in \citep{Rosolowsky2008}.
Uncertainties for the temperatures ($T_{\rm{kin}}$, $T_{\rm{ex}}$, $T_{\rm{rot}}$) and velocity
are relatively small, typically less than 0.5\,K and 0.1\,km\,s$^{-1}$, respectively.
The errors for the beam filling factor are also relativity low as the values are calculated from the ratio of the  excitation and rotational temperatures. As mentioned in Section~\ref{sect:column_densities}, errors for the majority of optical depth and column density calculations cause these values to be unreliable.

A number of fields within the sample show signs of multiple components: visual inspection indicates that five fields have at least two spectral components in either the \nhthree (1,1) or (2,2) transitions, as there is a clear difference between the average and peak spectra as shown in Figure~\ref{fig:dual_spectra}. The \nhthree (1,1) average spectrum appears coherent, whereas the peak spectrum shows an indication of multiple components at different velocities, which are unresolved, as the FWHM of each \nhthree spectral line component is not separated. Multiple components tend to increase the measured line-widths, and it is difficult to resolve both individual components without higher spectral and angular resolution. Affected regions can be found labeled with a star (*) in Table~\ref{table:fitted_parameters_table_sub}.

%{\color{red} [TM: not clear what this means -- is the difference between the mean and peak {\em velocities}?  How large does it have to be before declaring multiple components detected?]}. 

The statistical values for the luminosity and mass of all dust clumps can be found in Table~\ref{table:total_stats}, while the values for individual sources are given in Table~\ref{table:atlas_values}. Fitted parameter maps for all of the regions, along with the corresponding GLIMPSE images, can be found in the appendix.

\begin{figure}
  \includegraphics[width=0.45\textwidth]{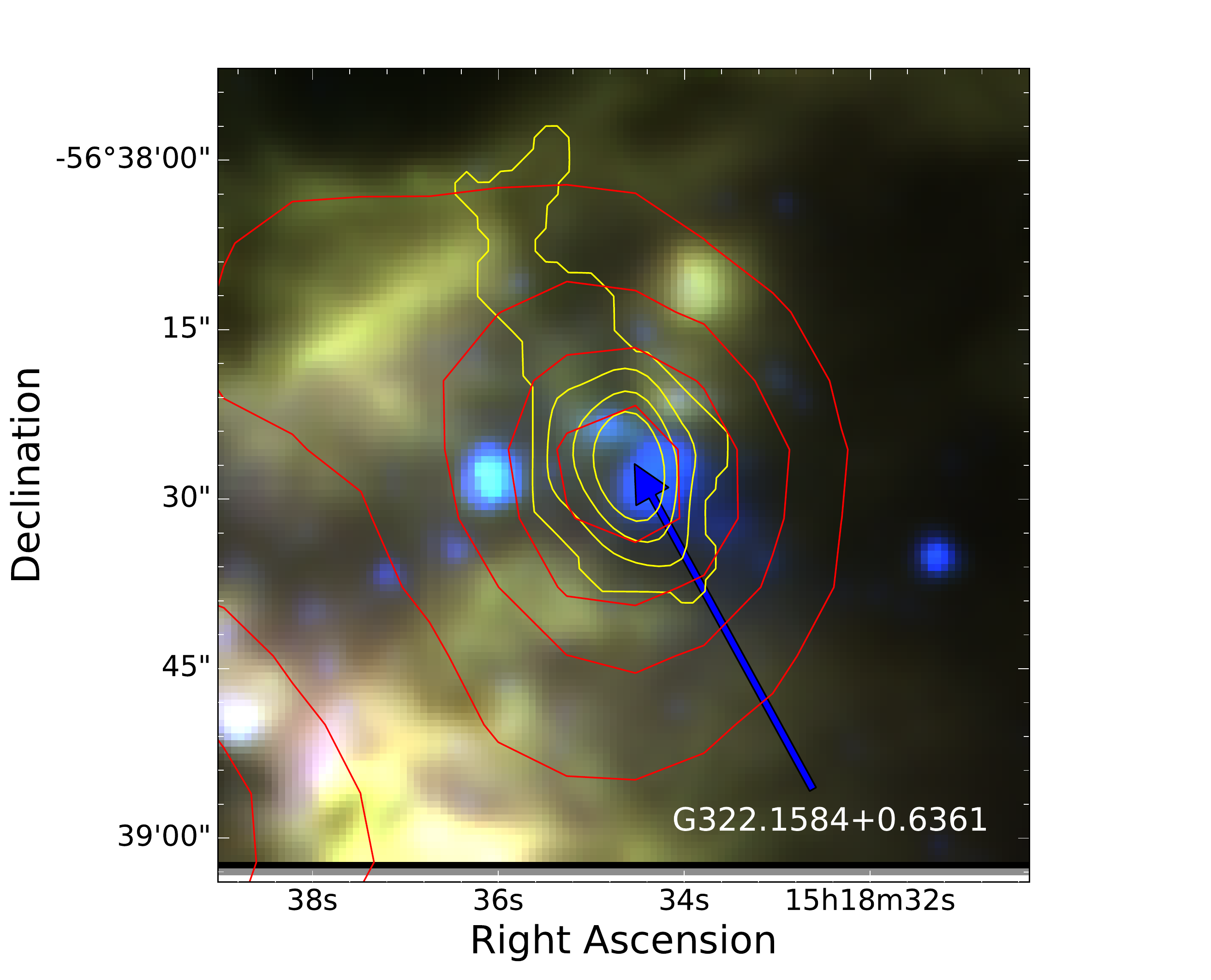}
  \includegraphics[width=0.45\textwidth]{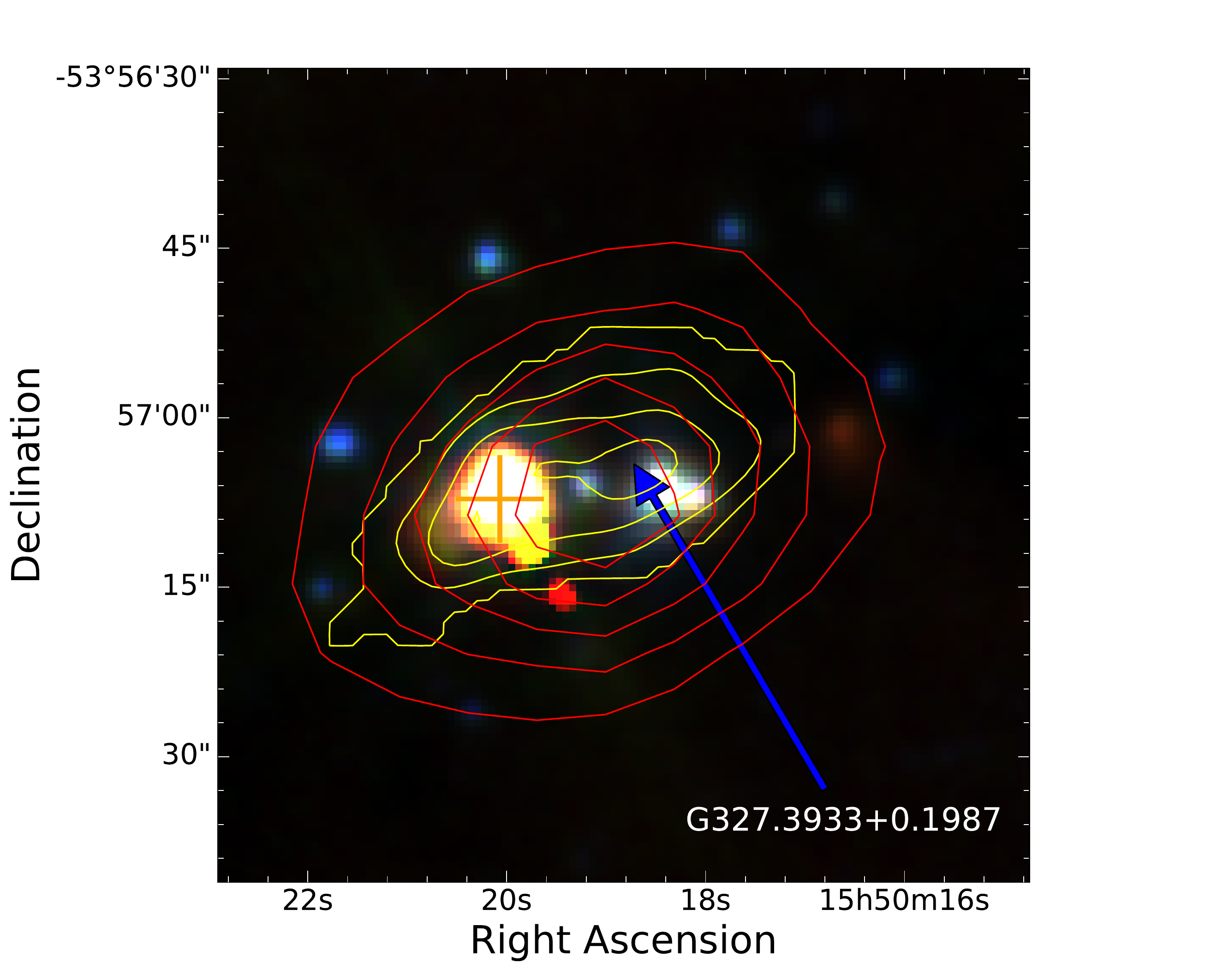}
  \includegraphics[width=0.45\textwidth]{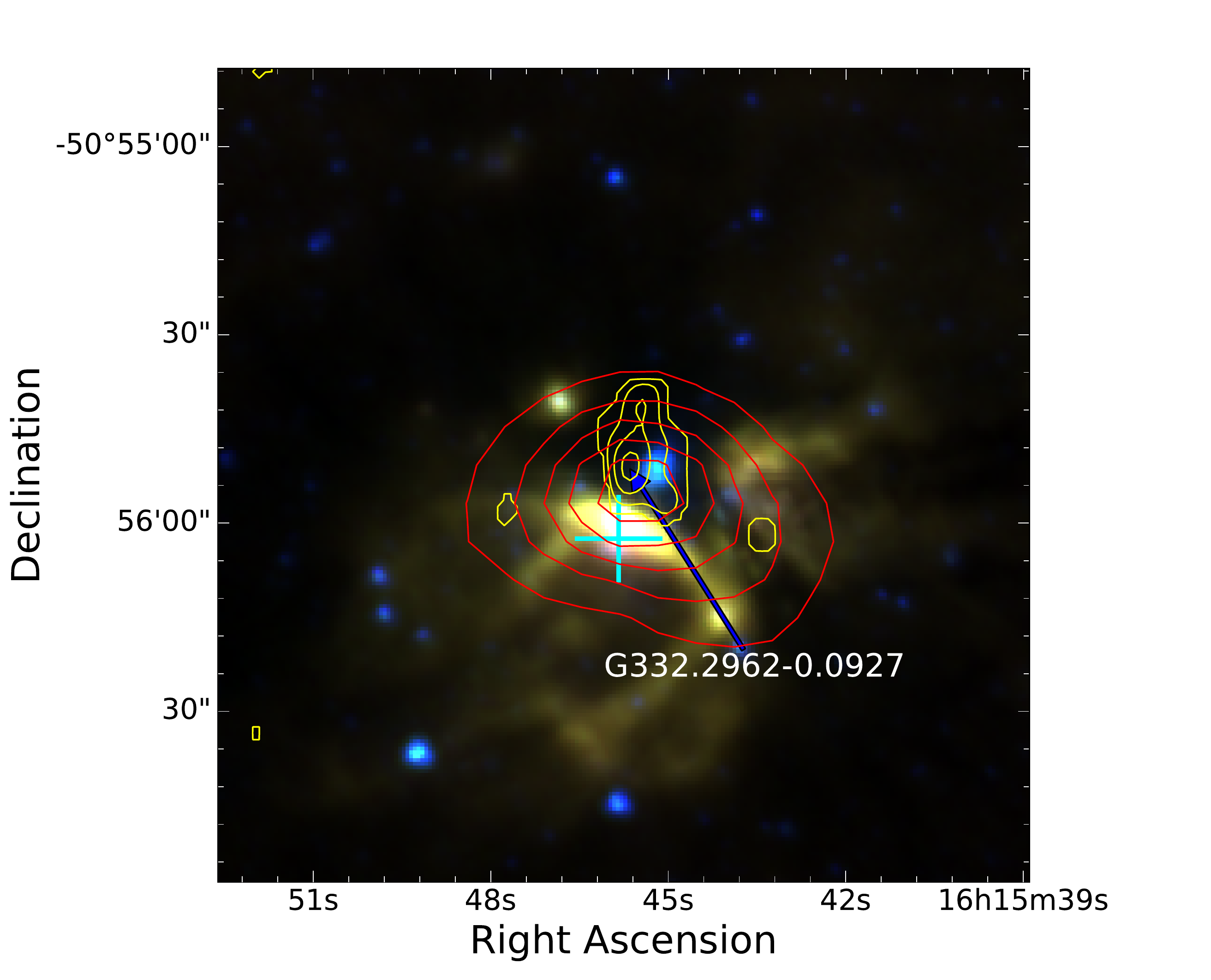}
\caption{Example GLIMPSE three-colour images of the morphological structure and differences between the three evolutionary stages outlined in this study. The upper panel presents a quiescent clump with no embedded or nearby RMS source, the middle panel presents an ESF field, with an embedded MYSOs, while the lower panel presents an LSF field, where a \hii region has developed and started to disrupt its environment. \nhthree (1,1) contours are shown in yellow starting at 3$\sigma$ increasing in steps of 0.5$\sigma$, equivalent to those in Figure~\ref{fig:8mu_examples}. RMS objects are marked with magenta and cyan crosses for MYSOs and \hii regions respectively.}
\label{fig:example_evolutionary_stage_maps}
\end{figure}

\section{Discussion of Physical Parameters}
\label{sect:physical_discussion}

\subsection{Overview of Derived Properties}

We have derived multiple properties from the \nhthree spectral line analysis (kinetic temperatures, FWHM line-widths, beam filling factors and pressure ratios), which have been supplemented with parameters from the ATLASGAL survey (luminosities and masses). Values for the temperatures and beam filling factors, which range between 10-40\,K and 0.09-0.21, along with our measured FWHM line-widths, are similar to previous studies \citep{Sridharan2002,Wienen2012}.

\subsection{Evolutionary Sequence}

As outlined in the introduction, the main purpose of this study is to investigate the initial physical conditions of the environments of high-mass stars in order to better understand where they form and how their feedback in turn affects their environments. We examine  morphology of the molecular gas with respect to the different type of embedded objects and investigate how they are affecting their local environment. The accumulated ATLASGAL survey data has aided in the understanding of these regions and the larger structure in the local environment and how it relates to the dense gas as mapped by the \nhthree emission.

It is possible to construct a rudimentary evolutionary track by incorporating all of this information and the three types of classifications for each field. An example of how this can be visualised is shown in Figure~\ref{fig:example_evolutionary_stage_maps}. A region begins in the quiescent/protostellar state with no observable ongoing formation and no visible embedded objects at the infrared wavelengths of the GLIMPSE survey (3.6, 4.5 \& 8\micron). These initial regions are already associated with methanol masers, and are therefore harbouring massive protostars (upper panel of Figure~\ref{fig:example_evolutionary_stage_maps}). The morphology of these clumps is relatively unbroken with only a single clump of emission. 
As accretion processes drive the embedded objects into the MYSO stage, the environments increase in temperature and turbulence as the MYSOs feed back into their local environment.
These early-stage clumps have high signal-to-noise and a well-defined structure with at least one RMS object embedded towards the centre of the emission (middle panel of Figure~\ref{fig:example_evolutionary_stage_maps}). Once the MYSO has reached maturity and it begins to produce an \hii region, it will disrupt its environment, fragmenting any dense material which could be observed via \nhthree emission (lower panel of Figure~\ref{fig:example_evolutionary_stage_maps}).

\begin{figure*}
  \includegraphics[width=0.45\textwidth]{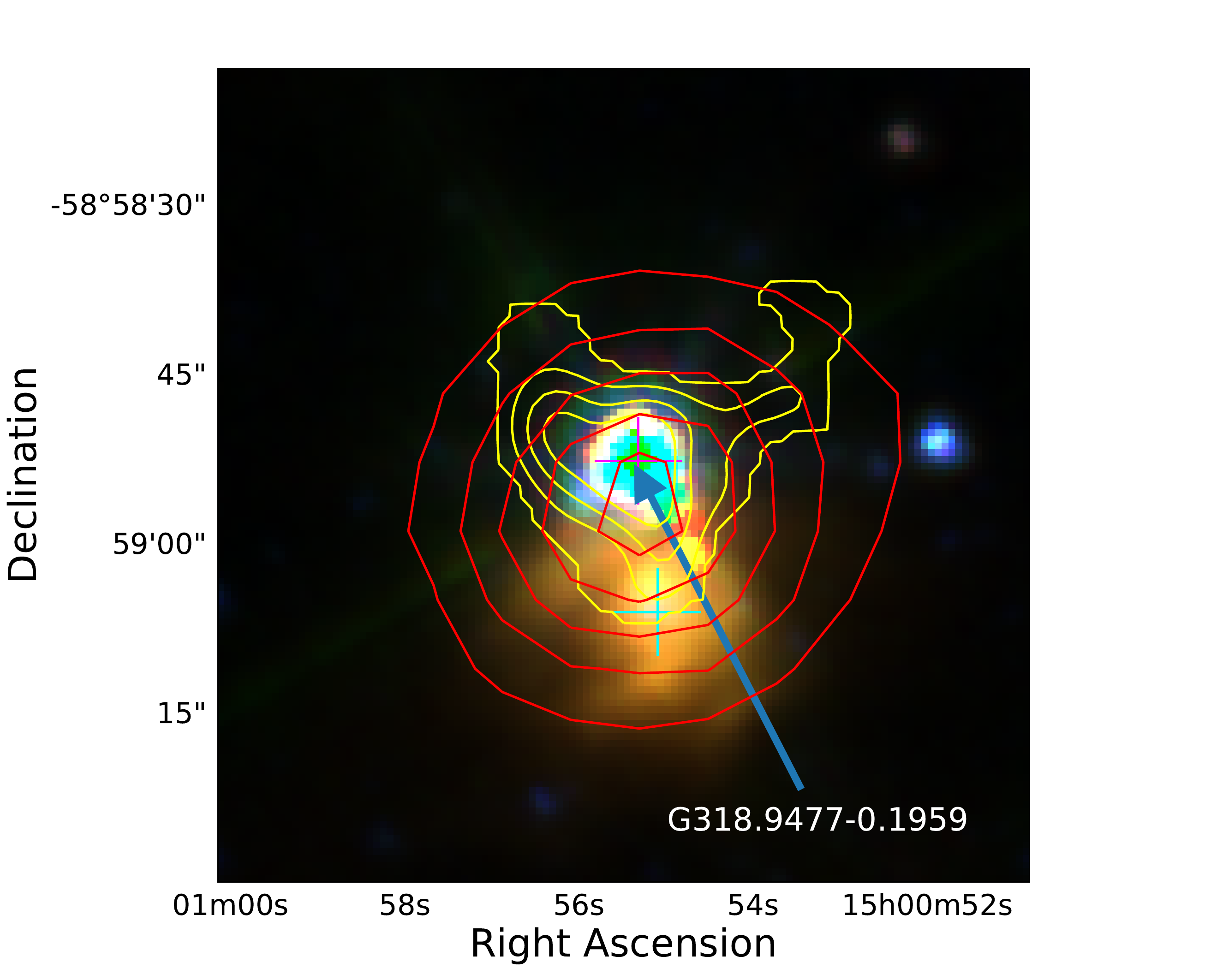}
  \includegraphics[width=0.45\textwidth]{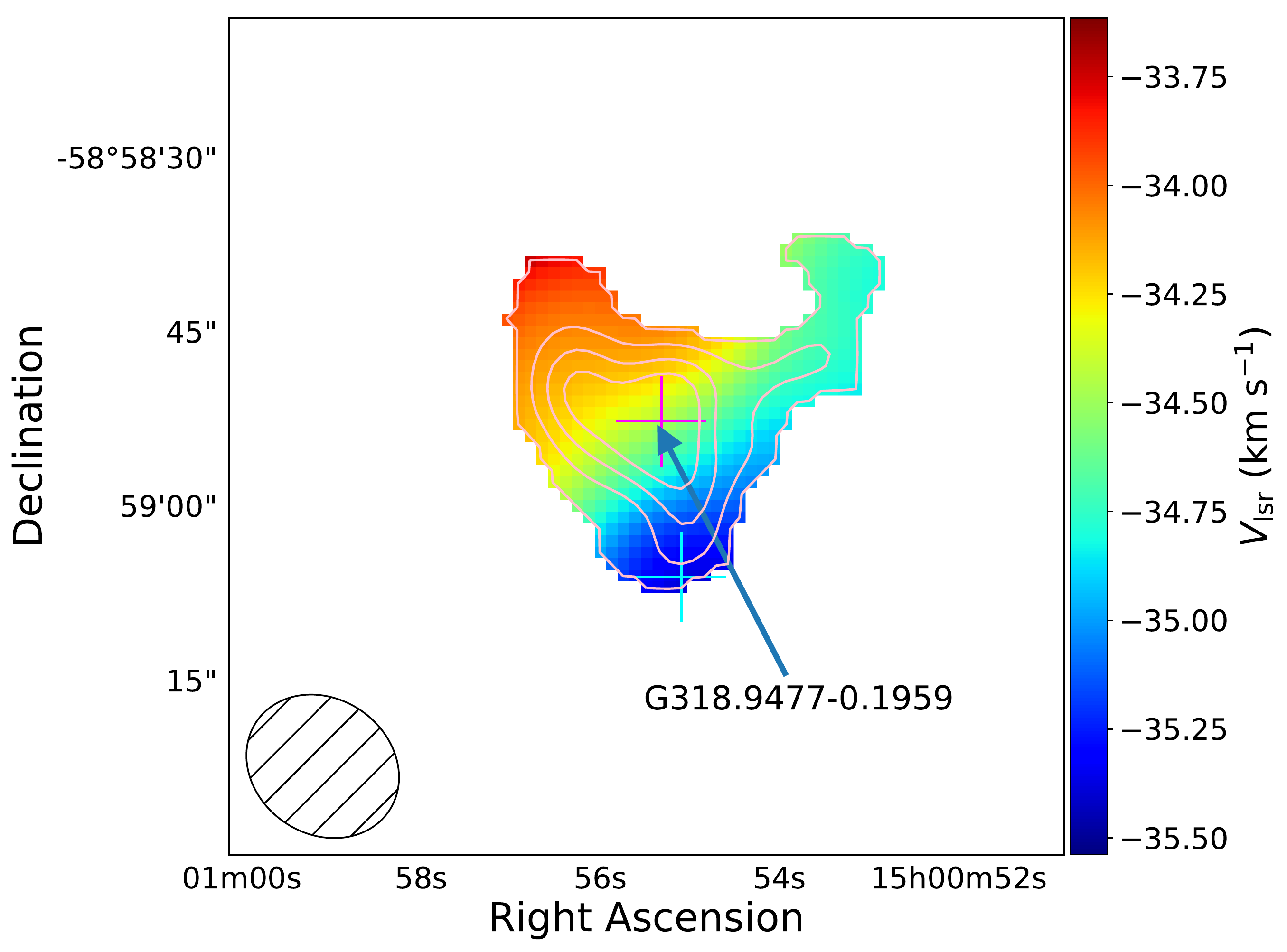} \\
  \includegraphics[width=0.45\textwidth]{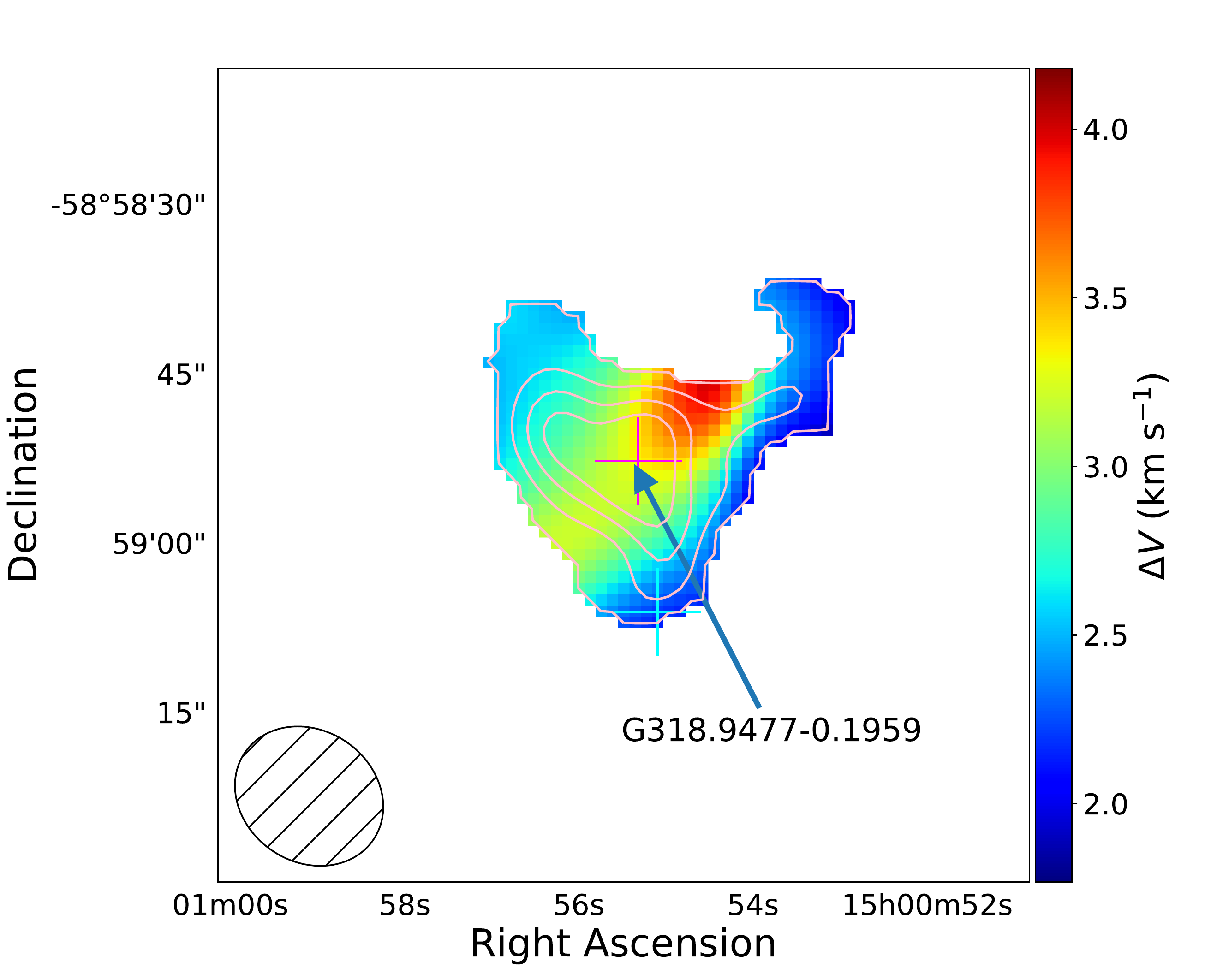} 
  \includegraphics[width=0.45\textwidth]{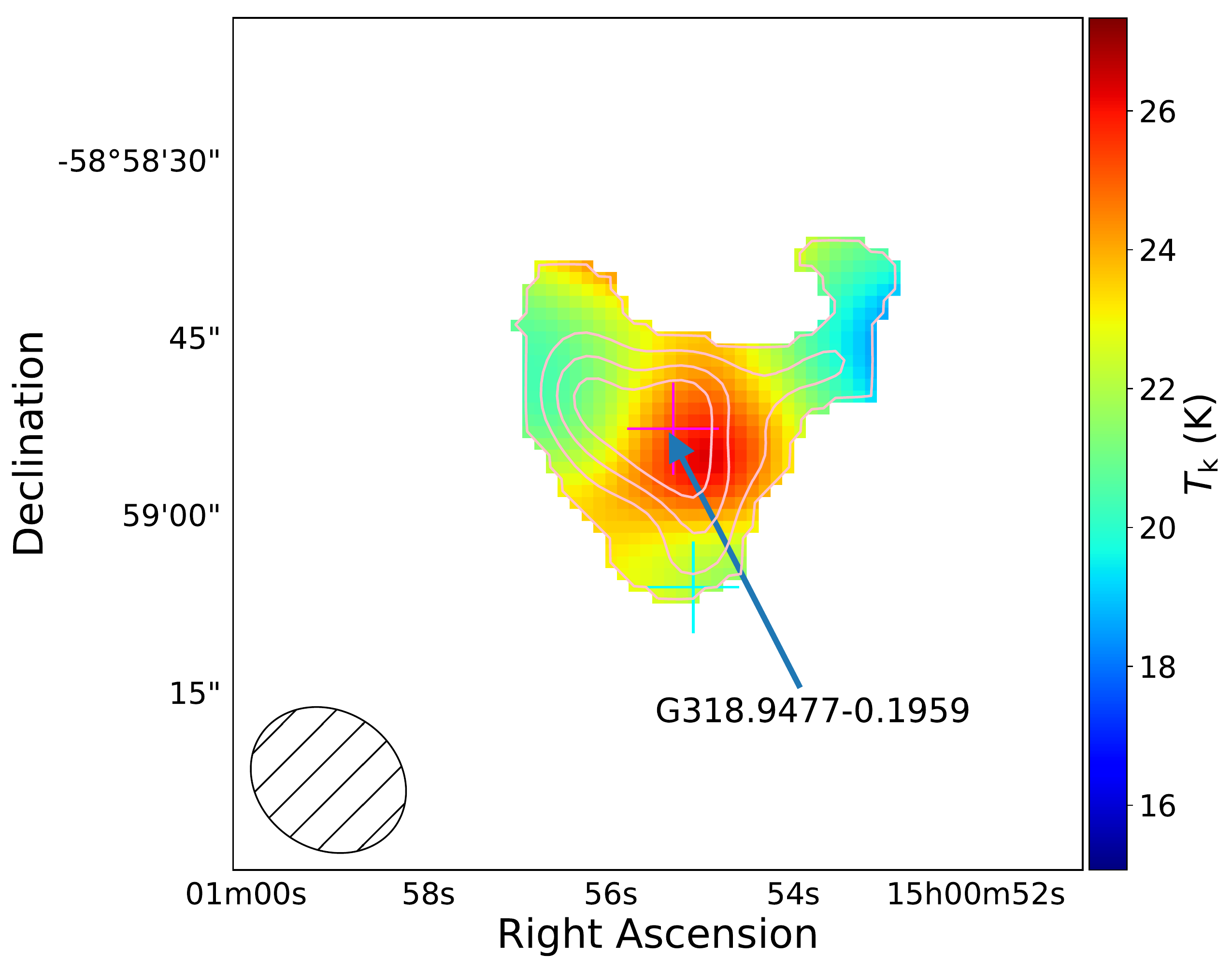}
\caption{Example maps showing field G318.9480-00.1969. Each panel presents, clockwise from the upper left, three-colour RGB GLIMPSE image overlaid with \nhthree (1,1) and ATLASGAL dust emission contours in yellow and red respectively, velocity, kinetic temperature and velocity dispersion. The contours levels are equivalent to those in Figure~\ref{fig:8mu_examples}. Magenta and cyan crosses mark MYSOs and \hii regions respectively.}
\label{fig:example_maps_318}
\end{figure*}

This proposed sequence of events is similar in nature to previous studies (e.g. \citealt{Zinnecker2007,Chambers2009,Battersby2014}), where similar evolutionary sequences are presented using a number of different tracers. \cite{Chambers2009} proposed an evolutionary sequence for infrared dark clumps (IRDCs), this begins with a quiescent clump transitioning into an active clump and finally into a more evolved "red" clump. These classifications were based solely on IR emission, 24\micron\ point sources and enhanced 8\micron\ emission. That evolutionary model is similar in nature to what is presented in this study, although the main focus for this work is on the distribution and offsets of \nhthree and 870\micron\ emission, relative to the positions of RMS sources. This work also gives a physical view to other large sample studies, such as (\citealt{Urquhart2014}; Figure 21), which includes ATLASGAL compact source catalogue data and shows two distinct phases, accretion and dispersion, which can clearly be seen in our observations. While this may not be a robust evolutionary sequence model for massive star formation, our data does nicely highlight differences between each evolutionary stage identified here.

\subsection{Comparison of Physical Parameters}

The statistical properties for the entire sample are given in Table~\ref{table:total_stats}, and a complete set of maps for each individual field can be found in the appendix.

Throughout this study we have classified each individual source as either early-phase star-forming (ESF) or late-phase star-forming (LSF), and have tried to identify differences between these two classes. The principle basis for this classification system is the difference in morphologies as shown in Figure~\ref{fig:esf_example_maps} and \ref{fig:lsf_example_maps}. The use of maser emission data has additionally allowed the reclassification of regions which appear to be quiescent but are likely to be in a protostellar stage, and therefore are in an early stage of star formation. Although the quiescent clumps are each associated with at least one methanol maser, which are almost exclusively associated with high-mass star formation (e.g. \citealt{breen2011_methanol}) and are therefore habouring a protostar, they are yet in an earlier evolutionary stage than the ESF clumps.

We have conducted analysis of the various fitted and derived parameters from the \nhthree observations, GLIMPSE and ATLAGSAL emissions. ESF and LSF region cumulative distributions were analysed for six of the parameters (luminosity, mass, column density, clump volume and clump velocity dispersion) in an effort to identify any significant differences and to compare the properties of the two evolutionary samples identified. We have employed the use of a two-sample KS-test for each distribution. The KS-test is used for calculating the probability that two samples are drawn from the same population, and produces a $p$-value which can be used to reject the null hypothesis. We use a 3$\sigma$ confidence threshold ($\alpha$ < 0.0013) to reject the null hypothesis that any two distributions are drawn from the same parent population. All associated $p$-values were found to be greater than this confidence value, and so no statistical difference can be seen between the ESF and LSF fields/clumps for any of these parameters.

\begin{figure*}
	\includegraphics[width=0.45\textwidth]{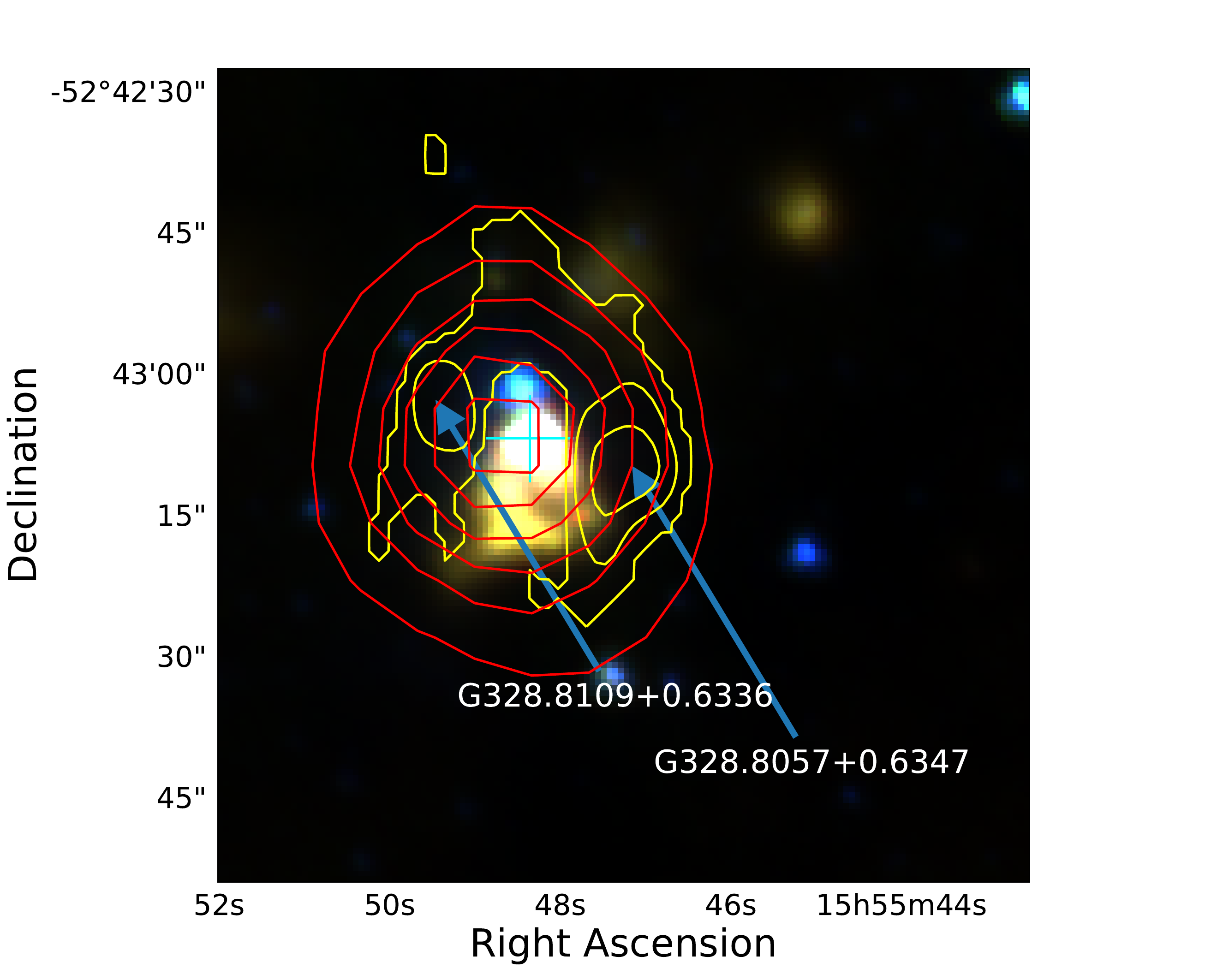}
	\includegraphics[width=0.45\textwidth]{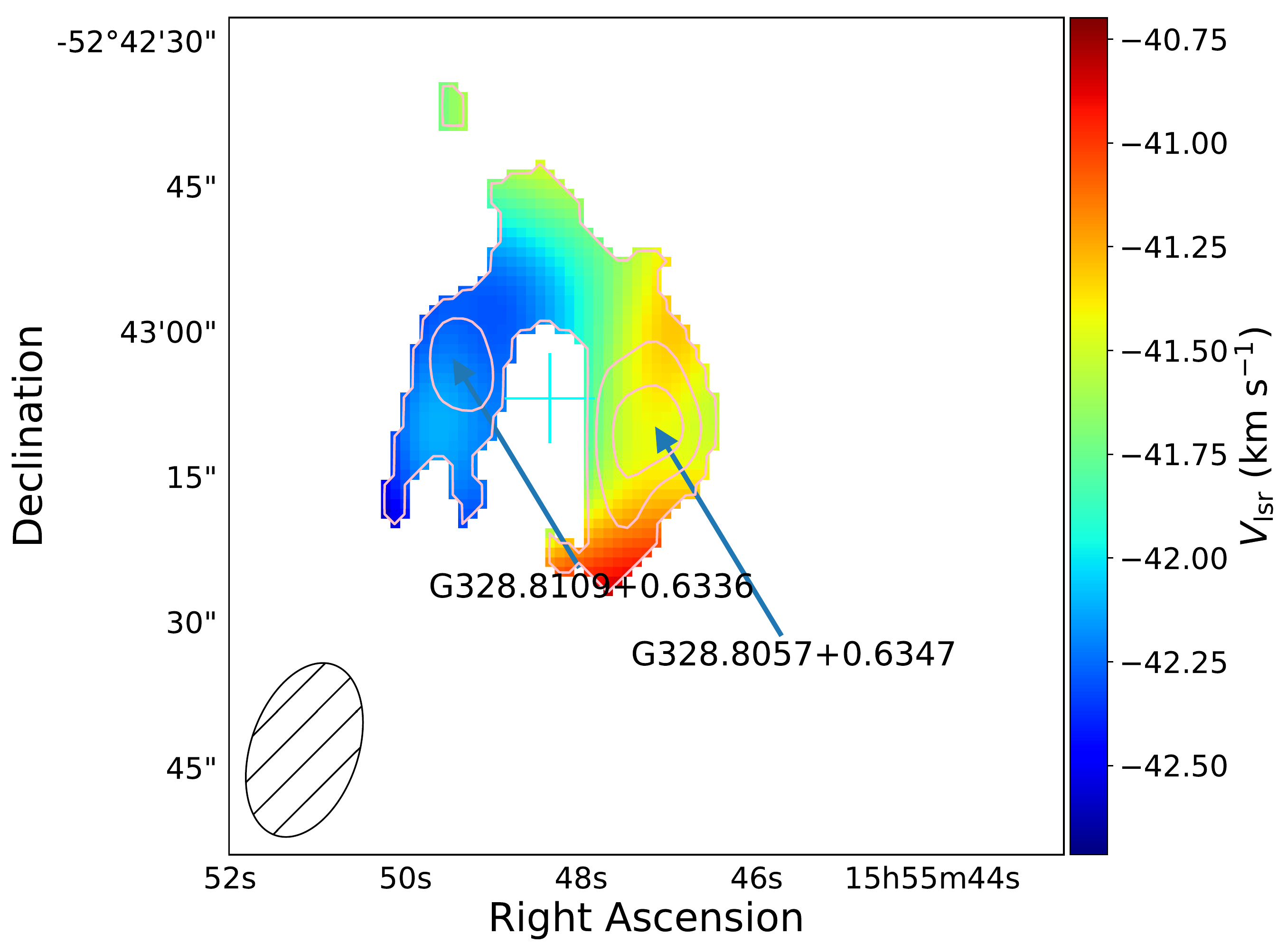} \\
	\includegraphics[width=0.45\textwidth]{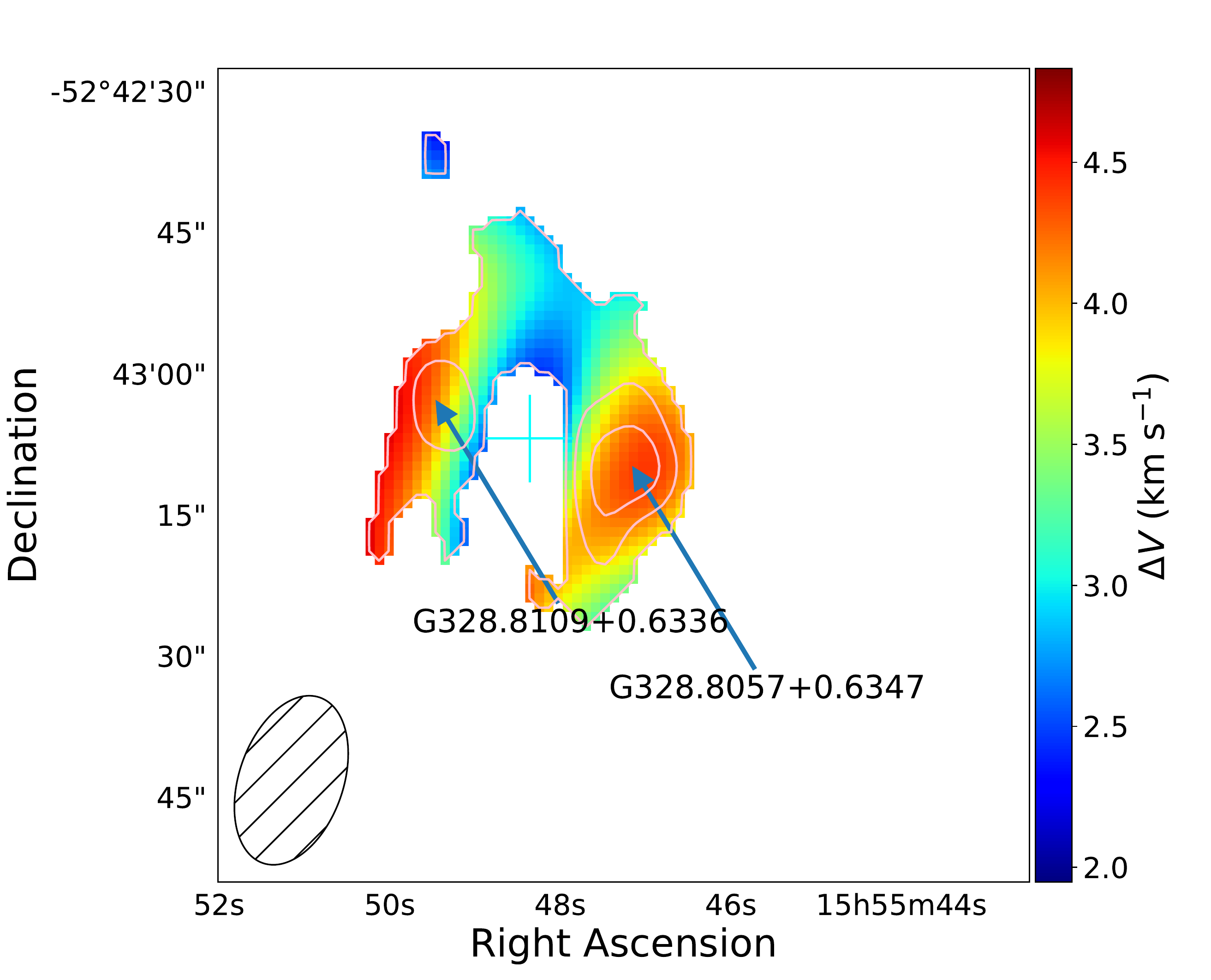} 
	\includegraphics[width=0.45\textwidth]{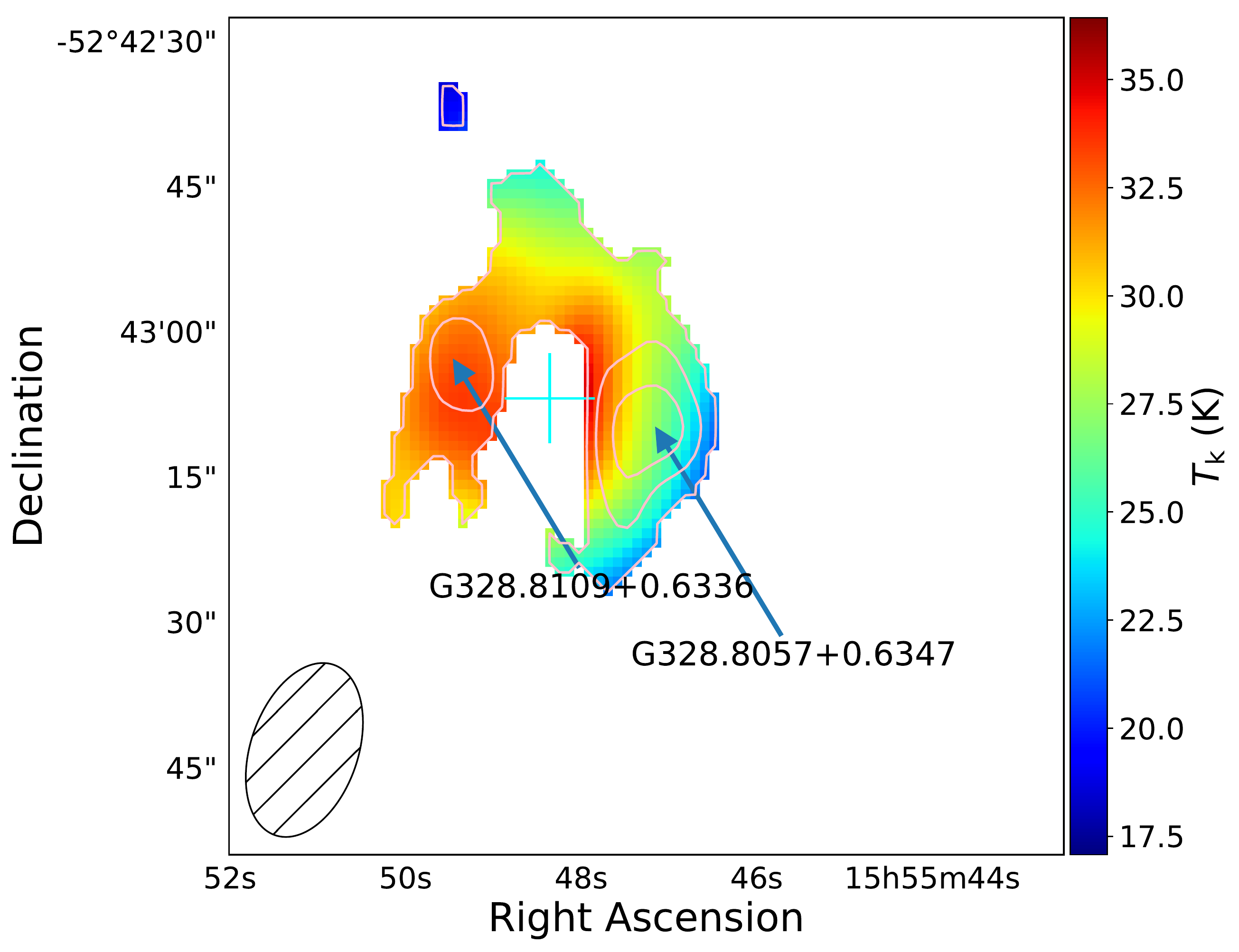}
	\caption{Examples maps showing field G328.8074+00.6324. Each panel presented is the same as Figure~\ref{fig:example_maps_318}.}
	\label{fig:example_maps_328}
\end{figure*}

Temperature and velocity dispersion gradients are common across the sample, ranging from $\sim$10 to $\sim$40\,K. The peak kinetic temperature measurements of star-forming clumps are in general coincident with an embedded object and decrease towards the edges for the ESF clumps. Many clumps located near LSF sources that have recently broken out of their natal clumps exhibit gradients, with values that peak at the edge closest to the RMS source and decrease with increasing distance from that source. Examples of these two gradient cases seen towards the ESF and LSF clumps are shown in Figures~\ref{fig:example_maps_318} and \ref{fig:example_maps_328}, respectively. These regions will be discussed in the following section.

Feedback from the RMS sources is therefore having a significant impact on the temperature and dynamics of their natal clumps regardless of whether they are still deeply embedded (ESF) or have already started to disrupt their local environment and are starting to emerge from their dust cocoons (LSF). Both classifications have similar median temperatures across the sample, so while the temperature gradients may differ as described above, the absolute values are similar (15-35\,K).

\begin{figure}
	\includegraphics[width=0.45\textwidth]{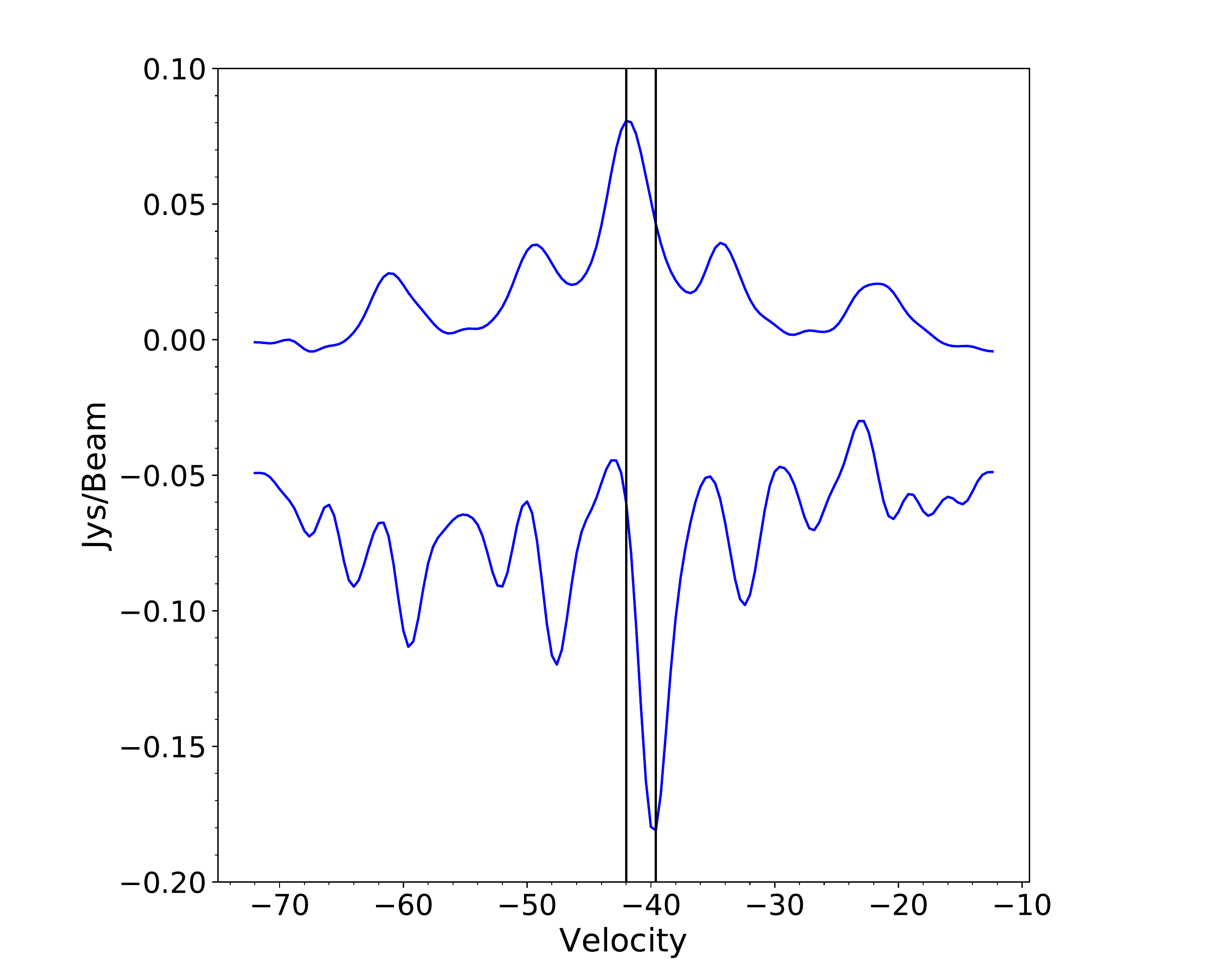}
	\caption{Absorption and emission spectra from field: G328.8074+00.6324. The absorption spectrum has been offset by -0.05\,Jy. The black lines represent the peaks from the two spectra. The spectra have been smoothed with a 1D gaussian kernel ($\sigma$ = 2).}
	\label{fig:G328_spectra}
\end{figure}

Region-wide parameter distributions (luminosity, masses and column densities) between the two classifications were also investigated for significant differences. It might be expected that the luminosity for LSF regions should be naturally higher as the embedded objects are more luminous whereas ESF clumps are likely to show higher masses and column densities since the natal environments have not started to be dispersed.
However, as previously noted, we find no statistical difference between the mass and luminosities of the two classifications in KS-tests. The sizes of individual clumps has also been compared, and we find that younger clumps tend to be larger than the more evolved material in terms of pixel area (arcsecond$^2$). The mean luminosity-to-mass (L/M) ratio for ESF sample is 1.33 with a standard error of 0.03, while the mean LSF ratio is found to be 1.4, with a standard error of 0.05. The median values do also differ by a similar amount. The L/M ratios are a good indicator of evolutionary stage (increasing L/M indicates more evolved regions), therefore, from this analysis, it appears that the entire sample covers a narrow evolutionary time scale. However, analysis of the morphological distribution of the dense gas (as traced by the ammonia) do reveal significant differences between the two evolutionary groups that are not observed in the dust emission maps, which is due to the relatively low spatial resolution of the ATLASGAL survey. The dust emission traces the total column density, rather than just the high-density regions, and so may smooth out any clumpy substructure that is present in the sources. 

\citet{urquhart2014_csc} found the L/M ratio to be similar for a large sample of MYSOs and \hii regions identified by the RMS survey and concluded that both evolutionary samples were towards the end of the main accretion phase, which is consistent with the finding here. It is also clear that although the \hii\ regions are having a significant impact on the internal structure of their natal clumps and so properties derived from dust emission alone are not very useful to investigate the evolution of these two stages.

\subsection{Interesting Sources}

In the following subsections we present three case studies of interesting regions, along with the associated three-colour RGB GLIMPSE image and parameter maps.

\subsubsection{G318.9477$-$0.1959}

Figure~\ref{fig:example_maps_318} shows a selection of parameter maps for a single star forming region (G318.9480-00.1969). This particular region has a \nhthree (1,1) SNR of 11.97 and is associated with two sources identified from the RMS survey, one of which is a MYSO (G318.9480$-$00.1969A) that is clearly embedded towards the peak of the ammonia emission.  The second is an \hii region (G318.9480$-$00.1969B) located towards the southern edge of the NH$_3$ clump. A single clump (G318.9477$-$0.1959) has been identified in this region by the \fw\ algorithm with an aspect ratio of 1.19, and so has a morphology that is fairly circular as seen by the \nhthree (1,1) emission contours. This clump is located towards the centre of the ATLASGAL dust emission, as expected for an embedded region in the early stages of star formation.

The GLIMPSE RGB image shows a central bright saturated source corresponding to the position of the MYSO, as well as revealing the presence of some diffuse MIR emission coincident with the \hii region. Given that the \hii region is not associated with any ammonia and is associated with diffuse MIR emission it would appear that the \hii region is quite evolved and has started to disperse its surroundings. These two sources dominate the mid-infrared emission in the map. The position of the central object is correlated with an increase in kinetic temperature as expected for an embedded MYSO (see lower right panel of Fig~\ref{fig:example_maps_318}); however, the \hii region to the south does not appear to be correlated with a locally elevated gas temperature.  This also indicates that it has already dispersed its natal material and is located either slightly in the foreground or background with respect to the clump hosting the MYSO. The line-width maximum is not coincident with either object (see lower left panel of Fig~\ref{fig:example_maps_318}), although the peak values do seem to form a elongated structure running north-west to south-east through the clump roughly centred on the position of the MYSO (see lower left panel of Fig~\ref{fig:example_maps_318}). This is somewhat suggestive of the presence of a bipolar molecular outflow. Similar types of this kind of structure were seen in Paper\,I in multiple fields (e.g., G010,47+00.03, G011.11-00.40, G013.33-00.03 and G014.61+00.02).

This source was studied as G318.9477$-$0.1959 by \cite{Navarete2015} as part of an investigation of the accretion processes in a sample of 353 MYSOs selected from the RMS survey. The study revealed that extended H$_2$ emission is a good tracer of outflow activity, which is itself a signpost of ongoing accretion processes. 

A comparison of this H$_2$ emission in Figure A296 from \cite{Navarete2015} to the \nhthree maps presented in this study shows that the outflow is directed towards the excavated cavity indicated towards the north west in Figure~\ref{fig:example_maps_318}; this is also in the direction of the maximum FWHM line width. Closer inspection indicates that there are in fact multiple outflows in this region, likely to be driven by a central cluster. None of these outflows is directly aligned with the FWHM maxima or the excavated cavity, however.

This field is also associated with one methanol and four water masers, all of which are offset less than 0.05\arcsec\ from the central RMS object (G318.9480$-$00.1969A). The presence of the methanol maser confirms that this region is indeed undergoing high-mass star formation, as these masers are thought to be excited in dense material in an accretion disk by mid-IR pumping from the central source (e.g. \citealt{DeBuizer2000a}).

\begin{figure*}
  \includegraphics[width=0.45\textwidth]{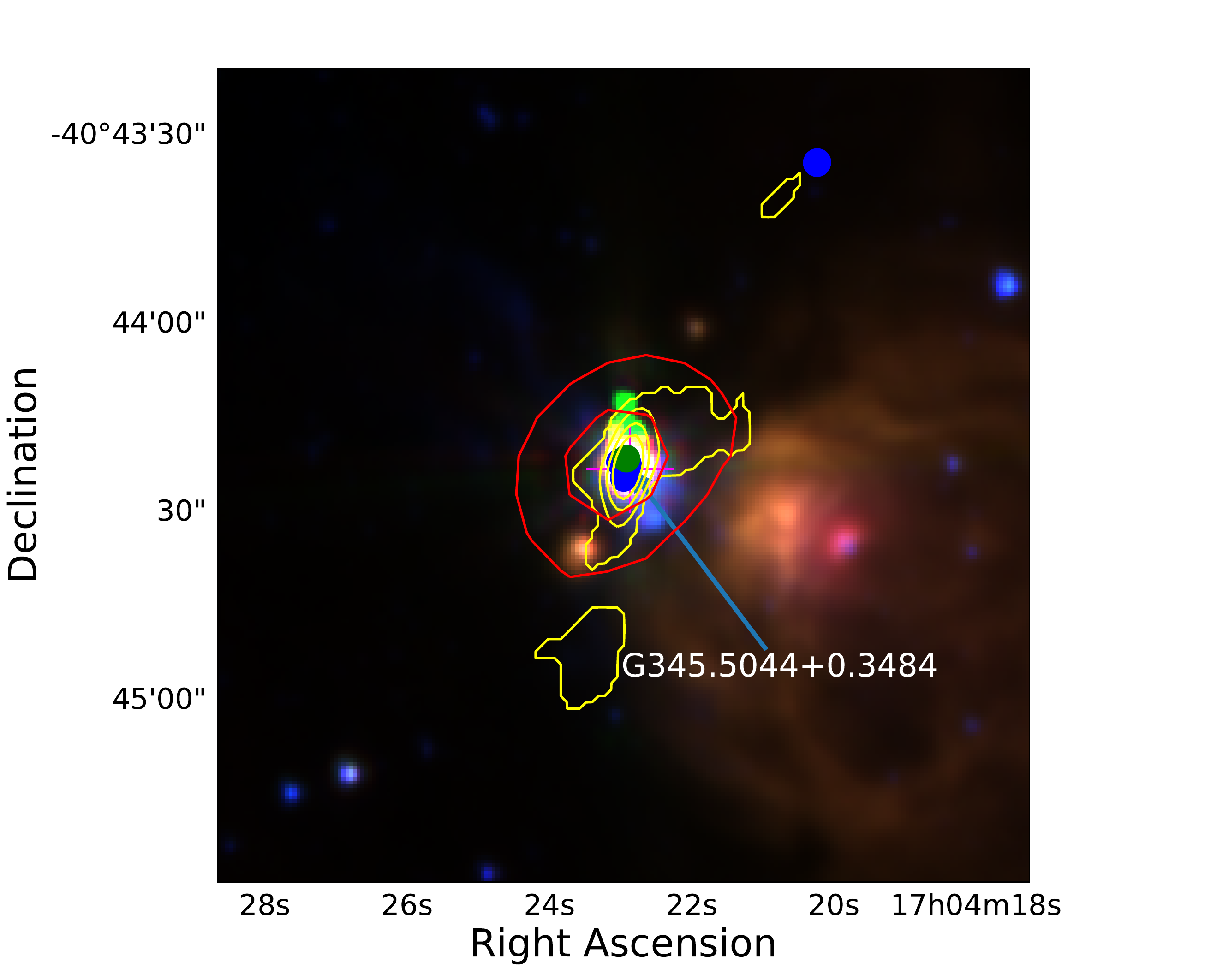}
  \includegraphics[width=0.45\textwidth]{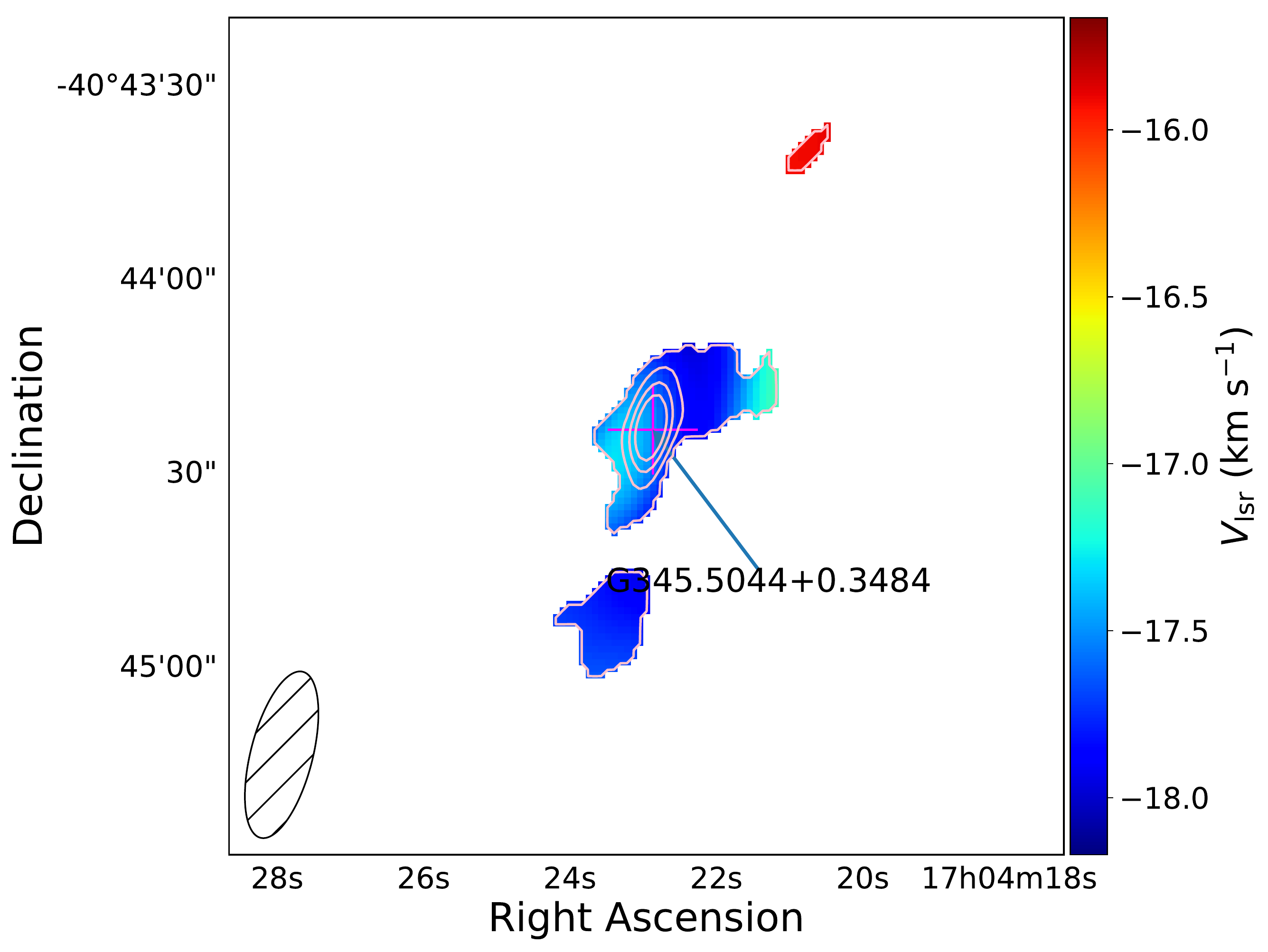} \\
  \includegraphics[width=0.45\textwidth]{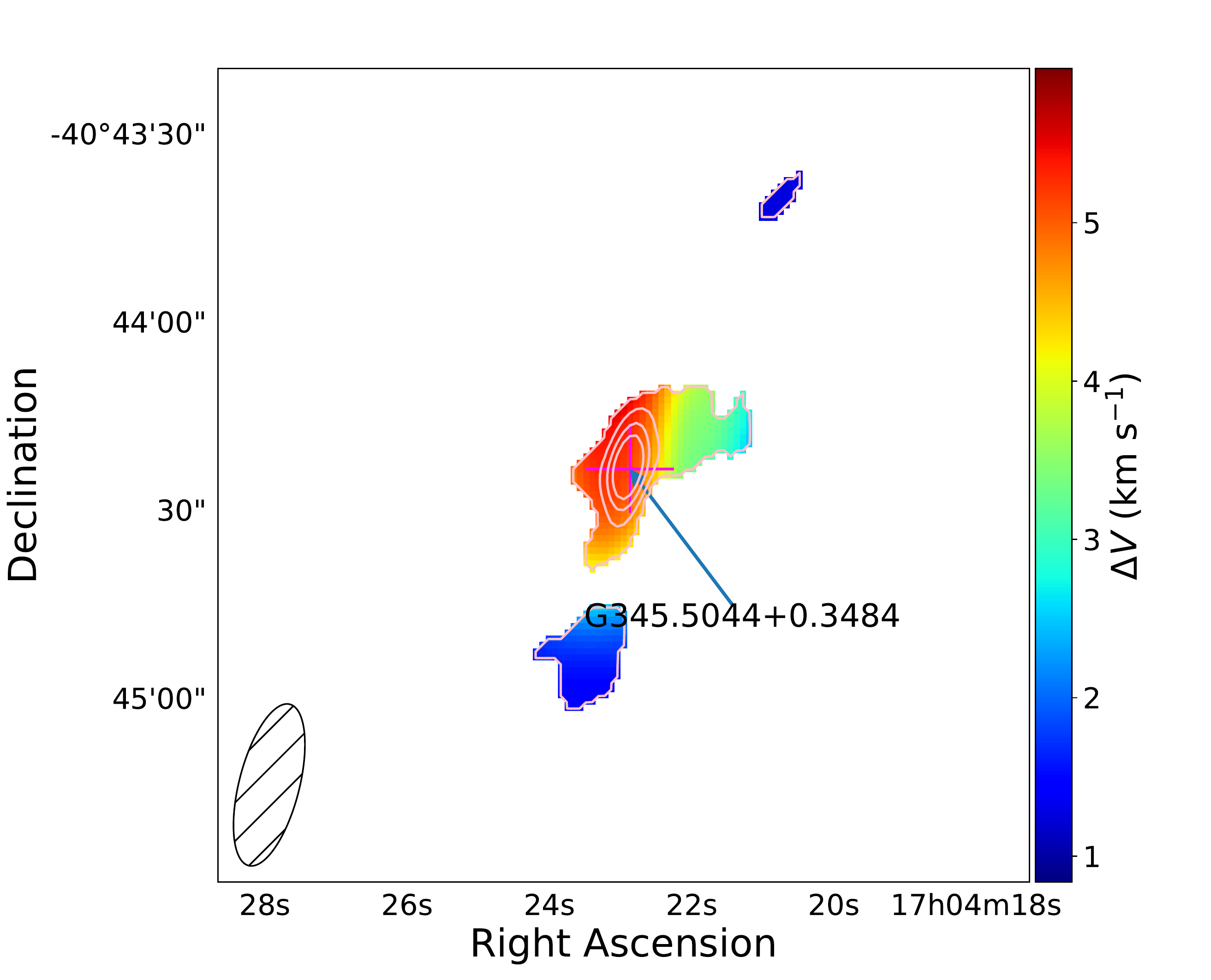} 
  \includegraphics[width=0.45\textwidth]{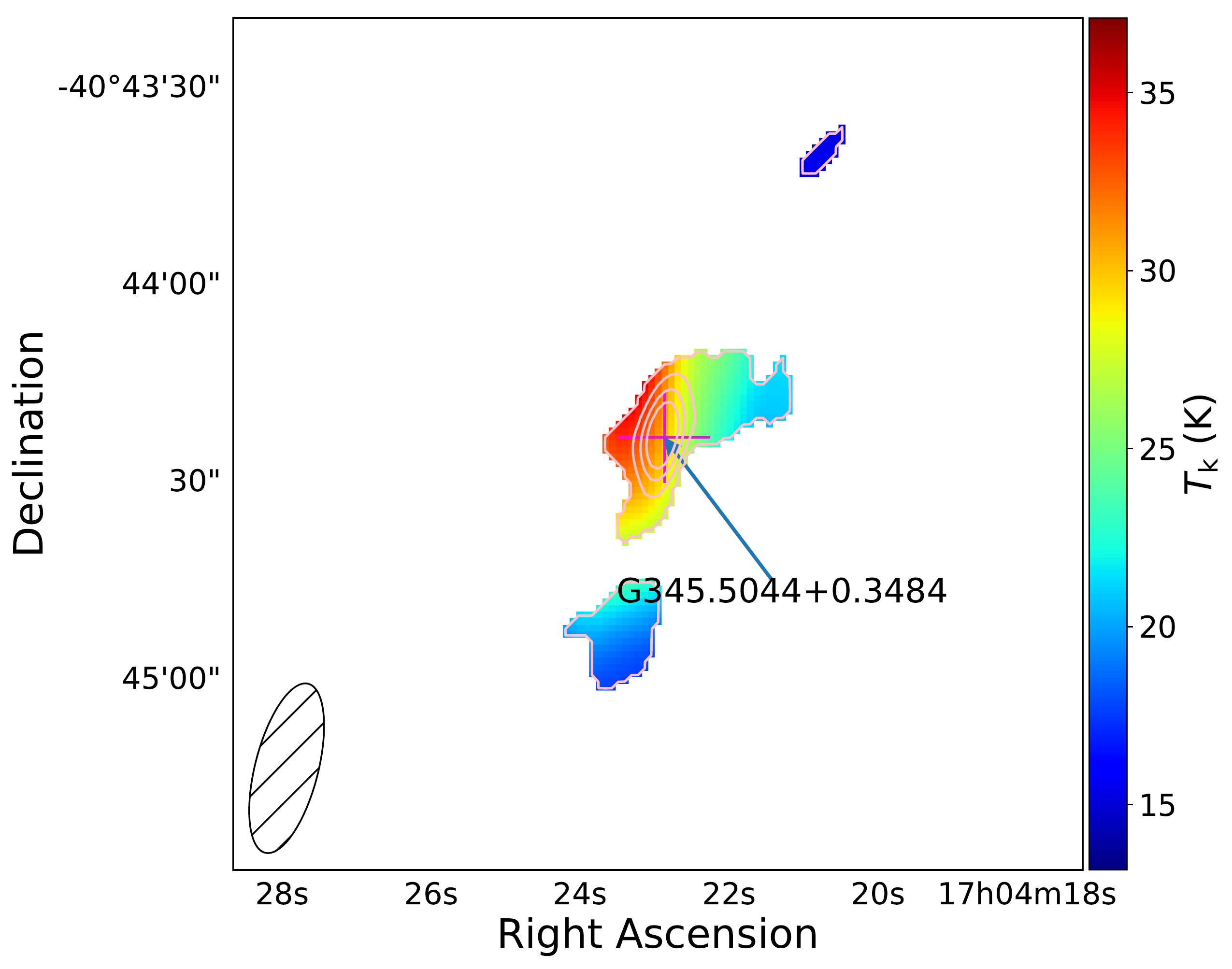}
\caption{Examples maps showing field G345.5043+00.3480. Methanol and \htwoo\ masers have been included in the three-colour GLIMPSE image in with green and blue filled circles respectively, otherwise each panel presented is the same as Figure~\ref{fig:example_maps_318}.}
\label{fig:example_maps_345}
\end{figure*}

\subsubsection{G328.8074+00.6324}

A second morphologically interesting example is region G328.8074+00.6324, shown in Figure~\ref{fig:example_maps_328}. This particular field involves two \nhthree-detected clumps which are connected via weaker \nhthree emission, forming a `horseshoe' shape. These clumps (G328.8057+0.6347 \& G328.8109+0.6336) have \nhthree (1,1) SNR values of 7.98 and 5.94, respectively, and aspect ratios of 1.96 and 1.78, so both clumps are fairly elongated in shape. Only one RMS source has been detected in this region (G328.8074+00.6324). This object has been classified as an \hii region by the RMS survey, and lies at the centre of the complex but appears to be removed from any \nhthree emission above the detection threshold. The ATLASGAL dust emission, as with the previous example, is well-correlated with the spatial position of the gas emission and completely encompasses it, with the position of the RMS source approximately coincident with the dust emission. The GLIMPSE image shows multiple evolved stars in the region and a large central IR object corresponding to the \hii region, which appears to be somewhat compact with a more diffuse, less intense, envelope. This region is defined as LSF, as the \hii region at the centre of the emission is disrupting its environment, and dispersing the dense gas.

The kinetic temperature map shown in Figure\,\ref{fig:example_maps_328} suggests central heating by the RMS source. The kinetic temperature is highest directly around the central \hii region and within the central area of the field, while the western edge of G328.8057+0.6347 is approximately 10\,K cooler, suggesting the feedback from the RMS source is not yet affecting this sector of gas. 

The line-width maxima do not correspond well with the temperature distribution: line-widths are dominated by non-thermal motions concentrated on the east-west extremities. As with G318.9480$-$00.1969, there appears to be a velocity gradient across the complex, however, a more detailed look at the \nhthree (1,1) data cube shows that there is \nhthree gas in front of the \hii region which is currently in absorption against the bright free-free continuum. Therefore, the \hii region is likely to be still enveloped in a shell of dense gas. 

Figure\,\ref{fig:G328_spectra} presents two spectra from this region, the first is the average emission spectrum for the \nhthree emission above the 3$\sigma$ threshold, while the second presents the average absorption averaged over the central nine pixels towards the embedded \hii region. The spectral peak velocities are offset by $-$2.4\,\kms, which indicates that the material in front of the \hii region is moving towards us slower than the systemic velocity of the natal clump. This infall motion of material suggests that the clump is still undergoing gravitational collapse. By assuming that the material surrounding the central object is a shell with an average density of 92.5\,\msun\,pc$^{-3}$ ($\frac{Mass_{ATLASGAL}}{Volume_{ATLASGAL}}$), we have calculated the infall rate for this particular source to be $9.6\times 10^{-3}$\,\msun\,yr$^{-1}$. \cite{Wyrowski2015} observed nine massive molecular clumps to search for infall signatures. The mean infall rate was found to be $6.33\times 10^{-3}$\,\msun\,yr$^{-1}$ with a standard deviation of $5.95\times 10^{-3}$\,\msun\,yr$^{-1}$. The value found for G328.8074+00.6324 is similar to this distribution.

We can investigate the stability of the clump using the virial parameter, which is the ratio of virial mass (the mass that can be supported by the internal energy of the clump) and the actual mass of the clump. This is defined as:

\begin{equation}
\alpha = \frac{M_{\rm vir}}{M_{\rm clump}},
\end{equation}

\noindent where

\begin{equation}
\left( \frac{M_{\rm vir}}{{\rm M_\odot}} \right) = 161 \left(\frac{R}{\rm pc}\right)\left(\frac{\Delta v}{{\rm km\,s^{-1}}}\right)^2
\end{equation}

\noindent where $R$ is the radius of the clump and $\Delta v$ is the line-width (Paper I).
 
An isothermal sphere in hydrostatic equilibrium, supported by equipartition of thermal and magnetic energy, has a critical value of $\alpha_{\rm{cr}}$ = 2 \citep{Kauffmann2013}. Clumps with a value above this are subcritical and will undergo expansion if not pressure-confined by their local environments. Clumps with values below $\alpha_{\rm{cr}}$ are supercritical; they are gravitationally unstable, and should be in a state of free-fall unless supported by strong magnetic fields. This particular region has a virial parameter of $\alpha$ = 2.04, placing it in the subcritical regime.  The error in this value for $\alpha$ is dominated by the ATLASGAL clump mass uncertainty, which is $\sim$20\% \citep{Urquhart2018}. The virial mass also depends on the source distance to calculate the spatial radius of the clump; the error in this distance is $\sim$17\%, obtained from the \cite{Reid2016} model. The difference between $\alpha$ and $\alpha_{\rm{cr}}$ is relatively small and due to the large errors associated with the virial parameter, the clump could either be sub or supercritical. As the region shows signs of global infall, this potentially means that the central object is beginning to stabilise the inner section of the clump, but the outer layers are undergoing collapse.

\subsubsection{G345.5043+00.3480}

Figure~\ref{fig:example_maps_345} presents 4 panels similar to the previous two examples. This source is classified as early star forming, and contains a single MYSO (G345.5043+00.3480), offset by less than one arcsecond (one pixel) from the centre of the \nhthree emission. Line-width and temperature maps have similar distributions, with the maxima located towards the centroid of the clump, with ranges of $\sim$15-35\,K and $\sim$1.2-5.6\,\kms, respectively.

One \nhthree clump has been detected through \fw, and although another two sources have been detected in the observations, both are smaller than the beam size.  The lack of ATLASGAL emission from around these two smaller sources suggests that these are therefore likely to be image artifacts (see Section~\ref{sect:atca_obs}). However, while there is no dust emission in this region beyond the main detected clump toward the centre of the field, the \nhthree emission to the north west of the main clump is coincident with an H$_2$O maser which can be seen in upper left panel of Fig.\,\ref{fig:example_maps_345} and so this source may in fact be a genuine detection.

The GLIMPSE image reveals a large, evolved \hii region that lies to the southwest of the main portion of gas emission (see upper left panel of Figure~\ref{fig:example_maps_345}). This \hii region has been identified as IRAS~15520-5234 \citep{Walsh1998, Garay2006} and is a known radio source. The shape of the clump would imply that the \hii region has compressed the dense gas, and subsequently a new star-forming region has been formed on the edge of the region, suggesting an instance of triggered star formation.

This region has the expected temperature and FWHM line-width peaks toward the embedded MYSO, and there is also a very small velocity difference across the central clump, with the smaller source to the north-west having a lower velocity. There appears to be no overlapping infrared or radio emission between these two clumps,  although all of the emission appears to be coincident with the edge of the evolved \hii region and has similar velocities, and so is likely to be associated.

\begin{figure}
\includegraphics[width=0.45\textwidth]{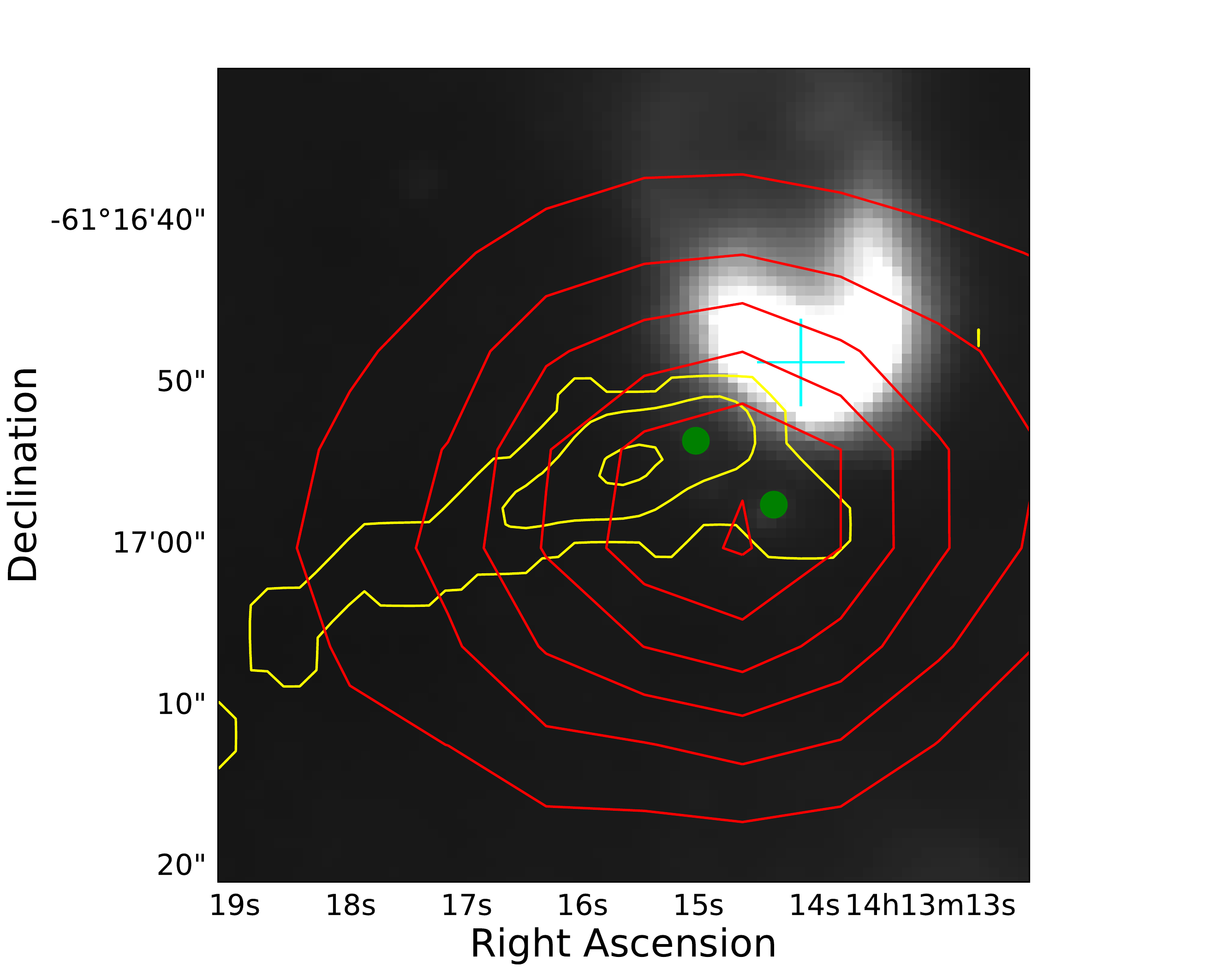} \\
\includegraphics[width=0.45\textwidth]{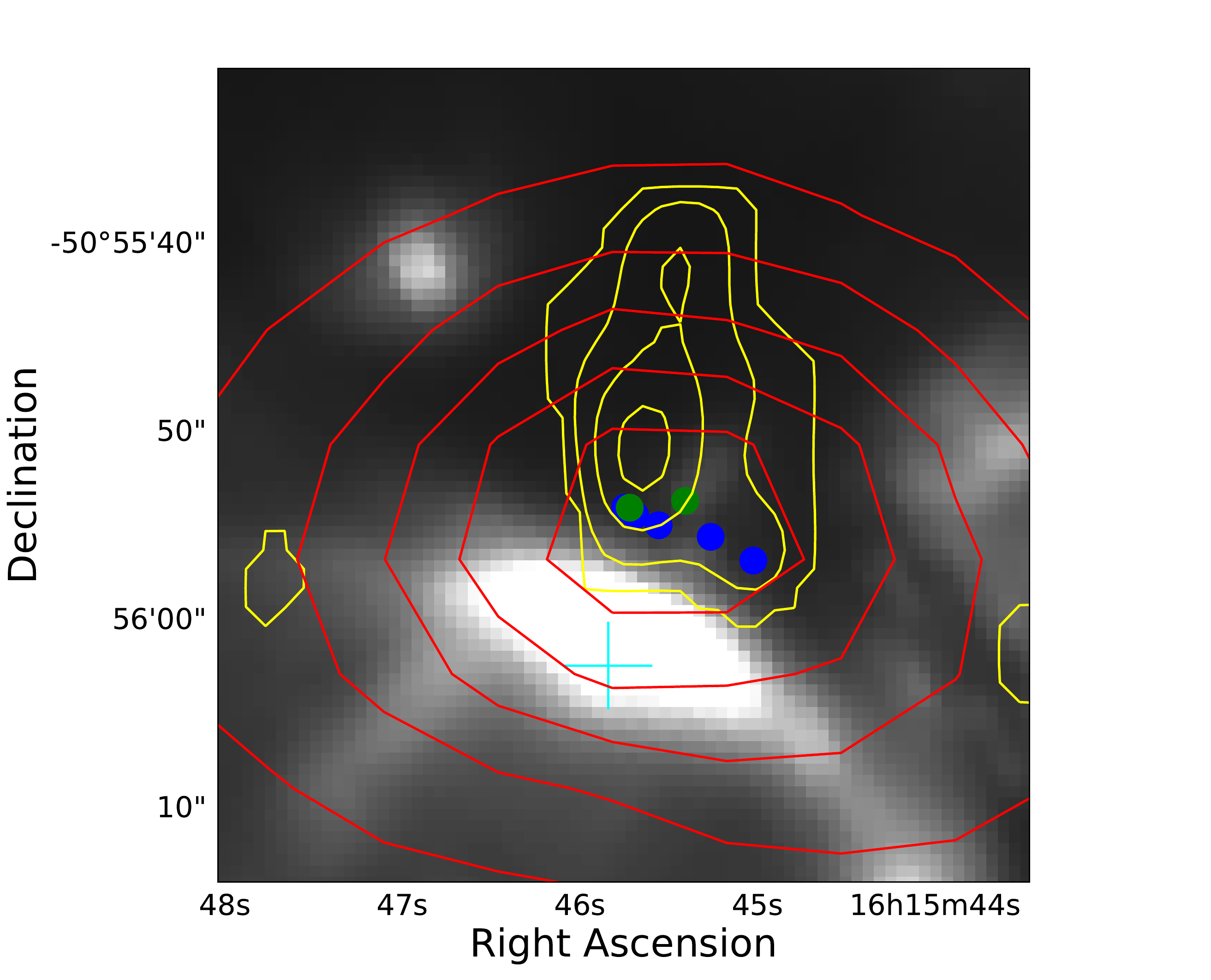} \\
\includegraphics[width=0.45\textwidth]{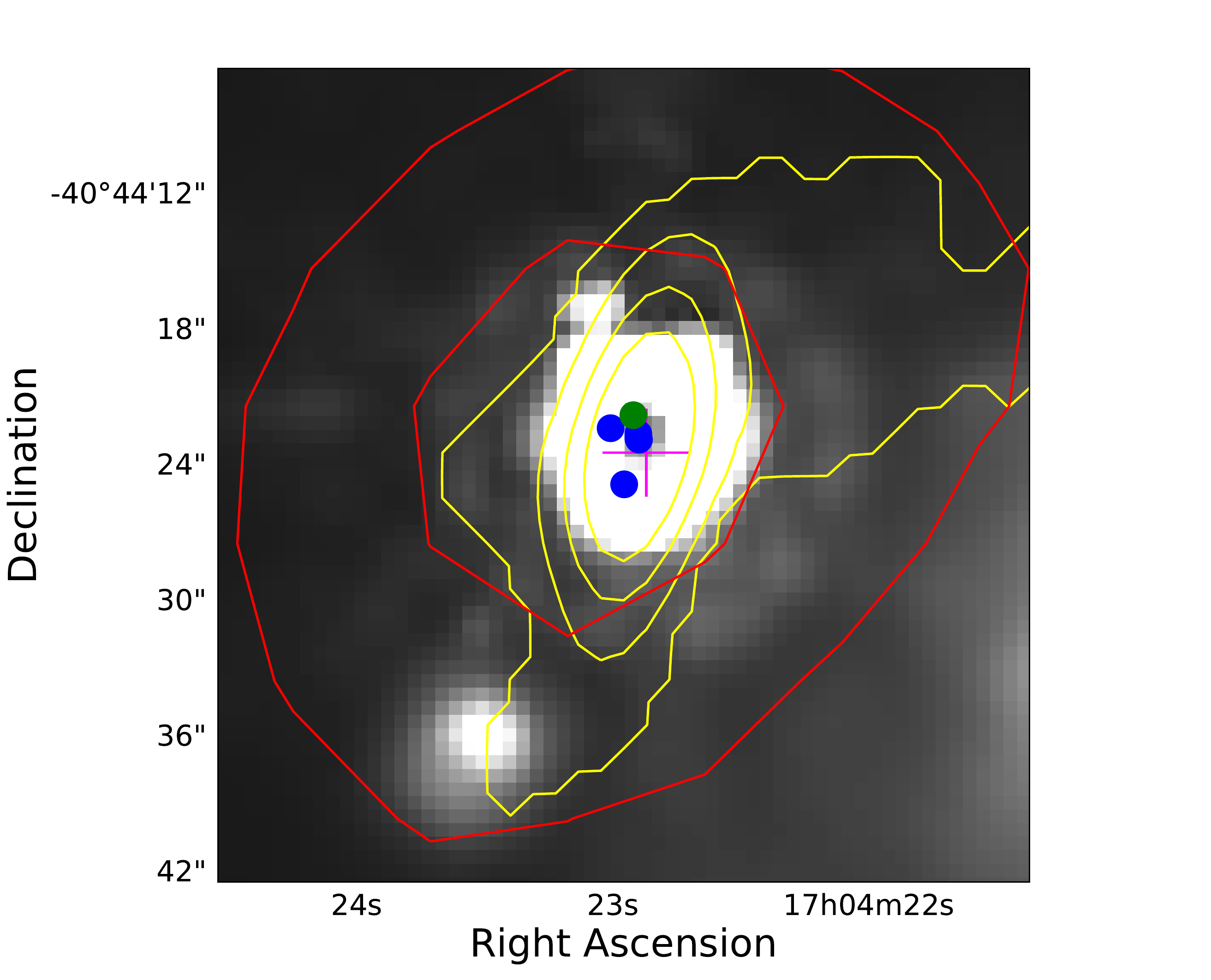} \\
\caption{Three GLIMPSE 8\,\micron\ images presenting regions of possible triggered star formation. Each panel presents an 8\,\micron\ image overlaid with \nhthree and ATLASGAL emission contours in yellow and red respectively. The contours levels are equivalent to those in Figure~\ref{fig:8mu_examples}. MYSOs and \hii regions identified by the RMS survey are show with magenta and cyan crosses respectively. Methanol and water maser positions are marked with green and blue filled circles respectively.}
\label{fig:tsf_example_maps}
\end{figure}

\subsection{Triggered Star Formation}

\cite{Elmegreen1992} proposed a model of triggered star formation, wherein an ionization shock front produced by the expansion of \hii\ regions created by one group of massive stars provides the external pressure to compress adjacent material in a molecular clouds, thereby inducing subsequent generations of massive stars to form. This mechanism is thought to play a significant role in the evolution of massive star forming regions (e.g. \citealt{thompson2012}).

As discussed previously, methanol masers are located at the sites of high-mass star formation, while H$_2$O masers are sensitive to shocked material. By using maser positions and the likely presence of an \hii region (which can be inferred through the morphology of extended 8\,\micron\ emission), multiple regions presented in this study are consistent with triggered star formation. As \hii regions expand into nearby molecular material, they begin to compresses this matter which could form future generations of high-mass stars. The incorporated maser emission data helps to refine this picture as both methanol and water maser emission is known to be coincident with star formation.

Figure~\ref{fig:tsf_example_maps} presents three regions which appear to contain examples of triggered star formation at different stages. The top panel of Figure~\ref{fig:tsf_example_maps} shows a relatively early and compact \hii region to the north-west of the \nhthree emission. It can be seen that two methanol masers are present within the dense material traced by the \nhthree in different locations, implying two protostellar cores are embedded within the environment. This is potentially due to the influence of the nearby \hii region which suggests that this material is undergoing shocks. The middle panel of Figure~\ref{fig:tsf_example_maps} presents the same effect more conclusively. The \hii region towards the south is clearly having an impact on the dense material and is dispersing its environment, indicated by the appearance of multiple \nhthree clumps on the edges of the radiation. The central clump also has a string of \htwoo\ masers along this boundary, which suggests that this material is undergoing shock. Two methanol masers are located behind these \htwoo\ masers, again implying the presence of high-mass protostellar cores, which are likely to have formed due to the compression of gas as the \hii region expands towards it. The lower panel of Figure~\ref{fig:tsf_example_maps} presents a more diffuse \hii region in the south-west which seems to be in a later evolutionary stage. A MYSO has already formed within the gas and dust emission, which has been subjected to the effects of the nearby \hii region. This MYSO is associated with two methanol masers and 10 \htwoo\ masers. 

These three regions show the effect that \hii regions have on their local environment and the likelihood that they can cause triggered star formation and produce future generations of stars. The use of the classification system presented in this paper creates a difficulty in assigning these types of regions, as they display both qualities of ESF and LSF fields. While \hii regions are present and are impacting on their local environment, the presence of a large number of maser detections implies protostellar objects in an initial stage of formation.

\section{Conclusions}
\label{sect:conclusions}

We have mapped 34 fields using the \nhthree (1,1) and (2,2) inversion transitions.  We have used the \fw\ algorithm with a threshold value of 3$\sigma$ to identify 44 clumps distributed among these regions. No statistical differences are found between clumps classified as early star-forming or late star-forming when using a two sample KS-test. The early star formation (ESF) star forming clumps do generally exhibit elevated temperatures, while the late star formation (LSF) counterparts have larger line-widths. Most ESF clumps have been previously identified by as MYSOs while the majority of LSF clumps are associated with \hii regions. Physical parameters have been extracted from complementary surveys such as ATLASGAL and GLIMPSE legacy surveys. Masses for the fields range between a few hundred to over 10,000\,M$_{\sun}$ and luminosities range between $\sim$10$^{3}$ to $\sim$10$^{6}$\,\lsun. 

We have used these data to construct a rudimentary evolutionary sequence of proposed massive star formation based on the morphology of their associated high density gas. ESF fields generally have RMS objects which are found towards the centre of centrally condensed coherent clumps, while the embedded objects within LSF regions are not coincident with dense gas but are generally found to be surrounded by multiple fragments. The higher level of fragmentation seen towards the LSF clumps is consistent with an expectation of association with more evolved and more luminous \hii regions that are disrupting and ``burning out from'' their natal clumps. The clump-averaged physical properties of these two samples are not statistically different from each other (mass, luminosity, temperatures and line-widths), although the structure of the dense gas is very different, which may indicate that the formation of the \hii region has a much more significant impact on the overall structure of natal clump than on its physical properties. 

We find that 31 methanol masers associated with 34 fields and 52 water masers are associated 21 fields. The majority of both types of masers are found towards ESF fields. The detection rate for water masers for the ESF and LSF field are 89\,per\,cent and 25\,per\,cent respectively, with a similar average number of spots detected in both ESF and LSF fields. The detection rates for methanol masers are 72 and 44\,per\,cent for the ESF and LSF fields, respectively. The reason for the significantly lower detection rates towards the LSF clumps is likely to be linked to the disruption of natal clumps by the expansion of the \hii regions.  

Triggered star formation appears to be present in three of the observed regions. The 8\,\micron\ images of these sources all show the presence of nearby \hii regions that show evidence of affecting the dense gas as traced by \nhthree emission seen towards these sources. A large number of methanol and water masers are found to be coincident with these regions revealing that these nearby \hii regions are not only having a significant impact on their local environments but are also appears to be having an a direct influence star formation in the surrounding dense gas.

\section*{Acknowledgments}

The authors would like to thank the staff of the ATCA for their assistance during the preparation and execution of these observations. SB wishes to acknowledge an STFC
PhD studentship for this work. We have used the collaborative tool Overleaf available at: https://www.overleaf.com/.

%%%%%%%%%%%%%%%%%%%%%%%%%%%%%%%%%%%%%%%%%%%%%%%%%%



\bibliographystyle{mn2e_new}
\bibliography{library,urquhart2016}

\begin{thebibliography}{82}
\expandafter\ifx\csname natexlab\endcsname\relax\def\natexlab#1{#1}\fi

\bibitem[{Battersby {et~al}\mbox{.}(2014)Battersby, Ginsburg, Bally, Longmore,
  Dunham, \& Darling}]{Battersby2014}
Battersby C., Ginsburg A., Bally J., Longmore S., Dunham M., Darling J., 2014,
  The Astrophysical Journal, 787, 113

\bibitem[{Benjamin {et~al}\mbox{.}(2003)Benjamin, Churchwell, Babler, Bania,
  Clemens, Cohen, Dickey, Indebetouw, Jackson, Kobulnicky, Lazarian, Marston,
  Mathis, Meade, Seager, Stolovy, Watson, Whitney, Wolff, \&
  Wolfire}]{benjamin2003}
Benjamin R.~A. {et~al.}, 2003, Publications of the Astronomical Society of
  Pacific, 115, 953

\bibitem[{Berry(2015)}]{Berry2015}
Berry D.~S., 2015, Astronomy and Computing, 10, 22

\bibitem[{{Breen} {et~al}\mbox{.}(2011){Breen}, {Ellingsen}, {Caswell},
  {Green}, {Fuller}, {Voronkov}, {Quinn}, \& {Avison}}]{breen2011_methanol}
{Breen} S.~L., {Ellingsen} S.~P., {Caswell} J.~L., {Green} J.~A., {Fuller}
  G.~A., {Voronkov} M.~A., {Quinn} L.~J., {Avison} A., 2011, \apj, 733, 80

\bibitem[{{Caswell}(2010)}]{caswell2010_mmb}
{Caswell}, J.~L. e., 2010, \mnras, 404, 1029

\bibitem[{{Chambers} {et~al}\mbox{.}(2009){Chambers}, {Jackson}, {Rathborne},
  \& {Simon}}]{Chambers2009}
{Chambers} E.~T., {Jackson} J.~M., {Rathborne} J.~M., {Simon} R., 2009, \apjs,
  181, 360

\bibitem[{Chira {et~al}\mbox{.}(2013)Chira, Beuther, Linz, Schuller, Walmsley,
  Menten, \& Bronfman}]{chira2013}
Chira R.-a., Beuther H., Linz H., Schuller F., Walmsley C.~M., Menten K.~M.,
  Bronfman L., 2013, Astronomy {\&} Astrophysics, 552, A40

\bibitem[{Churchwell {et~al}\mbox{.}(2009)Churchwell, Babler, Meade, Whitney,
  Benjamin, Indebetouw, Cyganowski, Robitaille, Povich, Watson, \&
  Bracker}]{churchwell2009}
Churchwell E. {et~al.}, 2009, Publications of the Astronomical Society of the
  Pacific, 121, 213

\bibitem[{Claussen {et~al}\mbox{.}(1996)Claussen, Wilking, Benson, Wootten,
  Myers, \& Terebey}]{Claussen1996}
Claussen M.~J., Wilking B.~A., Benson P.~J., Wootten A., Myers P.~C., Terebey
  S., 1996, Astrophysical Journal Supplement v.106, 106, 111

\bibitem[{Contreras {et~al}\mbox{.}(2013)Contreras, Schuller, Urquhart,
  Csengeri, Wyrowski, Beuther, Bontemps, Bronfman, Henning, Menten, Schilke,
  Walmsley, Wienen, Tackenberg, \& Linz}]{Contreras2013a}
Contreras Y. {et~al.}, 2013, Astronomy {\&} Astrophysics, 549, A45

\bibitem[{Cooper {et~al}\mbox{.}(2013)Cooper, Lumsden, Oudmaijer, Hoare,
  Clarke, Urquhart, Mottram, Moore, \& Davies}]{Cooper2013}
Cooper H. D.~B. {et~al.}, 2013, Monthly Notices of the Royal Astronomical
  Society, 430, 1125

\bibitem[{{Csengeri} {et~al}\mbox{.}(2014){Csengeri}, {Urquhart}, {Schuller},
  {Motte}, {Bontemps}, {Wyrowski}, {Menten}, {Bronfman}, {Beuther}, {Henning},
  {Testi}, {Zavagno}, \& {Walmsley}}]{csengeri2014}
{Csengeri} T. {et~al.}, 2014, \aap, 565, A75

\bibitem[{{Cutri} {et~al}\mbox{.}(2003){Cutri}, {Skrutskie}, {van Dyk},
  {Beichman}, {Carpenter}, {Chester}, {Cambresy}, {Evans}, {Fowler}, {Gizis},
  {Howard}, {Huchra}, {Jarrett}, {Kopan}, {Kirkpatrick}, {Light}, {Marsh},
  {McCallon}, {Schneider}, {Stiening}, {Sykes}, {Weinberg}, {Wheaton},
  {Wheelock}, \& {Zacarias}}]{cutri2003}
{Cutri} R.~M. {et~al.}, 2003, VizieR Online Data Catalog, 2246, 0

\bibitem[{Cyganowski {et~al}\mbox{.}(2008)Cyganowski, Whitney, Holden, Braden,
  Brogan, Churchwell, Indebetouw, Watson, Babler, Benjamin, Gomez, Meade,
  Povich, Robitaille, \& Watson}]{Cyganowski2008}
Cyganowski C.~J. {et~al.}, 2008, The Astronomical Journal, 136, 2391

\bibitem[{{De Buizer} {et~al}\mbox{.}(2000){De Buizer}, Pina, \&
  Telesco}]{DeBuizer2000a}
{De Buizer} J.~M., Pina R.~K., Telesco C.~M., 2000, The Astrophysical Journal
  Supplement Series, 130, 437

\bibitem[{{De Buizer} {et~al}\mbox{.}(2002){De Buizer}, {Watson}, {Radomski},
  {Pi{\~n}a}, \& {Telesco}}]{de-buizer2002}
{De Buizer} J.~M., {Watson} A.~M., {Radomski} J.~T., {Pi{\~n}a} R.~K.,
  {Telesco} C.~M., 2002, \apjl, 564, L101

\bibitem[{{Deharveng} {et~al}\mbox{.}(2010){Deharveng}, {Schuller}, {Anderson},
  {Zavagno}, {Wyrowski}, {Menten}, {Bronfman}, {Testi}, {Walmsley}, \&
  {Wienen}}]{Deharveng2010}
{Deharveng} L. {et~al.}, 2010, \aap, 523, A6

\bibitem[{Dickinson(1976)}]{Dickinson1976}
Dickinson D.~F., 1976, The Astrophysical Journal Supplement Series, 30, 259

\bibitem[{{Dunham} {et~al}\mbox{.}(2011){Dunham}, {Rosolowsky}, {Evans},
  {Cyganowski}, \& {Urquhart}}]{dunham2011b}
{Dunham} M.~K., {Rosolowsky} E., {Evans}, II N.~J., {Cyganowski} C., {Urquhart}
  J.~S., 2011, \apj, 741, 110

\bibitem[{Eden {et~al}\mbox{.}(2017)Eden, Moore, Plume, Urquhart, Thompson,
  Parsons, Dempsey, Rigby, Morgan, Thomas, Berry, Buckle, Brunt, Butner,
  Carretero, Chrysostomou, Currie, Devilliers, Fich, Gibb, Hoare, Jenness,
  Manser, Mottram, Natario, Olguin, Peretto, Pestalozzi, Polychroni, Redman,
  Salji, Summers, Tahani, Traficante, DiFrancesco, Evans, Fuller, Johnstone,
  Joncas, Longmore, Martin, Richer, Weferling, White, \& Zhu}]{Eden2017}
Eden D.~J. {et~al.}, 2017, MNRAS, 2183, 2163

\bibitem[{Elmegreen(1992)}]{Elmegreen1992}
Elmegreen B.~G., 1992, in Star Formation in Stellar Systems, p. 381

\bibitem[{Elmegreen \& Scalo(2004)}]{Elmegreen2004}
Elmegreen B.~G., Scalo J., 2004

\bibitem[{Fazio {et~al}\mbox{.}(1998)Fazio, Hora, Willner, Stauffer, Ashby,
  Wang, Tollestrup, Pipher, Forrest, McCreight, {Moseley, Jr.}, Hoffmann,
  Eisenhardt, \& Wright}]{Fazio1998}
Fazio G.~G. {et~al.}, 1998, in Proc. SPIE Vol. 3354, p. 1024-1031, Infrared
  Astronomical Instrumentation, Albert M. Fowler; Ed., Fowler A.~M., ed., Vol.
  3354, p. 1024

\bibitem[{Forster \& Caswell(1999)}]{Forster1996}
Forster J., Caswell J.~L., 1999, Astronomy {\&} Astrophysics Supplement Series,
  137, 43

\bibitem[{Friesen {et~al}\mbox{.}(2009)Friesen, {Di Francesco}, Shirley, \&
  Myers}]{Friesen2009}
Friesen R.~K., {Di Francesco} J., Shirley Y.~L., Myers P.~C., 2009, The
  Astrophysical Journal, 697, 1457

\bibitem[{Garay {et~al}\mbox{.}(2006)Garay, Brooks, Mardones, \&
  Norris}]{Garay2006}
Garay G., Brooks K.~J., Mardones D., Norris R.~P., 2006, ApJ, 651, 914

\bibitem[{{Ginsburg} \& {Mirocha}(2011)}]{Ginsburg2011}
{Ginsburg} A., {Mirocha} J., 2011, {PySpecKit: Python Spectroscopic Toolkit}.
  Astrophysics Source Code Library

\bibitem[{Green {et~al}\mbox{.}(2009)Green, Caswell, Fuller, Avison, Breen,
  Brooks, Burton, Chrysostomou, Cox, Diamond, Ellingsen, Gray, Hoare, Masheder,
  McClure-Griffiths, Pestalozzi, Phillips, Quinn, Thompson, Voronkov, Walsh,
  Ward-Thompson, Wong-Mcsweeney, Yates, \& Cohen}]{Green2009}
Green J.~A. {et~al.}, 2009, Monthly Notices of the Royal Astronomical Society,
  392, 783

\bibitem[{Ho {et~al}\mbox{.}(1983)Ho, Haschick, Vogel, \& Wright}]{Ho1983}
Ho P., Haschick A., Vogel S., Wright M., 1983, The Astronomical Journal, 295

\bibitem[{Kauffmann {et~al}\mbox{.}(2013)Kauffmann, Pillai, \&
  Goldsmith}]{Kauffmann2013}
Kauffmann J., Pillai T., Goldsmith P.~F., 2013, Astrophysical Journal, 779

\bibitem[{{Kennicutt}(2005)}]{kennicutt2005}
{Kennicutt} R.~C., 2005, in IAU Symposium, Vol. 227, Massive Star Birth: A
  Crossroads of Astrophysics, {Cesaroni} R., {Felli} M., {Churchwell} E.,
  {Walmsley} M., eds., pp. 3--11

\bibitem[{{Kurtz} {et~al}\mbox{.}(1994){Kurtz}, {Churchwell}, \&
  {Wood}}]{kurtz1994}
{Kurtz} S., {Churchwell} E., {Wood} D.~O.~S., 1994, \apjs, 91, 659

\bibitem[{Longmore {et~al}\mbox{.}(2017)Longmore, Walsh, Purcell, Burke,
  Henshaw, Walker, Urquhart, Barnes, Moore, Thompson, \&
  Voronkov}]{longmore2017}
Longmore S.~N. {et~al.}, 2017, MNRAS, 470, 1462

\bibitem[{Lumsden {et~al}\mbox{.}(2013)Lumsden, Hoare, Urquhart, Oudmaijer,
  Davies, Mottram, Cooper, \& Moore}]{Lumsden2013}
Lumsden S.~L., Hoare M.~G., Urquhart J.~S., Oudmaijer R.~D., Davies B., Mottram
  J.~C., Cooper H. D.~B., Moore T. J.~T., 2013, The Astrophysical Journal
  Supplement Series, 208, 11

\bibitem[{Mangum {et~al}\mbox{.}(1992)Mangum, Wootten, \& Mundy}]{Mangum1992}
Mangum J.~G., Wootten A., Mundy L.~G., 1992, The Astrophysical Journal, 388,
  467

\bibitem[{Minier {et~al}\mbox{.}(2003)Minier, Ellingsen, Norris, \&
  Booth}]{Minier2003}
Minier V., Ellingsen S.~P., Norris R.~P., Booth R.~S., 2003, Astronomy {\&}
  Astrophysics, 403, 1095

\bibitem[{Miranda {et~al}\mbox{.}(2001)Miranda, G{\'{o}}mez, Anglada, \&
  Torrelles}]{Miranda2001}
Miranda L.~F., G{\'{o}}mez Y., Anglada G., Torrelles J.~M., 2001, Nature, 414,
  284

\bibitem[{Moore {et~al}\mbox{.}(2015)Moore, Plume, Thompson, Parsons, Urquhart,
  Eden, Dempsey, Morgan, Thomas, Buckle, Brunt, Butner, Carretero,
  Chrysostomou, DeVilliers, Fich, Hoare, Manser, Mottram, Natario, Olguin,
  Peretto, Polychroni, Redman, Rigby, Salji, Summers, Berry, Currie, Jenness,
  Pestalozzi, Traficante, Bastien, DiFrancesco, Davis, Evans, Friberg, Fuller,
  Gibb, Gibson, Hill, Johnstone, Joncas, Longmore, Lumsden, Martin, {Nguyẽn
  Lu'o'ng}, Pineda, Purcell, Richer, Schieven, Shipman, Spaans, Taylor, Viti,
  Weferling, White, \& Zhu}]{Moore2015}
Moore T.~J. {et~al.}, 2015, Monthly Notices of the Royal Astronomical Society,
  453, 4264

\bibitem[{{Mottram} {et~al}\mbox{.}(2011{\natexlab{a}}){Mottram}, {Hoare},
  {Davies}, {Lumsden}, {Oudmaijer}, {Urquhart}, {Moore}, {Cooper}, \&
  {Stead}}]{mottram2011b}
{Mottram} J.~C. {et~al.}, 2011{\natexlab{a}}, \apjl, 730, L33+

\bibitem[{Mottram {et~al}\mbox{.}(2007)Mottram, Hoare, Lumsden, Oudmaijer,
  Urquhart, Sheret, Clarke, \& Allsopp}]{Mottram2007c}
Mottram J.~C., Hoare M.~G., Lumsden S.~L., Oudmaijer R.~D., Urquhart J.~S.,
  Sheret T.~L., Clarke A.~J., Allsopp J., 2007, Astronomy and Astrophysics,
  476, 1019

\bibitem[{{Mottram} {et~al}\mbox{.}(2011{\natexlab{b}}){Mottram}, {Hoare},
  {Urquhart}, {Lumsden}, {Oudmaijer}, {Robitaille}, {Moore}, {Davies}, \&
  {Stead}}]{mottram2011a}
{Mottram} J.~C. {et~al.}, 2011{\natexlab{b}}, \aap, 525, A149

\bibitem[{Navarete {et~al}\mbox{.}(2015)Navarete, Damineli, Barbosa, \&
  Blum}]{Navarete2015}
Navarete F., Damineli A., Barbosa C.~L., Blum R.~D., 2015, Monthly Notices of
  the Royal Astronomical Society, 450, 4364

\bibitem[{Perault {et~al}\mbox{.}(1996)Perault, Omont, Simon, Seguin, Ojha,
  Blommaert, Felli, Gilmore, Guglielmo, Habing, Price, Robin, de~Batz,
  Cesarsky, Elbaz, Epchtein, Fouque, Guest, Levine, Pollock, Prusti,
  Siebenmorgen, Testi, \& Tiphene}]{Perault1996}
Perault M. {et~al.}, 1996, $\backslash$Aap, 315, L165

\bibitem[{Pillai {et~al}\mbox{.}(2006)Pillai, Wyrowski, Carey, \&
  Menten}]{pillai2006}
Pillai T., Wyrowski F., Carey S.~J., Menten K.~M., 2006, Astronomy {\&}
  Astrophysics, 450, 569

\bibitem[{Price {et~al}\mbox{.}(2001)Price, Egan, Carey, Mizuno, \&
  Kuchar}]{Price2001}
Price S.~D., Egan M.~P., Carey S.~J., Mizuno D.~R., Kuchar T.~A., 2001, The
  Astronomical Journal, 121, 2819

\bibitem[{{Purcell} {et~al}\mbox{.}(2012){Purcell}, {Longmore}, {Walsh},
  {Whiting}, {Breen}, {Britton}, {Brooks}, {Burton}, {Cunningham}, {Green},
  {Harvey-Smith}, {Hindson}, {Hoare}, {Indermuehle}, {Jones}, {Lo}, {Lowe},
  {Phillips}, {Thompson}, {Urquhart}, {Voronkov}, \& {White}}]{purcell2012}
{Purcell} C.~R. {et~al.}, 2012, \mnras, 426, 1972

\bibitem[{Ragan {et~al}\mbox{.}(2011)Ragan, Bergin, \& Wilner}]{ragan2011}
Ragan S.~E., Bergin E.~A., Wilner D., 2011, The Astrophysical Journal, 736

\bibitem[{Reid {et~al}\mbox{.}(2016)Reid, Dame, Menten, \&
  Brunthaler}]{Reid2016}
Reid M.~J., Dame T.~M., Menten K.~M., Brunthaler A., 2016, The Astrophysical
  Journal, 823, 1

\bibitem[{{Reid} {et~al}\mbox{.}(2009){Reid}, {Menten}, {Zheng}, {Brunthaler},
  {Moscadelli}, {Xu}, {Zhang}, {Sato}, {Honma}, {Hirota}, {Hachisuka}, {Choi},
  {Moellenbrock}, \& {Bartkiewicz}}]{reid2009}
{Reid} M.~J. {et~al.}, 2009, \apj, 700, 137

\bibitem[{Rosolowsky {et~al}\mbox{.}(2010)Rosolowsky, Ginsburg, \&
  Ellsworth-Bo}]{Rosolowsky2010}
Rosolowsky E., Ginsburg A., Ellsworth-Bo, 2010, 438, 76

\bibitem[{Rosolowsky {et~al}\mbox{.}(2008)Rosolowsky, Pineda, Foster, Borkin,
  Kauffmann, Caselli, Myers, Goodman, Green, \& Telescope}]{Rosolowsky2008}
Rosolowsky E.~W. {et~al.}, 2008, The Astrophysical Journal Supplement Series,
  175, 509

\bibitem[{{Sault} {et~al}\mbox{.}(1995){Sault}, {Teuben}, \&
  {Wright}}]{sault1995}
{Sault} R.~J., {Teuben} P.~J., {Wright} M.~C.~H., 1995, in ASP Conf. Ser. 77:
  Astronomical Data Analysis Software and Systems IV, {Shaw} R.~A., {Payne}
  H.~E., {Hayes} J.~J.~E., eds., pp. 433--+

\bibitem[{Schuller {et~al}\mbox{.}(2009)Schuller, Menten, Contreras, Wyrowski,
  Schilke, Bronfman, Henning, Walmsley, Beuther, Bontemps, Cesaroni, Deharveng,
  Garay, Herpin, Lefloch, Linz, Mardones, Minier, Molinari, Motte, Nyman,
  Reveret, Risacher, Russeil, Schneider, Testi, Troost, Vasyunina, Wienen,
  Zavagno, Kovacs, Kreysa, Siringo, \& Wei{\ss}}]{Schuller2009a}
Schuller F. {et~al.}, 2009, Astronomy and Astrophysics, 504, 415

\bibitem[{Sridharan {et~al}\mbox{.}(2002)Sridharan, Beuther, Schilke, Menten,
  \& Wyrowski}]{Sridharan2002}
Sridharan T.~K., Beuther H., Schilke P., Menten K.~M., Wyrowski F., 2002, The
  Astrophysical Journal, 566, 931

\bibitem[{Swift {et~al}\mbox{.}(2005)Swift, Welch, \& {Di
  Francesco}}]{Swift2005}
Swift J.~J., Welch W.~J., {Di Francesco} J., 2005, The Astrophysical Journal,
  620, 823

\bibitem[{{Thompson} {et~al}\mbox{.}(2012){Thompson}, {Urquhart}, {Moore}, \&
  {Morgan}}]{thompson2012}
{Thompson} M.~A., {Urquhart} J.~S., {Moore} T.~J.~T., {Morgan} L.~K., 2012,
  \mnras, 421, 408

\bibitem[{{Urquhart} {et~al}\mbox{.}(2010){Urquhart}, {Hoare}, {Moore},
  {Morgan}, \& {Figura}}]{2010atnf}
{Urquhart} J., {Hoare} M., {Moore} T., {Morgan} L., {Figura} C., 2010, {Probing
  the environments of young massive stars}. ATNF Proposal

\bibitem[{{Urquhart} {et~al}\mbox{.}(2007{\natexlab{a}}){Urquhart}, {Busfield},
  {Hoare}, {Lumsden}, {Clarke}, {Moore}, {Mottram}, \&
  {Oudmaijer}}]{urquhart_radio_south}
{Urquhart} J.~S., {Busfield} A.~L., {Hoare} M.~G., {Lumsden} S.~L., {Clarke}
  A.~J., {Moore} T.~J.~T., {Mottram} J.~C., {Oudmaijer} R.~D.,
  2007{\natexlab{a}}, \aap, 461, 11

\bibitem[{{Urquhart} {et~al}\mbox{.}(2007{\natexlab{b}}){Urquhart}, {Busfield},
  {Hoare}, {Lumsden}, {Oudmaijer}, {Moore}, {Gibb}, {Purcell}, {Burton}, \&
  {Marechal}}]{urquhart_13co_south}
{Urquhart} J.~S. {et~al.}, 2007{\natexlab{b}}, \aap, 474, 891

\bibitem[{{Urquhart} {et~al}\mbox{.}(2008){Urquhart}, {Busfield}, {Hoare},
  {Lumsden}, {Oudmaijer}, {Moore}, {Gibb}, {Purcell}, {Burton}, {Mar{\'e}chal},
  {Jiang}, \& {Wang}}]{urquhart_13co_north}
{Urquhart} J.~S. {et~al.}, 2008, \aap, 487, 253

\bibitem[{{Urquhart} {et~al}\mbox{.}(2014{\natexlab{a}}){Urquhart}, {Csengeri},
  {Wyrowski}, {Schuller}, {Bontemps}, {Bronfman}, {Menten}, {Walmsley},
  {Contreras}, {Beuther}, {Wienen}, \& {Linz}}]{urquhart2014_csc}
{Urquhart} J.~S. {et~al.}, 2014{\natexlab{a}}, \aap, 568, A41

\bibitem[{Urquhart {et~al}\mbox{.}(2015)Urquhart, Figura, Moore, Csengeri,
  Lumsden, Pillai, Thompson, Eden, \& Morgan}]{Urquhart2015}
Urquhart J.~S. {et~al.}, 2015, Monthly Notices of the Royal Astronomical
  Society, 452, 4029

\bibitem[{{Urquhart} {et~al}\mbox{.}(2014{\natexlab{b}}){Urquhart}, {Figura},
  {Moore}, {Hoare}, {Lumsden}, {Mottram}, {Thompson}, \&
  {Oudmaijer}}]{urquhart2014_rms}
{Urquhart} J.~S., {Figura} C.~C., {Moore} T.~J.~T., {Hoare} M.~G., {Lumsden}
  S.~L., {Mottram} J.~C., {Thompson} M.~A., {Oudmaijer} R.~D.,
  2014{\natexlab{b}}, \mnras, 437, 1791

\bibitem[{{Urquhart} {et~al}\mbox{.}(2012){Urquhart}, {Hoare}, {Lumsden},
  {Oudmaijer}, {Moore}, {Mottram}, {Cooper}, {Mottram}, \&
  {Rogers}}]{urquhart2012_hiea}
{Urquhart} J.~S. {et~al.}, 2012, \mnras, 420, 1656

\bibitem[{{Urquhart} {et~al}\mbox{.}(2009){Urquhart}, {Hoare}, {Purcell},
  {Lumsden}, {Oudmaijer}, {Moore}, {Busfield}, {Mottram}, \&
  {Davies}}]{urquhart_radio_north}
{Urquhart} J.~S. {et~al.}, 2009, \aap, 501, 539

\bibitem[{Urquhart {et~al}\mbox{.}(2018)Urquhart, K{\"{o}}nig, Giannetti,
  Leurini, Moore, Eden, Pillai, Thompson, Braiding, Burton, Csengeri, Dempsey,
  Figura, Froebrich, Menten, Schuller, Smith, \& Wyrowski}]{Urquhart2018}
Urquhart J.~S. {et~al.}, 2018, MNRAS, 473, 1059

\bibitem[{Urquhart {et~al}\mbox{.}(2014)Urquhart, Moore, Csengeri, Wyrowski,
  Schuller, Hoare, Lumsden, Mottram, Thompson, Menten, Walmsley, Bronfman,
  Pfalzner, K{\"{o}}nig, \& Wienen}]{Urquhart2014}
Urquhart J.~S. {et~al.}, 2014, Monthly Notices of the Royal Astronomical
  Society, 443, 1555

\bibitem[{{Urquhart} {et~al}\mbox{.}(2015){Urquhart}, {Moore}, {Menten},
  {K{\"o}nig}, {Wyrowski}, {Thompson}, {Csengeri}, {Leurini}, \&
  {Eden}}]{urquhart2015_mathanol}
{Urquhart} J.~S. {et~al.}, 2015, \mnras, 446, 3461

\bibitem[{{Urquhart} {et~al}\mbox{.}(2013{\natexlab{a}}){Urquhart}, {Moore},
  {Schuller}, {Wyrowski}, {Menten}, {Thompson}, {Csengeri}, {Walmsley},
  {Bronfman}, \& {K{\"o}nig}}]{urquhart2013_methanol}
{Urquhart} J.~S. {et~al.}, 2013{\natexlab{a}}, \mnras, 431, 1752

\bibitem[{{Urquhart} {et~al}\mbox{.}(2011){Urquhart}, {Morgan}, {Figura},
  {Moore}, {Lumsden}, {Hoare}, {Oudmaijer}, {Mottram}, {Davies}, \&
  {Dunham}}]{urquhart2011_nh3}
{Urquhart} J.~S. {et~al.}, 2011, \mnras, 418, 1689

\bibitem[{Urquhart {et~al}\mbox{.}(2009)Urquhart, Morgan, \&
  Thompson}]{Urquhart2009b}
Urquhart J.~S., Morgan L.~K., Thompson M.~a., 2009, Astronomy and Astrophysics,
  497, 789

\bibitem[{{Urquhart} {et~al}\mbox{.}(2013{\natexlab{b}}){Urquhart}, {Thompson},
  {Moore}, {Purcell}, {Hoare}, {Schuller}, {Wyrowski}, {Csengeri}, {Menten},
  {Lumsden}, {Kurtz}, {Walmsley}, {Bronfman}, {Morgan}, {Eden}, \&
  {Russeil}}]{urquhart2013_cornish}
{Urquhart} J.~S. {et~al.}, 2013{\natexlab{b}}, \mnras, 435, 400

\bibitem[{Urquhart {et~al}\mbox{.}(2007)Urquhart, Thompson, Morgan, Pestalozzi,
  White, \& Muna}]{Urquhart2007b}
Urquhart J.~S., Thompson M.~a., Morgan L.~K., Pestalozzi M.~R., White G.~J.,
  Muna D.~N., 2007, Astronomy and Astrophysics, 467, 1125

\bibitem[{Walmsley \& Ungerechts(1983)}]{Walmsley1983}
Walmsley C.~M., Ungerechts H., 1983, Astronomy {\&} Astrophysics, 122, 164

\bibitem[{Walsh {et~al}\mbox{.}(2011)Walsh, Breen, Britton, Brooks, Burton,
  Cunningham, Green, Harvey-Smith, Hindson, Hoare, Indermuehle, Jones, Lo,
  Longmore, Lowe, Phillips, Purcell, Thompson, Urquhart, Voronkov, White, \&
  Whiting}]{walsh2011}
Walsh A.~J. {et~al.}, 2011, Monthly Notices of the Royal Astronomical Society,
  416, 1764

\bibitem[{{Walsh} {et~al}\mbox{.}(1998){Walsh}, {Burton}, {Hyland}, \&
  {Robinson}}]{Walsh1998}
{Walsh} A.~J., {Burton} M.~G., {Hyland} A.~R., {Robinson} G., 1998, \mnras,
  301, 640

\bibitem[{{Walsh} {et~al}\mbox{.}(2014){Walsh}, {Purcell}, {Longmore}, {Breen},
  {Green}, {Harvey-Smith}, {Jordan}, \& {Macpherson}}]{walsh2014}
{Walsh} A.~J., {Purcell} C.~R., {Longmore} S.~N., {Breen} S.~L., {Green} J.~A.,
  {Harvey-Smith} L., {Jordan} C.~H., {Macpherson} C., 2014, \mnras, 442, 2240

\bibitem[{Wienen {et~al}\mbox{.}(2012)Wienen, Wyrowski, Schuller, Menten,
  Walmsley, Bronfman, \& Motte}]{Wienen2012}
Wienen M., Wyrowski F., Schuller F., Menten K.~M., Walmsley C.~M., Bronfman L.,
  Motte F., 2012, Astronomy {\{}{\&}{\}} Astrophysics, 544, A146

\bibitem[{{Wilson} {et~al}\mbox{.}(2011){Wilson}, {Ferris}, {Axtens}, {Brown},
  {Davis}, {Hampson}, {Leach}, {Roberts}, {Saunders}, {Koribalski}, {Caswell},
  {Lenc}, {Stevens}, {Voronkov}, {Wieringa}, {Brooks}, {Edwards}, {Ekers},
  {Emonts}, {Hindson}, {Johnston}, {Maddison}, {Mahony}, {Malu}, {Massardi},
  {Mao}, {McConnell}, {Norris}, {Schnitzeler}, {Subrahmanyan}, {Urquhart},
  {Thompson}, \& {Wark}}]{wilson2011}
{Wilson} W.~E. {et~al.}, 2011, \mnras, 416, 832

\bibitem[{{Wood} \& {Churchwell}(1989)}]{wood1989a}
{Wood} D.~O.~S., {Churchwell} E., 1989, \apj, 340, 265

\bibitem[{Wyrowski {et~al}\mbox{.}(2015)Wyrowski, G{\"{u}}sten, Menten,
  Wiesemeyer, Csengeri, Heyminck, Klein, K{\"{o}}nig, \&
  Urquhart}]{Wyrowski2015}
Wyrowski F. {et~al.}, 2015, 149, 1

\bibitem[{Zinnecker \& Yorke(2007)}]{Zinnecker2007}
Zinnecker H., Yorke H.~W., 2007, 45, 1

\end{thebibliography}


%%%%%%%%%%%%%%%%%%%%%%%%%%%%%%%%%%%%%%%%%%%%%%%%%%

%%%%%%%%%%%%%%%%% APPENDICES %%%%%%%%%%%%%%%%%%%%%]
%\clearpage

%\appendix

%\input{appendix.tex}

%%%%%%%%%%%%%%%%%%%%%%%%%%%%%%%%%%%%%%%%%%%%%%%%%%

% Don't change these lines
\bsp	% typesetting comment
\label{lastpage}
\end{document}